\documentclass[a4paper,12pt]{article}

\usepackage{epsfig,graphicx,amsmath,amssymb}
\usepackage{url}
\usepackage{hyperref}
\usepackage{subcaption}
\usepackage{siunitx,mathtools}
\usepackage{mhchem,booktabs}
\usepackage{relsize}
\RequirePackage{xspace}
\newcommand{\gev}{\ensuremath{\mathrm{\,Ge\kern -0.1em V}}\xspace}
\def\babar{\mbox{\slshape B\kern-0.1em{\smaller A}\kern-0.1em
    B\kern-0.1em{\smaller A\kern-0.2em R}}}
\def\KS    {\ensuremath{K^0_{\scriptscriptstyle S}}\xspace}
\def\KL    {\ensuremath{K^0_{\scriptscriptstyle L}}\xspace}
\def\piz   {\ensuremath{\pi^0}\xspace}
\def\pipi  {\ensuremath{\pi^+\pi^-}\xspace}
\def\en         {\ensuremath{e^-}\xspace}
\def\ep         {\ensuremath{e^+}\xspace}
\def\Y#1S{\ensuremath{\Upsilon{(#1S)}}\xspace}
\def\FourS {\Y4S}
\def\gev{\mathrm{Ge\kern -0.1em V}}

\newcommand{\ee}{e^{+} e^{-}}
\def\amu{a_\mu}

\def \ee {e^+e^-}

\newcommand{\pip}{\pi^{+}}
\newcommand{\pim}{\pi^{-}}

\newcommand{\bc}{\begin{center}}
\newcommand{\ec}{\end{center}}

\DeclareSIUnit[number-unit-product = \,]{\promille}{\textperthousand}
\DeclareSIUnit\ch{ch}
\newlength{\mylinewidth}
\begin{document}

\thispagestyle{empty}

$\phantom{.}$

\begin{flushright}
{\sf  MITP/14-045 \\
  } 
\end{flushright}

\hfill

\begin{center}
{\Large {\bf Hadronic contributions to the muon anomalous magnetic moment: strategies for improvements of the accuracy of the theoretical prediction}} \\

\vspace{0.5cm}

{\large April 1--5, 2014 in Waldthausen Castle, Mainz, Germany}\\
\vspace{1cm}
AND
\vspace{1cm}
\begin{center}
{\Large {\bf $(g-2)_{\mu}$: Quo vadis?}} \\
\vspace{0.5cm}
{\large April 7--10, 2014 in Mainz, Germany}\\
\vspace{1cm}
{\large Mini Proceedings}
\vspace{2cm}

{\it Editors:}
Tom Blum$^1$, Pere Masjuan$^2$, and Marc Vanderhaeghen$^2$\\
\vspace{0.5cm}
$^1$ Physics Department, University of Connecticut, Storrs, CT 06269-3046, USA\\  
$^2$PRISMA Cluster of Excellence, Institut f\"ur Kernphysik, Johannes Gutenberg-Universit\"at Mainz, Germany\\
\end{center}

\vspace{1.5cm}

ABSTRACT

\end{center}

\vspace{0.3cm}

\noindent
We present the mini-proceedings of the workshops \emph{Hadronic contributions to the muon anomalous magnetic moment: strategies for improvements of the accuracy of the theoretical prediction} and \emph{$(g-2)_{\mu}$: Quo vadis?} held in Mainz from April 1$^{\rm rst}$ to 5$^{\rm th}$ and from April 7$^{\rm th}$ to 10$^{\rm th}$, 2014, respectively.

\medskip\noindent
The web page of the conferences, which contains all talks, can be found at
\begin{itemize}
\item \emph{Hadronic contributions to the muon anomalous magnetic moment}: \url{https://indico.mitp.uni-mainz.de/conferenceDisplay.py?confId=13}
\item \emph{$(g-2)_{\mu}$: Quo vadis?:} \url{https://indico.cern.ch/event/284012/}
\end{itemize}

\vspace{0.5cm}

\newpage

{$\phantom{=}$}

\vspace{0.5cm}

\tableofcontents

\newpage

\section{Introduction to the Workshops}

\addtocontents{toc}{\hspace{1cm}{\sl }\par}

\vspace{5mm}

\noindent
Tom Blum$^1$, Achim Denig$^2$, Simon Eidelman$^3$, Fred Jegerlehner$^4$, Pere Masjuan$^2$, Dominik  St\"ockinger$^5$ and Marc Vanderhaeghen$^2$
\vspace{0.5cm}

\noindent
$^1$ Physics Department, University of Connecticut, Storrs, CT 06269-3046, USA\\  
$^2$ PRISMA Cluster of Excellence, Institut f\"ur Kernphysik, Johannes Gutenberg-Universit\"at Mainz, Germany\\
$^3$ Budker Institute of Nuclear Physics SB RAS and Novosibirsk State University, Novosibirsk, Russia\\
$^4$ Humboldt-Universit\"at zu Berlin, Institut f\"ur Physik, Berlin and Deutsches Elektronen-Synchrotron (DESY), Zeuthen, Germany\\
$^5$ Institut f\"ur Kern und Teilchenphysik, TU Dresden, Dresden, D-01062, Germany\\

\vspace{5mm}

The muon anomalous magnetic moment is one of the most precisely measured observables in particle physics, which at the same time can be predicted by theory with the required accuracy. It is simultaneously a monitor for in depth testing the Standard Model as well as for finding deviations from it caused by unknown physics.

The aim of the Workshop \emph{Hadronic contributions to the muon anomalous magnetic moment: strategies for improvements of the accuracy of the theoretical prediction} was to gather leading experts as well as new faces having new ideas to work on the improvement of the predictions of the hadronic contributions to the muon anomalous magnetic moment, in particular on the challenging hadronic light-by-light (HLbL) scattering problem. The goal was to trigger new activities which should lead to the theoretical improvements required for a successful confrontation of theory and experiment once the results from the next generation of experiments at Fermilab and at J-PARC are available. With such purpose, the Workshop contained plenary talks during the morning sessions and discussion talks during the afternoons in the format of four working groups (Lattice QCD, Experimental input, Hadronic models, and Others). 

The topics covered by the working groups were:
\begin{itemize}
\item Perspectives for reducing the hadronic vacuum polarization (HVP) error by new cross-section measurements (Novosibirsk, Frascati, Beijing, Belle, BaBar). Theory issues here are the necessary radiative corrections calculations required for the extraction of the cross sections from the experimental data.
\item Exploiting low-energy effective theories in conjunction with experimental data (including hadron production in gamma gamma physics) as required for the calculation of the HLbL contribution or for including tau-decay spectra and pi-pi scattering phase shifts to improve the HVP contribution. New developments on dispersion relation approaches aimed at a data driven approach of the HLbL contribution were presented. General theory tools, resonance Lagrangian approach, Schwinger-Dyson approach etc.
\item Perspectives for improvements in lattice QCD calculations of the HVP and HLbL contributions. Participants discussed how to reduce systematic errors in the HVP contribution to $a_\mu^{\rm HLO}$, including fitting and related systematics arising in the very low momentum regime, quark-loop disconnected diagrams, and charm quark contributions with an aim towards $1\%$ accuracy in lattice calculations. New methods based on taking moments of correlation functions and computing in the time-like region were also discussed. The status and prospects for HLbL lattice calculations were also reviewed.
\end{itemize}

The web page of the conference, which contains all talks, can be found at
\begin{center}
\url{https://indico.mitp.uni-mainz.de/conferenceDisplay.py?confId=13}.
\end{center}

\vspace{1cm}

The theory workshop was followed by the Workshop \emph{$(g-2)$: Quo Vadis?} of the Mainz Collaborative Research Center SFB-1044 ``The Low-Energy Frontier of the Standard Model".

The goal of this second workshop was to review recent developments in experiment and theory regarding the anomalous magnetic moment of the muon. Main topics included the future direct measurements of the muon anomaly at FNAL and JPARC, measurements of the $\ee$ hadronic cross section as well as transition form factors.

The web page of the conference, which contains all talks, can be found at
\begin{center}
\url{https://indico.cern.ch/event/284012/}.
\end{center}

This workshop was followed by the 15$^{th}$ RadioMonteCarLOW-Meeting \emph{Radiative Corrections and Generators for Low Energy Hadronic Cross Section and Luminosity}, [\url{https://agenda.infn.it/conferenceDisplay.py?confId=7800}]~\cite{vanderBij:2014mxa}.

Related workshops under the SFB-1044 Collaborative Research Center can be found in Refs.~\cite{Czyz:2013zga,Bijnens:2014fya}.

Both workshops where held in Mainz, the theoretical one in the Waldthausen Castle, and the experimental one in the campus of the Johannes Gutenberg-Universit\"at Mainz, during the first and second weeks of April, 2014, respectively, enjoying the hospitality from Institute of Nuclear Physics, Mainz, Germany.

The present document contains the mini-proceedings of both conferences, chapters 2 and 3 respectively, with 20+17 oral presentations.

We acknowledge the support of the PRISMA Cluster of Excellence, the Mainz Institute for Theoretical Physics MITP, and the Deutsche Forschungsgemeinschaft DFG through the Collaborative Research Center ``The Low-Energy Frontier of the Standard Model" (SFB 1044).

\vspace{5mm}
\noindent
This work is a part of the activity of the MITP:
\begin{center}
[\url{http://www.mitp.uni-mainz.de}]
\end{center}
and part of the activity of the SFB 1044:
\begin{center}
[\url{http://sfb1044.kph.uni-mainz.de/sfb1044/}]
\end{center}

\newpage

\section{Summaries of the talks \emph{Hadronic contributions to the muon anomalous magnetic moment} Workshop}

\subsection{Effective Lagrangian approach to estimating the hadronic vacuum polarization contribution to the muon $g-2$ }
\addtocontents{toc}{\hspace{2cm}{\sl M.~Benayoun}\par}

\vspace{5mm}

M.~Benayoun

\vspace{5mm}

\noindent
LPNHE des Universit\'es Paris VI et VII, CRRS-IN2P3, Paris, France\\

\vspace{5mm}

The anomalous magnetic moment of the muon $a_\mu$ is a physics quantity
measured with a very high accuracy ($\amu^{EXP}=116 592 08.9(6.3)\times 10^{-10}$)
\cite{Bennett:2006fi}. New experiments are foreseen in a
near future which should improve its precision by a factor of 4. Most of its
ingredients are theoretically known with a precision of a few $10^{-11}$
or better, except for the photon hadronic vacuum polarization (HVP)
and the hadronic light--by--light (HLbL) contributions which each carries 
an uncertainty of the order of $4\times 10^{-10}$. This leads to a discrepancy 
between the measured and the predicted values for $a_\mu$
ranging between 3 and 4 $\sigma$ -- depending on estimation method  
(among recent studies, see for instance,
\cite{Jegerlehner:2009rya,Hagiwara:2011af1}).

 Unlike the  HLbL contribution, HVP is related with the
 annihilation cross sections $e^+e^- \rightarrow $ hadrons through
 an integral containing a kernel which sharply enhances the very
 low energy region. It is the reason why the contribution of the 
 annihilation channels $e^+e^- \rightarrow \pi^+ \pi^- /\pi^+ \pi^- \pi^0/
 K^+ K^-/ K_L K_S/\pi^0 \gamma /\eta \gamma $ from their thresholds up 1.05
 GeV (including the $\phi(1020)$ region) represents more than 80 \% of the
 total HVP; this region also contributes an important amount to the 
 theoretical uncertainty on  $a_\mu$.  It is therefore of special concern to
 find methods allowing an improved knowledge of these contributions to
 the muon $g-2$ in order to guaranty at best their central values and their
 uncertainties.
 
 The model described in \cite{BDDJ} relies on the Effective Lagrangian named
Hidden Local Symmetry Model (HLS) as can be found in \cite{HLS}. This
model, appropriately broken,  has been proved to provide a quite
successful $simultaneous$ fit of the existing data samples covering the 
annihilation channels quoted above \cite{BDDJ}. Within this framework, a 
limited number (3) of the ($\simeq 40-45$) experimental spectra 
have been shown to exhibit -- within the global fit -- a behavior
in contradiction with the other data (covering the 6 channels involved)
 and, consequently, should be discarded for consistency.

The most
important result derived is an improvement by the a factor $\simeq 2$
of the uncertainty on the HVP integrated up to 1.05 GeV. Additionally,
the central value for the HVP is slightly reduced, exhibiting the
influence of the recently produced KLOE dipion spectra. 
All this sums up
into an estimate of the muon $g-2$ discrepancy ranging in $ (4 \div 4.5 )\sigma$.

\newpage

\subsection{Hadronic light-by-light: the (resonance) Lagrangian approach}
\addtocontents{toc}{\hspace{2cm}{\sl J.~Bijnens}\par}

\vspace{5mm}

J.~Bijnens

\vspace{5mm}

\noindent
Department of Astronomy and Theoretical Physics, Lund University, Sweden\\
\vspace{5mm}

The main theoretical uncertainty in the future for the muon anomaly,
 $a_\mu=(g-2)/2$,  will be hadronic light-by-light (HLbL).
Most summaries give values in the range 10--14 and errors in 2.6--4 in units of $10^{-10}$.

The underlying hadronic object is the Green function of four
electromagnetic currents, integrated over two photon momenta 
and the third ($p_3$) set to zero after taking the derivative
$\partial/\partial p_{3\mu}$.
This object has 138 Lorentz-structures. 28 contribute to $a_\mu$.
They are a function of the off-shellness
or mass of the three photons connecting to the muon line.
The mixing of long and short distances turns possible double counting of
quark-gluon and hadron contributions into a difficult problem.
Using $N_c$ and chiral counting
as a guide \cite{deRafael}, two groups did a full estimate with similar final
numbers \cite{HKS,BPP}. A sign mistake was found in both by
\cite{KN} and the main contribution,
$\pi^0,\eta,\eta^\prime$-exchange, has been recalculated many times with all
results fitting in the range (8--10)$\,10^{-10}$, for references see
\cite{PdRV,BP,JNreview}. The short distance constraint found in \cite{MV}
increased the result. Recent additions are
the resonance chiral theory estimates \cite{KNprague,RGL}, but note
that models with a finite number of states
need to compromise between QCD constraints \cite{BGLP}.

My own new results discussed concern the pion loop contribution.
The models used earlier were the hidden local symmetry (HLS) model
and the ENJL model where all photon propagators are modulated with a factor
resembling $m_V^2/(m_V^2+Q^2)$. The full VMD model uses exactly that factor.
These gave $-0.45\cdot\,10^{-10}$,$-1.9\cdot\,10^{-10}$, $-1.6\cdot\,10^{-10}$ respectively.
The large difference between the first and latter two is disturbing.
In \cite{talk,Mehranthesis,MesonNet13,BR}
this was studied further. The HLS model has contributions of the opposite
sign at higher photon masses. \cite{Ramsey-Musolf1,Ramsey-Musolf2}
suggested that pion polarizability
effects might be important. Pure ChPT can only be used here up to
500 MeV or so, so to fully study the effect models with an $a_1$ are needed.
Even with including many more couplings,
no satisfying model with the $a_1$ that gives a finite result
for the muon $g-2$ was found \cite{BR}. However all models that gave a
reasonable low-energy behaviour, when integrated up to about 1~GeV gave similar
answers: $a_\mu^{LbL\,\pi\mathrm{-loop}} = (-2.0\pm0.5)\,10^{-10}$
is the new preliminary result for this contribution.

Plots showing the contributions at the different values of the photon masses
as introduced in \cite{BP} can be extremely useful in comparing different
estimates of the various contributions.

\newpage

\subsection{Two-pion low-energy contribution to the muon $g-2$ 
with improved precision from analyticity and unitarity }
\addtocontents{toc}{\hspace{2cm}{\sl  I.~Caprini}\par}

\vspace{4mm}

 I.~Caprini

\vspace{4mm}

\noindent
Institute for Physics and Nuclear Engineering\\
P.O.B. MG-6, 077125 Bucharest-Magurele, Romania
\vspace{4mm}

The hadronic vacuum polarization contribution to the muon  $g-2$ is dominated by the two-pion channel, which is expressed to Leading Order (LO) in terms of the modulus $|F(t)|$ of the pion electromagnetic form factor. The low-energy contribution has a relative large uncertainty, since the experimental data in this region have large errors, which are further enhanced by the kernel of the $a_\mu$  integral \cite{Davier:2009}. 
In the present talk, based on \cite{ Caprini:2014}, I present an attempt to improve the precision by exploiting analyticity and unitarity.

 Since the modulus of the form  factor is poorly known at low energies, we use instead the phase, known with high precision below the $\omega\pi$ inelastic threshold from Fermi-Watson theorem and Roy equations for $\pi\pi$ scattering.  Above the inelastic threshold, where the phase is not known, we use measurements of the modulus  by BABAR  experiment  \cite{BABAR} up to 3 GeV, together with weak assumptions about the asymptotic behaviour, expressed as an integral condition on the modulus squared. We use also several values of the modulus from the region 0.65 - 0.70 GeV, measured with higher precision by the  $e^+e^-$ experiments 
SND \cite{SND},  CMD2 \cite{CMD22}, BABAR \cite{BABAR} and  KLOE \cite{KLOE3}. The input is introduced in a suitable functional extremal problem of Meiman type, which leads to upper and lower bounds on the modulus at other energies. 
It can be shown rigorously \cite{Abbas:2010EPJA} that the bounds are optimal  for a given input, do not depend on the unknown phase of $F(t)$  above the inelastic threshold and satisfy a monotonicity property, which allows a good control of the errors. 

\begin{figure}[htb]\vspace{0.4cm}
\begin{center}
\includegraphics[width = 7.2cm]{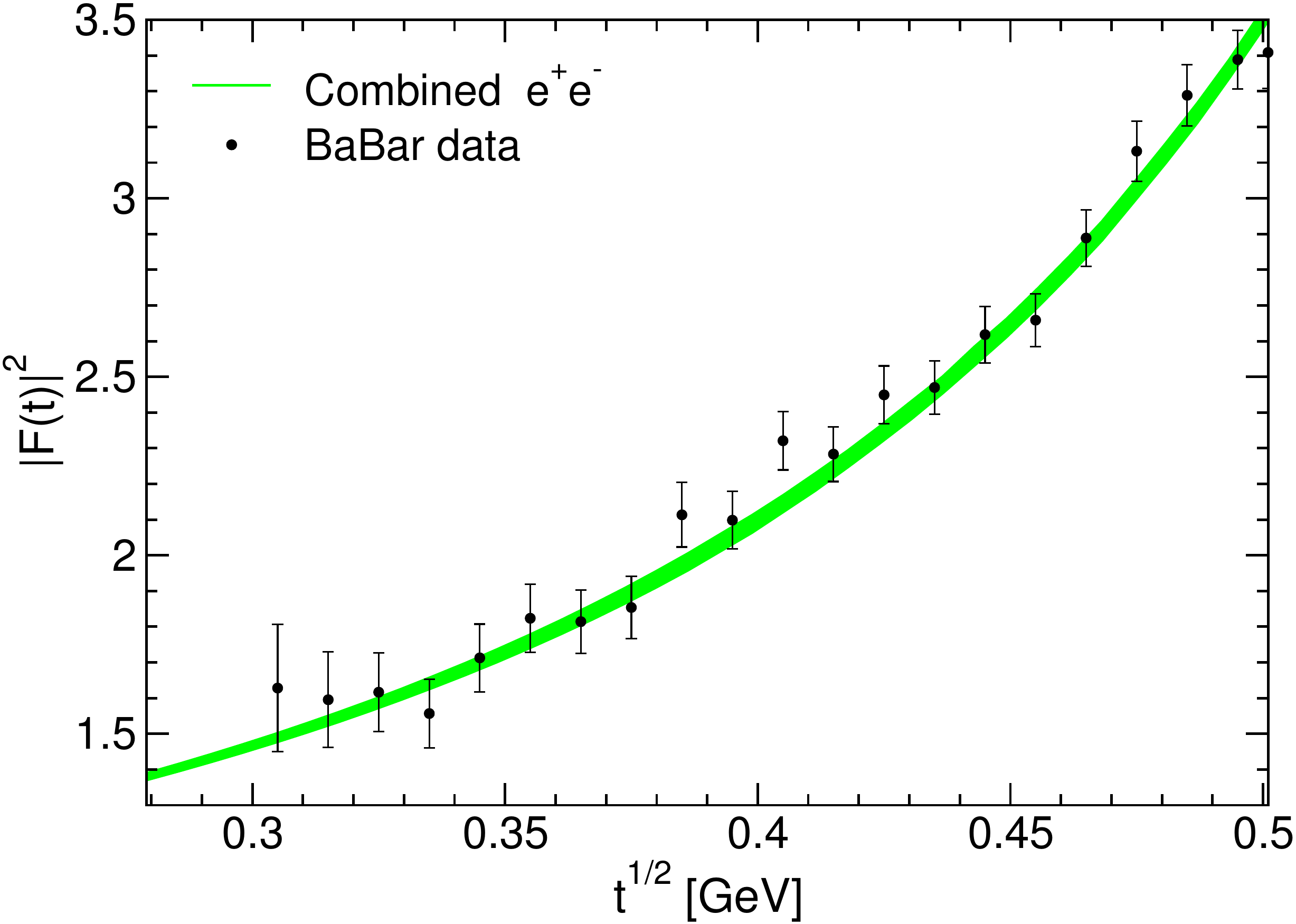}\hspace{0.3cm} \includegraphics[width =7.2cm]{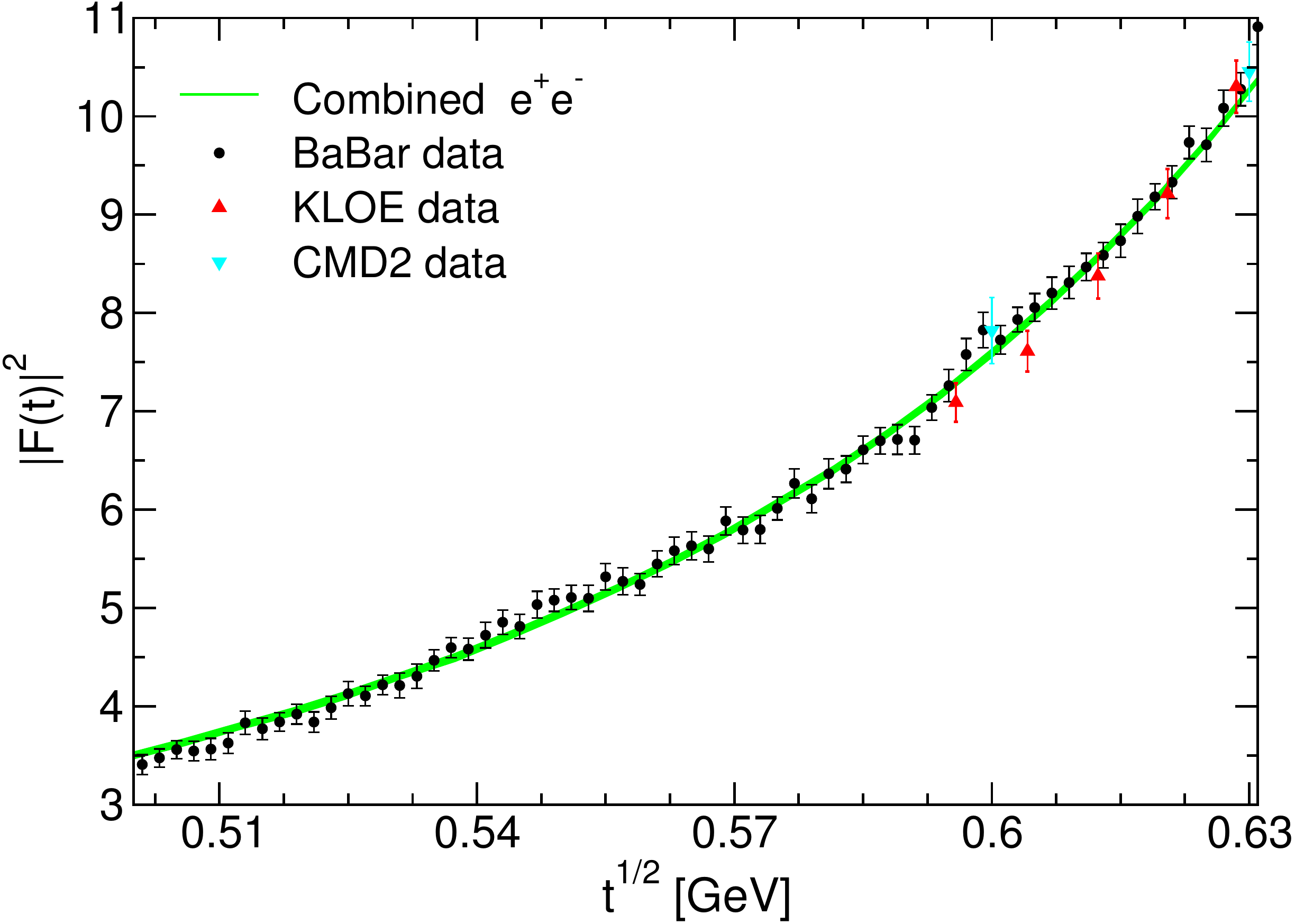}
\end{center}
\end{figure}

 The method can be viewed therefore as a parametrization-free analytic extrapolation from higher energies to the low energy region of interest for the improved calculation of $a_\mu$. The above Figure shows the allowed band for the modulus squared values below 0.63 GeV, obtained by combining the upper and lower bounds derived using the modulus measured at higher energies by the $e^+e^-$ experiments \cite{BABAR, SND, CMD2, KLOE3}. The precision of the extrapolation appears to be slightly better than that of the  direct measurements performed at low energies.

We expect therefore a better accuracy of the low energy contribution to $a_\mu$  calculated with the extrapolated values. We consider the quantity
\begin{equation}
a_\mu^{\pi\pi, {\rm LO}} [\sqrt{t_l}, \sqrt{t_u}] = \frac{\alpha^2 m_\mu^2}{12 \pi^2}\int_{t_l}^{t_u} \frac{dt}{ t}  \, K(t)\, \beta^3_\pi(t) \,   
|F(t)|^2 \left(1+\frac{\alpha}{\pi}\,\eta_\pi(t)\right),
\end{equation}
where $\beta_\pi(t)=(1- 4 m_\pi^2/t)^{1/2}$,  $K(t) = \int_0^1 du(1-u)u^2(t-u+m_\mu^2u^2 )^{-1}$ 
and the last factor  accounts
for the final state radiation (FSR) calculated in scalar QED. The form factor modulus $|F(t)|$ does not include vacuum polarization. As the formalism described above is valid in the isospin limit, we removed from the input modulus the main  isospin breaking factor occuring in this channel, due to $\omega-\rho$ interference. The factor was finally reintroduced in the upper and lower bounds used in the calculation of $a_\mu$.

By combining the results obtained with input from  $e^+e^-$ experiments SND,  CMD2, BABAR and  KLOE,  we obtained for the contribution to $a_\mu$ of the energies from 0.30 to  0.63 GeV  the prediction
\begin{equation}
a_\mu^{\pi\pi, {\rm LO}}\,[0.30 \,\gev,\, 0.63\, \gev]=(132.673 \pm 0.866) \times 10^{-10},
\end{equation}
to be compared with the value $ (132.6 \pm 1.3) \times 10^{-10}$, derived in \cite{Davier:2009} for the same quantity by direct integration of the  $e^+e^-\to \pi^+\pi^-$  cross sections.  A slight improvement of the precision of the $a_\mu$ determination was therefore obtained. The method can be applied also to  $\tau$ decays and for testing the consistency of different data sets with analyticity and unitarity. 

\vspace{0.2cm}\noindent {\bf Acknowledgements:}  I thank the Workshop  organizers for the invitation and  the Mainz Institute for Theoretical Physics (MITP) for hospitality and support.

\newpage

\subsection{Dispersive approach to hadronic light-by-light}
\addtocontents{toc}{\hspace{2cm}{\sl G.~Colangelo }\par}

\vspace{5mm}

\underline{G.~Colangelo}, M.~Hoferichter, M.~ Procura,
P.~Stoffer

\vspace{5mm}

\noindent
Albert Einstein Center for Fundamental Physics,
Institute for Theoretical Physics, \\
University of Bern, Sidlerstrasse 5, CH--3012 Bern, Switzerland
\vspace{5mm}
\setcounter{equation}{0}

In this talk I have presented a dispersive approach to hadronic
light-by-light which has been recently proposed in
\cite{Colangelo:2014dfa}. This approach aims to take into account only the
cuts in the hadronic tensor which are due to single- or double-pion
intermediate states -- this approximation is justified by the fact that
in explicit calculations higher-lying singularities (like the one due to
two kaons) give small contributions \cite{BPP95}. Further, we split the
hadronic tensor as follows:
\begin{equation}
\Pi_{\mu\nu\lambda\sigma}=\Pi_{\mu\nu\lambda\sigma}^{\pi^0\mathrm{-pole}}
+\Pi_{\mu\nu\lambda\sigma}^{\mathrm{FsQED}}+\bar \Pi_{\mu\nu\lambda\sigma}+\cdots,
\end{equation}
where the first term takes into account the one-pion pole, the second one
two-pion intermediate states with simultaneous cuts in the $s$ and $t$
channel (and all possible cyclic permutations including $u$), and the third
one is the one for which we write down a dispersion relation.

A central result of our analysis is that after separating
$\Pi_{\mu\nu\lambda\sigma}^{\pi^0\mathrm{-pole}}$ and
$\Pi_{\mu\nu\lambda\sigma}^{\mathrm{FsQED}}$ from the rest, we have been able
to derive explicit unitarity relations for the remainder $\bar
\Pi_{\mu\nu\lambda\sigma}$ and relate the imaginary parts to the helicity
amplitudes for $\gamma^* \gamma^* \to \pi \pi$. We then write down the
corresponding dispersion relations. In a properly chosen basis for the
Lorentz structure and thanks to the separation of
$\Pi_{\mu\nu\lambda\sigma}^{\mathrm{FsQED}}$, which has a double-spectral
region~\cite{Mandelstam}, the dispersion relations we derive for the scalar
functions are in a single Mandelstam variable. Our representation for $\bar 
\Pi_{\mu\nu\lambda\sigma}$ can be viewed as a generalization of the
reconstruction theorem~\cite{Stern:1993rg} originally derived for the $\pi
\pi$ scattering amplitude to the hadronic light-by-light tensor.

On the basis of this dispersive representation we have then expressed the
hadronic light-by-light contribution to $(g-2)_\mu$ as integrals over the
dispersive integrals discussed above. This will allow an evaluation of this
contribution based on measured helicity amplitudes of the subprocess
$\gamma^* \gamma^* \to \pi \pi$. Further work in this direction, and
towards a numerical evaluation of these contributions, is in progress.

\newpage

\subsection{Charm-quark contribution to g-2 of the muon: An entirely theoretical QCD calculation}
\addtocontents{toc}{\hspace{2cm}{\sl C.A.~Dominguez}\par}

\vspace{5mm}

C.A.~Dominguez

\vspace{5mm}
\noindent
Centre for Theoretical Physics and Astrophysics and Department of Physics, University of
Cape Town, Rondebosch 7700, South Africa
\vspace{5mm}

In recent work the hadronic contribution to g-2 of the muon was determined  by using Cauchy's theorem in the complex squared energy plane, and (i) using hadronic models for the low energy contribution \cite{1}, and (ii) using the operator product expansion (OPE) of current correlators at short distances \cite{2}. In the latter, a purely theoretical QCD calculation of the heavy-quark contribution to g-2 was performed for the first time. Subsequently, lattice QCD determinations \cite{3}-\cite{4} found excellent agreement with the charm-quark result. A sketch of this determination follows. The first step is to fit the kernel $K(s)$, entering the integral expression of $a_\mu$, with a function $K_2(s) = a_1/s + a_2/s^2$ in the heavy-quark region, with $a_1= 0.003712 \; {\mbox{GeV}}^2$, $a_2= - 0.0005122 \; {\mbox{GeV}}^4$ for charm, and $a_1= 0.003719 \; {\mbox{GeV}}^2$, $a_2= - 0.0007637 \; {\mbox{GeV}}^4$ for bottom. This parametrization is done in order to be able to use Cauchy's theorem as follows
\begin{eqnarray}
\oint_{|s|=s_0} \frac{ds}{s} \; K_{2}(s) \; \Pi_{Q}(s)|_{PQCD}
&+& 2\,i 
\int_{s_{th}}^{s_0} \frac{ds}{s}  K_{2}(s) \, \mbox{Im} \,\Pi_{Q}(s)|_{HAD} \nonumber \\
 &=&  2\, \pi\, i \; \mbox{Res} \;\left[ \frac{K_{2}(s)}{s} \, \Pi_{Q}(s) \right]_{s=0} \nonumber\; , 
\end{eqnarray}
where the residues can be computed using the well known QCD low energy expansion of the correlator, and the contour integral over the circle of radius $|s_0|$ is computed using the high energy QCD expansion, thus leading to
\begin{equation}
a_\mu^{HAD}|_c = 14.41 (1) \,\times\, 10^{-10}, \;\;\;\;
a_\mu^{HAD}|_b = 0.29 (1) \,\times\, 10^{-10} \nonumber \;.
\end{equation}
This result agrees quite well with the lattice QCD determinations $a_\mu^{HAD}|_c = 14.1(6)  \,\times\, 10^{-10}$ \cite{3}, and  $a_\mu^{HAD}|_c = 14.42(39)  \,\times\, 10^{-10}$ \cite{4}.

\newpage

\subsection{Hadronic vacuum polarization (experiment)}
\addtocontents{toc}{\hspace{2cm}{\sl S.~Eidelman}\par}

\vspace{5mm}

S.~Eidelman

\vspace{5mm}
\noindent
Budker Institute of Nuclear Physics SB RAS and  \\
Novosibirsk State University, Novosibirsk, Russia 
\vspace{5mm}
 Experiments at various $e^+e^-$ colliders are continuing to provide 
information on single-photon annihilation into hadrons that remains a
main source of data needed to calculate the leading-order hadronic 
contribution to the muon $(g-2)$. While BaBar~\cite{babar}, 
Belle~\cite{bellea} and KLOE~\cite{kloe} are using initial-state 
radiation~\cite{ISRe} to measure exclusive cross sections, three experiments in
Novosibirsk (CMD-3, SND and KEDR) and BESIII in Beijing are using
a scan of the accessible energy range to measure hadronic  $e^+e^-$
annihilation.
  
Two detectors, CMD-3 and SND, are now operated 
at the VEPP-2000 $e^+e^-$ collider at BINP, Novosibirsk with a goal of
high-precision measurements of exclusive multihadronic 
cross sections~\cite{fedor}. In 2011-2013 both detectors collected
data samples of about 60 pb$^{-1}$ each in the center-of-mass energy
range from 0.32 GeV to 2.0 GeV. Analysis is in progress and first results 
on $e^+e^- \to \omega\pi^0 \to \pi^0\pi^0\gamma$ from SND~\cite{snde} 
and  $e^+e^- \to 3\pi^+3\pi^-$ from CMD-3~\cite{cmd3}
have already been published with many more expected by
summer conferences. The statistics already achieved for most of the
channels including  $e^+e^- \to \pi^+\pi^-$ are comparable to or
better than those achieved with ISR at BaBar. After reaching the
designed luminosity of $10^{32}~{\rm cm}^{-2}{\rm s}^{-1}$  one hopes
to collect about 1 fb$^{-1}$ between 1 and 2 GeV.

The data from 2 to 5 GeV are dominated by the previous BES 
measurements~\cite{ebes1,ebes2,ebes3,ebes4} with a typical accuracy 
of (3-5)\% (statistical) and (5-8)\% (systematic). It can be improved
by BESIII, which recently completed a fine scan with ~100 energy points 
between 3.8 and 4.6 GeV, each with ~8 pb$^{-1}$ of data. Finally, KEDR 
at the VEPP-4M collider in Novosibirsk
is completing data processing for an experiment at eight energy points
between $J/\psi$ and $\psi(2S)$. They selected about 2000 events per point
and hope to have a 4\% systematic error. With additional 2 pb$^{-1}$ planned
to be collected in 2014 the total uncertainty of 3\% 
can be achieved.

This work is supported by the Ministry of Education and Science of the
Russian Federation, the RFBR grants 12-02-01032, 13-02-00215   and 
the DFG grant HA 1457/9-1.

\newpage

\subsection{Hadronic light-by-light from Dyson--Schwinger equations }
\addtocontents{toc}{\hspace{2cm}{\sl C.~S. Fischer}\par}

\vspace{5mm}

 \underline{C.~S. Fischer}, R.~Williams

\vspace{5mm}

\noindent
Institute of Theoretical Physics, Justus-Liebig University of Giessen, \\ 
Heinrich-Buff-Ring 16, 35392, Giessen, Germany\\

\vspace{5mm}

\bigskip\noindent{\bf Dyson--Schwinger equations}\\
The Dyson--Schwinger equations (DSEs) furnish relations between the Green's functions of QED and QCD, in which all (non)-perturbative information is contained.
Being exact, symmetries such as EM gauge-invariance and chiral symmetry are encoded; further constraints can be elucidated by, e.g., the (axial)-vector Ward--Takahashi identites. Mesons and baryons appear dynamically as $s$-channel poles in $4$- and $6$-point functions and can be described covariantly by Bethe--Salpeter and Faddeev type of equations.  Phenomena such as dynamical chiral symmetry breaking and the pseudo-Goldstone nature of the pion are manifest.

This infinite tower of Green's functions must be truncated, wherein higher $n$-point functions are provided by Ansatz in accordance to symmetries. 
The simplest viable truncation is that of Rainbow-Ladder (RL).
Though constructed phenomenologically, it has been tested for a wide range of meson and baryon observables and is found to be remarkably effective, especially in the context of electromagnetic processes.

\bigskip\noindent{\bf Hadronic vacuum polarisation (HVP)}\\
Applying the DSEs to the hadronic vacuum polarisation contribution to the muon $g-2$ requires evaluation of the photon polarization tensor in terms of the dressed quark propagator and quark-photon vertex. These are calculated from their own DSEs with the quark-photon vertex dynamically generating vector meson poles at time-like momenta.
At the space-like momenta relevant for HVP the off-shell behaviour of the vertex is
treated correctly as can be inferred e.g. by the agreement of the pion-electromagnetic 
form factor with experiment. For HVP we find
$a_\mu^{\textrm{(LO)HVP}}=676(34)\times 10^{-10}$ with individual contributions 
$600(30)$, $60(3)$, $15.0(0.8)$ and $1.0(0.1)$ for the $u/d$, $s$, $c$ and $b$ 
quark flavours, respectively \cite{Goecke:2011pe}. The behaviour of HVP for non-physical quark masses agrees
with corresponding lattice calculations \cite{Goecke:2013fpa}.
%

\bigskip\noindent{\bf Hadronic light-by-light scattering (HLBL)}\\
The DSE for the photon four-point function can be derived from gauge invariance, consistent with the RL truncation.
Diagrams are arranged by the topology of their resummations into the quark-loop and $T$-matrix contributions; short-distance constrains apply to the sum of these. Resonant
expansion of the $T$-matrix suggests the pseudoscalar exchange dominates; this must be continued off-shell. The DSE formalism gives \cite{Fischer:2010iz,Goecke:2010if} 
$a_\mu =8.1 (1.2) \times 10^{-10}$, comparable to other approaches. Consistency with 
gauge-invariance and off-shell momenta will be achieved 
by a future calculation involving the full $T$-matrix.

This calculation should be accompanied by a complete DSE calculation of the quark-loop
contribution, which is work in progress. Our results from a partial calculation, however,
already show an important effect \cite{Goecke:2012qm}: as a multi-scale hadronic problem, 
the clear separation 
into short and long distance physics in HLBL is obviated. This leads to potentially large 
uncertainties when momentum dependences are neglected, as is the case in the model 
calculations performed previously. The DSE-approach, taking all momentum dependencies 
into account therefore has the potential to overcome these shortcomings. In the light of our
findings, the error estimates of previous approaches should be treated with caution.

\newpage

\subsection{Hadronic vacuum polarization: Initial state radiation results at flavor factories}
\addtocontents{toc}{\hspace{2cm}{\sl A.~Hafner}\par}

\vspace{5mm}

A.~Hafner

\vspace{5mm}

\noindent
Institut f\"ur Kernphysik, Johannes Gutenberg Universit\"at Mainz, \\
          J.-J.-Becher-Weg 45, 55099 Mainz, Germany.

\vspace{5mm}

The uncertainty on the theoretical prediction of the anomalous magnetic moment of the muon $\amu^{theo}$ is dominated by two contributions: The hadronic Vacuum Polarization (VP), $a_\mu^{VP}=(692.3 \pm 4.2)\cdot 10^{-10}$~\cite{davier}, and the hadronic Light-by-Light terms, $a_\mu^{LbL}=(10.5 \pm 2.6)\cdot 10^{-10}$~\cite{prades}. This note is focused on the hadronic VP term, the largest contribution to the uncertainty of $\amu^{theo}$.

From causality and analyticity of the VP amplitude a dispersion relation for the VP contribution to $\amu^{theo}$ can be derived~\cite{brodsky}. This relation requires the inclusive hadronic cross section as input. The largest weight is given to low energy contributions, and thus the region below $2\gev$ dominates the contribution and the uncertainty to the hadronic contribution of $\amu$. 

The standard experimental approach is to measure the required hadronic cross sections exclusively at $e^+e^-$ energy scan experiments. Since the last decade, the method of Initial State Radiation (ISR) is used as an alternative approach to measure cross sections of exclusive final states at high luminosity flavor factories, running at a fixed center-of-mass energy. The emmitance of a high energy photon from initial state opens the window to low energy hadron physics. KLOE, running on the $\phi$ resonance, measured the $\ep\en\to\pipi$ final state~\cite{kloe:all} with a precision of better than $1\%$ in the peak region. \babar, running on the \FourS resonance, has an extensive ISR-scan program with various final states up to six hadrons from energy threshold up  to $4.5\gev$~\cite{babar:data}. The \babar\ measurement of the $\pipi$ final state shows a discrepancy in and above the $\rho$ region of up to 2-3 standard deviations to the KLOE measurement. Due to this difference, the resulting uncertainty for $\amu^{theo}$ is similar to the uncertainties of the individual measurements.

Energy scan measurements of the $\pipi$ cross section by CMD-3 and SND experiments are expected in the near future with an aimed uncertainty of below $1\%$ and $0.5\%$ respectively. In addition, currently a new ISR measurement at BES-III, running in the charmonium region, is performed. The aim is a precision of below $1\%$. These measurements will hopefully shed light into the \babar-KLOE discrepancy.

Recent results have been published for the $\KS\KL$ final state by \babar ~\cite{babar:kskl} with an uncertainty of $2.9\%$ in the peak region, dominated by the trigger uncertainty. The cross section is consistent with the existing data from CMD-2~\cite{cmd2:kskl}. \babar\ also published the final state of $\KS\KL\pipi$, $\KS\KS\pipi$, and $\KS\KS K^+ K^-$. These cross sections are measured for the first time with systematic uncertainties of $10\%$, $5\%$, and $5\%$, respectively, dominated by background subtraction. The $\KS\KS K^+ K^-$ final state is dominated by statistical uncertainties. Since these measurements are performed for the first time, as to date, isospin relations have been used to estimate their contribution to g-2. Thus, these measurements allow a reduction of the systematic uncertainties and a cross-check for the isospin estimates.

The remaining uncertainty is dominated by the $\pipi$ \babar -KLOE discrepancy and the uncertainties of the $\pipi\piz$ and $\pipi\piz\piz$ final states. Additional measurements of \babar, BES-III, CMD-3, and SND are expected in the near future to further reduce the existing systematic uncertainties.

I want to thank the organisation commitee of the MITP workshop for the warm hospitality.

\newpage

\subsection{Dispersion theory to connect  $\eta\to \pi \pi \gamma$ to $\eta \to \gamma^* \gamma$}
\addtocontents{toc}{\hspace{2cm}{\sl C.~Hanhart}\par}

\vspace{5mm}

C.~Hanhart

\vspace{5mm}

\noindent
IKP, IAS and JCHP, Forschungszentrum J\"ulich, Germany\\

\vspace{5mm}

Dispersion theory holds the promise to not only control model--independently
the hadronic vacuum polarization but also 
hadronic-light-by-light scattering~\cite{gilberto}. What is needed
as input for this analysis are amplitudes for $\gamma^*\gamma^*\to h$,
where $h$ denotes the lightest hadronic states, namely $\pi^0$, $\eta$, $\pi\pi$
and possibly also $3\pi$. In this work we outline the path to eventually
get a model--independent access to $\gamma^*\gamma^*\to \eta$ based
on yet another dispersion integral.

As a first step in this direction
a dispersion integral was derived that connects data on $\eta\to \pi^+\pi^-\gamma$
to the isovector part of the $\eta\to \gamma\gamma^*$ transition form factor~\cite{eta2gammagamma}. 
At least for small virtualities the isoscalar contribution turned out to be negligibly small. 
It is demonstrated that both reactions are controlled by two scales: a universal one
driven by the $\pi\pi$-final state interactions (and of the order of the lightest vector
meson mass) and one that is reaction specific~\cite{stollenwerk}. 
The available high accuracy data for  $\eta\to \pi^+\pi^-\gamma$~\cite{kloe}
enables one to predict the shape of  $\eta\to \gamma\gamma^*$ without free parameters 
with an accuracy better than the newest available direct measurement~\cite{mainz}.

The same method in principle allows one to   control the isovector
part of $\eta\to\gamma^*\gamma^*$ from data on $e^+e^-\to\eta\pi^+\pi^-$.
To perform the calculation differential data on the two--pion spectra for the
latter reaction at various total energies are needed. Such a spectrum is
already published by BaBar~($cf$. Fig. 9 of Ref.~\cite{babar}) and additional
data is expected from Novosibirsk soon~\cite{simon_private}.

\newpage

\subsection{Positronium resonance contribution to the electron $g-2$}
\addtocontents{toc}{\hspace{2cm}{\sl M.~Hayakawa}\par}

\vspace{5mm}

M.~Hayakawa

\vspace{5mm}

\noindent
Department of Physics, Nagoya University, Japan\\

\vspace{3mm}

Recently, it was pointed out \cite{Mishima:2013ama}
that the electron $g-2$, $a_e$, gets 
the additional QED contribution from the vector-like positroniums
(ortho-positroniums) 
that cannot be captured by the perturbation theory, 
resulting in the size comparable to
the tenth-order perturbative contribution \cite{Aoyama:2012wj}.
 The effect could modify the value of the fine structure constant 
$\alpha(a_e)$ which is inferred by equating 
the theory and the experiment of the electron $g-2$.

 Ref.~\cite{Mishima:2013ama} also carried out the similar analysis 
for the direct effect on the muon $g-2$, $a_\mu$,  
and found that it is much smaller than the accuracy 
$10^{-11}$ of our interest in view of the next-generation 
muon $g-2$ experiments.
 It should be remembered that 
the latest completion of tenth-order QED contribution 
to the muon $g-2$ eliminates the uncertainty of the order $10^{-11}$, 
and that the largest uncertainty now stems 
from the well-known second-order contribution 
$a_\mu^{(2)} = \alpha /(2\pi)$  
through the uncertainty of $\alpha$
\footnote{
 I recall that the dominant QED contribution
to the muon $g-2$ higher than $4$th-order comes
from the the sixth-order light-by-light scattering type diagrams 
and those with the photon propagators corrected by the second-order
vacuum polarizations.
 As a consequence, the direct contribution to the muon $g-2$
from positronium resonances is sufficiently smaller than $10^{-11}$. 
}.
 Since this dominant uncertainty is smaller than 
$10^{-11}$, the indirect effect on $a_\mu$ from 
the above additional contribution to $a_e$ 
is also negligibly small even if it exists.

 There has been continued discussions
after the report of Ref.~\cite{Mishima:2013ama}; 
one paper \cite{Fael:2014nha} supporting the existence 
of new contribution with refinement of the result; 
three papers \cite{Melnikov:2014lwa,Eides:2014swa,Hayakawa:2014tla} 
reaching a negative conclusion.
 The former four papers 
\cite{Mishima:2013ama,Fael:2014nha,Melnikov:2014lwa,Eides:2014swa}
neglect the property of instability  
of positroniums and argue if the bound states give 
extra and sizable contribution to $a_e$. 
 However, Ref.~\cite{Hayakawa:2014tla} deals with 
the positroniums just as resonances 
and analyzes the problem based on the state space of full QED, 
because it is not possible to switch off 
the interaction responsible to the decay of the positroniums
while keeping the dynamics binding $e^-$ and $e^+$
in the framework of quantum field theory.

\newpage

\subsection{Lattice QCD studies of the Adler function}
\addtocontents{toc}{\hspace{2cm}{\sl  G.~Herdo\'iza}\par}

\vspace{5mm}

A.~Francis\,$^{a}$,~V.~G\"ulpers\,$^{a,b}$,~\underline{G.~Herdo\'iza}\,$^{a}$,~H.~Horch\,$^{a}$,~B.~J\"ager\,$^{c}$,~H.~Meyer\,$^{a,b}$,~H.~Wittig\,$^{a,b}$

\vspace{5mm}

\noindent
$^a$\,PRISMA Cluster of Excellence and Institut f\"ur Kernphysik,\\
Johannes Gutenberg-Universit\"at Mainz, Mainz, Germany\\
$^b$\,Helmholtz Institute Mainz, Johannes Gutenberg-Universit\"at Mainz,
Mainz, Germany\\
$^c$\,Department of Physics, Swansea University, Swansea, United Kingdom\\

\vspace{5mm}

QCD effects appearing in the photon vacuum polarisation function
(VPF), $\Pi(Q^2)$, induce the largest fraction of the theoretical
uncertainties in the anomalous magnetic moment of the muon $a_\mu$ or
in the running of the QED coupling constant $\Delta\alpha_{\rm
  QED}(Q^2)$. The Adler function, $D(Q^2)=12\,\pi^2\,\,
d\Pi(Q^2)/d\log(Q^2)$, is a physical quantity depending on the
momentum transfer $Q^2$ in the Euclidean region. By its direct
relation to the VPF, $D(Q^2)$ provides an alternative way to determine
the leading-order hadronic contributions to $a_\mu$ and
$\Delta\alpha_{\rm QED}$. Furthermore, in the large $Q^2$ regime,
$D(Q^2)$ is described by perturbative QCD due to the absence of
resonance effects~\cite{Shintani:2010ph,Herdoiza:2014jta}.

Lattice QCD provides a first principles determination of the Adler
function in a large interval of $Q^2$
values~\cite{Horch:2013lla,Francis:2013fzp}. The low $Q^2$ regime
around the muon mass, where long-distance QCD effects induce large
uncertainties on the lattice calculations, is most important for
$a_\mu$. On the other hand, the statistical precision on $D(Q^2)$ is
higher for larger $Q^2 \sim 1\,{\rm GeV}^2$. This permits a detailed
study of $\Delta\alpha_{\rm QED}(Q^2)$ in an energy region where
determinations based on the dispersion relation
approach~\cite{Jegerlehner:2011mw,Davier:2010nc,Hagiwara:2011afg} are
affected by larger uncertainties thus limiting their impact on
electroweak precision tests. A further motivation to consider
intermediate $Q^2$ values is to establish a region of parameter space
where a comparison of computations from different lattice groups can
be best performed, due to a more favourable control of statistical and
systematic uncertainties than in the case of $a_\mu$.

The calculation of $D(Q^2)$ is based on a set of lattice ensembles with
two flavours of improved Wilson fermions, including three values of
the lattice spacing in an interval, $a\in[0.05,0.08]$\,fm. The
pseudoscalar meson masses $M_{\rm PS}$ are varied from $450$\,MeV down
to $190$\,MeV while also keeping $M_{\rm PS}\,L \geq 4$ to reduce
finite size effects. The use of twisted boundary conditions~\cite{DellaMorte:2011aa1}
significantly increases the number of accessible $Q^2$ values and
allows to construct the Adler function from the numerical derivative
of the VPF.

The lattice data for $D(Q^2)$ is parametrised by a fit ansatz that
simultaneously describes the $Q^2$ behaviour through Pad\'e
approximants and the continuum and chiral extrapolations. With respect
to the more conventional approach where $\Pi(Q^2)$ is directly used to
determine $a_\mu$ or $\Delta\alpha_{\rm QED}$, the use of the
Adler function allows to avoid the inclusion of a large set of fit
parameters $\Pi(0)$ (one for each lattice ensemble).

The determination of the Adler function and of $\Delta\alpha_{\rm
  QED}(Q^2)$ including light ($u$,$d$) and strange quark contributions
demonstrates that the statistical accuracy of the lattice calculation
is comparable to that of phenomenological results in the region $Q^2
\in [1,4]\, {\rm GeV}^2$. An extension of our study to include the
valence contribution from the charm quark as well as a detailed
analysis of lattice artifacts is currently being carried out and will
allow for a direct comparison to phenomenology.

\newpage

\subsection{Analytic continuation method for the hadronic vacuum polarization function}
\addtocontents{toc}{\hspace{2cm}{\sl K.~Jansen}\par}

\vspace{5mm}
X.~Feng$^1$, S.~Hashimoto$^2$, G.~Hotzel$^3$, \underline{K.~Jansen}$^4$, M.~Petschlies$^5$, D.~Renner$^6$

\vspace{5mm}

\noindent
$^1$Physics Department, Columbia University, New York, NY 10027, USA\\
$^2$High Energy Accelerator Research Organization (KEK), Tsukuba 305-0801, Japan\\
$^3$Humboldt University Berlin; NIC, Desy Zeuthen\\
$^4$NIC, Desy Zeuthen, Platanenallee 6, 15738 Zeuthen, Germany\\
$^5$The Cyprus Institute, P.O. Box 27456, 1645 Nicosia, Cyprus\\
$^6$Jefferson Lab, 12000 Jefferson Avenue, Newport News VA 23606, USA 
\vspace{5mm}

The calculation of the leading order hadronic contribution of the
muon anomalous magnetic moment, $a_\mu^{\rm had}$, is one of the prime targets of
lattice QCD activities presently. However, in such a computation
there is a
generic problem to reach small momenta, dominating the weight function,
on the lattice which are needed
to evaluate the
hadronic vacuum polarization (HVP) function from which
$a_\mu^{\rm had}$ is derived.
Present approaches to circumvent this problem
\cite{Blum:2002ii,Gockeler:2003cw,Aubin:2006xv,Feng:2011zk,Boyle:2011hu,DellaMorte:2011aa2,Aubin:2012me1,deDivitiis:2012vs}
design appropriate fit functions for the HVP function, employ
twisted boundary conditions, take
model independent Pad\'e polynomials or compute
the derivative of the vector current correlation function.

An alternative approach is to use
the method of
{\em analytic continuation} \cite{Feng:2013xsa} which is closely
related to the work in Refs.~\cite{Ji:2001wha,Meyer:2011um}.
This method allows, in principle, to compute the HVP function at small
space-like momenta and even at time-like momenta. The feasibility of the method
has been demonstrated at the examples of leading order hadronicn contribution 
to the muon anomalous magnetic moement, $a_\mu^{\rm hvp}$, and the Adler function in 
Refs.~\cite{Feng:2013xsa,Feng:2013xqa}.
When comparing to the standard method to compute $a_{\mu}^{\rm hvp}$ a full
agreement was found, but the analytic continuation method leads to noisier results.
Still, we believe that the analytic continuation method is a valuable alternative
which has, moreover, the
potential to address other quantities
where small or zero momenta are needed.

\newpage

\subsection{Towards a dispersive analysis of the $\pi^0$ transition form factor}
\addtocontents{toc}{\hspace{2cm}{\sl B.~Kubis}\par}

\vspace{5mm}
B.~Kubis
\vspace{5mm}

\noindent
Helmholtz-Institut f\"ur Strahlen- und Kernphysik (Theorie) and\\ Bethe Center for Theoretical Physics,
Universit\"at Bonn, Germany

\vspace{5mm}

The $\pi^0$ pole term is the largest individual piece in the hadronic light-by-light scattering contribution
to the anomalous magnetic moment of the muon.  Its strength is determined by the singly- and doubly-virtual
$\pi^0$ transition form factor, the momentum dependence of which can be analyzed in dispersion theory.
In the most important energy range (roughly up to 1\,GeV), the isovector and isoscalar part of 
$\gamma^* \to \pi^0 \gamma^{(*)}$ are dominated by two- and three-pion intermediate states, respectively.
While the dispersion relation for two pions require the (charged) pion vector form factor and 
the anomalous amplitude $\gamma^{(*)}\pi \to \pi\pi$ as input, three pions can be simplified due to the 
dominance of the narrow isoscalar resonances $\omega$ and $\phi$ and demand an understanding of the 
vector-meson transition form factors $\omega,\,\phi\to\pi^0\gamma^*$.  All of these components can in turn be 
reconstructed dispersively.

The process $\gamma\pi\to\pi\pi$ at zero energy and pion masses is determined---as is the decay of the 
$\pi^0$ into two real photons---by the Wess--Zumino--Witten anomaly.  A dispersive representation~\cite{g3pi} 
can be used to extract the anomaly from data in the full elastic region.  A similar analysis provides
decay amplitudes for $\omega,\,\phi\to3\pi$~\cite{V3pi}, which have been shown to reproduce high-statistics
data for the $\phi\to3\pi$ Dalitz plot~\cite{KLOE:phi} with excellent accuracy.  The corresponding partial waves,
again combined with the pion vector form factor, yield a dispersive representation of the above-mentioned 
vector-meson transition form factors~\cite{omegaTFF}.  Sum rules for the decays $\omega,\,\phi\to\pi^0\gamma$
work rather well, although the description of data on $\omega\to\pi^0\mu^+\mu^-$~\cite{NA60} remains problematic.
As a final step, a parametrization of the cross section data for $e^+e^-\to3\pi$ allows for
a full reconstruction of $\pi^0\to\gamma^*\gamma^*$; first comparisons to data on the singly-virtual form factor
in $e^+e^-\to\pi^0\gamma$ are very promising~\cite{pi0TFF}.

\newpage

\subsection{Mainz Workshop report on  the muon anomalous magnetic moment}
\addtocontents{toc}{\hspace{2cm}{\sl W.~Marciano}\par}

\vspace{5mm}

W.~Marciano

\vspace{5mm}

\noindent
Brookhaven National Laboratory, Upton, New York 11973 \\
\vspace{5mm}
\setcounter{equation}{0}

        The April 2014 muon $g-2$ Workshop in Mainz, Germany focused on hadronic loop corrections.  In that regard, a recent preprint by Kurz, Liu, Marquard and Steinhauser was especially timely~\cite{r1}.  It estimated the next to next to leading hadronic vacuum polarization corrections to be $+12.4(1)\times 10^{-11}$.  That reduces somewhat the discrepancy between experiment and theory to $276(63)(49) \times 10^{-11}$, now about 3.5 sigma~\cite{r2}. That is still a significant deviation, particularly when one recalls that the electroweak contribution \cite{r3} to the anomaly is only $+154 \times 10^{-11}$, i.e. about half that discrepancy. What could be responsible for such a large effect?
	Possible sources of the discrepancy include: 1) Hadronic Loop Theory, 2) Experiment and 3) "New Physics".  Currently, dispersion relations and lattice QCD calculations are refining the theory prediction and hope to reduce the estimated uncertainty by about a factor of $1/2$. Such an improvement would nicely complement the expected experimental improvement at Fermilab by a factor of $1/4$. Together, they have the potential to provide an overall $9+$ sigma effect if the central experimental and theoretical values remain unchanged.   
	A discussion topic at the Workshop addressed speculations regarding additional systematic effects due to the bound state storage ring environment of the muons and the muon bunch density.  Those somewhat vague issues are sometimes raised as a potential source of the discrepancy.  However, it appears, based on simple estimates, that such electromagnetic effects are likely to be negligible in comparison to the small errors in the corrections already applied to the data.
	In the case of "New Physics" solutions to the discrepancy, several viable options remain open.  The leading candidate explanation is supersymmetry loop effects, with a relatively light stau loop dominant.  The scale of susy particle masses in such a scenario is in the several hundred GeV range, causing some tension with the failure of the LHC to find supersymmetry.  In that regard, the next LHC run should prove to be more definitive. A second solution involves new muon mass generating dynamics at a scale of order $1-2$ TeV~\cite{r4}.  Again, no sign of such an underlying effect has yet to be seen at the LHC.
	A low mass solution to the anomaly discrepancy is the "dark" photon \cite{r5}.  Often invoked to explain astrophysical phenomena, it can lead, via small kinetic mixing with the ordinary photon, to a loop correction of the right sign and magnitude.  Currently, extensive searches for the "dark" photon have been carried out or are planned.  Much of the parameter space needed for the anomaly solution has been ruled out in the simplest model; so, its likelihood has been substantially reduced.  Nevertheless, even as a long shot, it remains interesting and well motivated.  Discovery of the "dark" photon would revolutionize particle physics.

\newpage

\subsection{The role of experimental data on the hadronic light-by-light of the muon $g-2$}
\addtocontents{toc}{\hspace{2cm}{\sl P.~Masjuan}\par}

\vspace{5mm}

\underline{P.~Masjuan} and P.~Sanchez-Puertas

\vspace{5mm}

\noindent
PRISMA Cluster of Excellence, Institut f\"ur Kernphysik, Johannes Gutenberg-Universit\"at Mainz,
D-55099 Mainz, Germany \\

\vspace{5mm}
\setcounter{equation}{0}

One of the open questions concerning the Hadronic Light-by-Light scattering contribution to the muon $g-2$ (HLBL) is the role of experimental data.

Part of the difficulty of including experimental data in the HLBL is due to the particular framework where the main calculations are done~\cite{Jegerlehner:2009ryc}, the large-$N_c$ of QCD~\cite{tHooft:1973jzc}. In such limit, one uses the resonance saturation scheme~\cite{nyffelerc} to reproduce the pseudoscalar transition form factor (TFF) that appears in the dominant piece of the HLBL, the pseudscalar-exchange contribution~\cite{Jegerlehner:2009ryc}. The main inputs are, then, the pion decay constant and the values of the resonance masses. On top, even though data on the TFF is willing to be included, one still faces the problem on how to link the different kinematic regimes between the experiment for the TFF and the kinematics for the pseudoscalar-exchange diagram. A direct fit to such TFF cannot be used for computing the HLBL~\cite{Jegerlehner:2009ryc}. 

In this work we provide an answer to that question in a model-independent fashion~\cite{ourpaper}, an approach compatible with the recent dispersion relations attempt~\cite{colangelo} with the advantage of having larger photon energy range of applicability (in practice, the full energy range), and based on the low-energy properties of the TFF.

It was pointed out in Ref.~\cite{Masjuan:2007ayc} that, in the large-$N_c$ framework, the resonance saturation can be understood from the mathematical theory of Pad\'e Approximants (PA) to meromorphic functions~\cite{Bakerc}, where one can compute the desired quantities in a model-independent way and even be able to ascribe a systematic error to the approach \cite{Masjuan:2008cpc}. 

For the discussion we use the models from Ref.~\cite{Knecht:2001qfc}. The inputs for the models can have two different sources: first, a pure theoretical origin based on large-$N_c$ and chiral limits (inputs are resonance masses within the \textit{half-width rule}~\cite{Masjuan:2012gcc} and the meson decay constant in the chiral $SU(3)$ limit~\cite{Eckerc}); second, a reconstruction of the models based on a matching with the TFF low-energy constants~\cite{Masjuan:2012wyc}, i.e, \textit{\'a la} PA~\cite{Masjuan:2007ayc,Masjuan:2008cpc} minimizing in such a way the model dependence~(see \cite{Masjuan:2012wyc,Masjuan:2012qnc} for details). 

Notice, nevertheless, that the standard procedure~\cite{Jegerlehner:2009ryc,Knecht:2001qfc} to treat the TFF  is through a factorization approach, e.g., define $F(Q_1^2,Q_2^2)=F(Q_1^2,0)\times F(0,Q_2^2)$ where $F(Q^2,0)$ is the measured quantity. The non-factorizing terms might yield effects not negligible (see~\cite{Pabloc}).

We found that the pure theoretical calculation referred before yields a final error of the pion contribution to HLBL to $15\%$ ($5\%$ from $F_0$ and $10\%$ from the masses). Applying the Pad\'e method to the current models for the HLBL~\cite{Knecht:2001qfc} yield higher central values of about $20\%$. Not only that, but also this method provides a rule-of-thumb for estimating the impact of experimental uncertainties, a point never discussed before. In fact, we found a similar $15\%$  provided that the $13\%$ error on the slope ($25\%$ on curvature) implies an error of $10\%$ ($5\%$) in the pion contribution; the impact of $F_{\pi}$ is more dramatic since $1\%$ error implies a $2\%$ error on HLBL. Using the prescription of Ref.~\cite{Melnikov:2003xdc}, the errors grow up to $30\%$.

In conclusion, we remark the important role of experimental data to determine the dominant pieces of the HLBL (i.e., $\pi^0,\eta,\eta'$). We argue that the way of including such information should be based on Pad\'e approximants which provides first a systematic error, and second a simple rule for estimating the impact of experimental uncertainties. We notice, finally, that the errors discussed above have been unfortunately ignored in the main reviews (no error for $F_0$ or resonance masses have been properly estimated, neither the possibility to match with experimental low-energy description of the TFF) and that posses a warrant on the reliability of the current error estimates for the HLBL.

\newpage

\subsection{On the disconnected diagram contribution to $a_{\mu}^{\rm HLO}$}
\addtocontents{toc}{\hspace{2cm}{\sl H.B.~Meyer}\par}

\vspace{5mm}

A.\ Francis, V.\ G\"ulpers, G.\ von Hippel, \underline{H.B.\ Meyer}, H.\ Wittig

\vspace{5mm}

\noindent
PRISMA Cluster of Excellence, Institut f\"ur Kernphysik and Helmholtz~Institut~Mainz, Johannes Gutenberg-Universit\"at Mainz,
D-55099 Mainz, Germany \\

\vspace{5mm}
\setcounter{equation}{0}
In the hadronic vacuum polarization contribution to $a_\mu$, many effects become important at the few-percent level.  
Here we discuss the contribution of the `disconnected diagrams', i.e.\ contributions
that for $N_f$ degenerate flavors are proportional to $N_f^2$.
We assume isospin symmetry and therefore decompose the e.m.\ current into an 
isovector and an isoscalar part,
\begin{equation}\nonumber
j_\mu^\gamma = \underbrace{{\textstyle\frac{1}{2}}(j_\mu^u-j_\mu^d)}_{\equiv j^\rho,~(I=1)} 
+ \underbrace{{\textstyle\frac{1}{6}}  (j_\mu^u + j_\mu^d - 2j_\mu^s)}_{I=0}.
\end{equation}
In the mixed representation,
$
G^{\gamma\gamma}(x_0) = - \frac{1}{3}\int d^3\vec x\; \Big\langle j_k^\gamma(x)j_k^\gamma(0)\Big\rangle
$,
we have the decomposition
\begin{eqnarray}\label{eq:ggFlStr}\nonumber
 G^{\gamma\gamma}(t) 
  &=& {\textstyle\frac{10}{9}}  G^{\rho\rho}(t)
  +{\textstyle\frac{1}{9}} G^{s}_{\rm conn}(t) + {\textstyle\frac{1}{9}} G^{ls}_{\rm disc}(t).
\end{eqnarray}

In this representation,  $a_\mu^{\rm HLO}$ is obtained by integrating
$G^{\gamma\gamma}(t)$ with an appropriate kernel~\cite{Bernecker:2011gh}.
A simple phenomenological analysis shows that the contribution from 0 to 1fm is $41\%$,
from 1fm to 2fm $45\%$, from 2 to 3fm $11\%$ and beyond 3fm $3\%$.

At short distance, the perturbative result for the disconnected diagram contribution 
of the light quarks shows that it is extremely small. At late time $t$ however,
the fact that $G^{\gamma\gamma}(t) = G^{\rho\rho}(t)(1+{\rm O}(e^{-m_\pi t}))$ implies that 
\begin{equation}\label{eq:GlsEst}
G^{ls}_{\rm disc}(t) \stackrel{t\to\infty}{=} - (G^{\rho\rho}(t) + G^{s}_{\rm conn}(t)).
\end{equation}

At what distance does this asymptotic estimate becomes a good approximation? Writing
\begin{equation}\label{eq:pred1}\nonumber
\frac{1}{9} \frac{G_{\rm disc}^{ls}(x_0)}{G^{\rho\rho}(x_0)} = 
 \frac{G^{\gamma\gamma}(x_0)- G^{\rho\rho}(x_0)}{G^{\rho\rho}(x_0)}
- \frac{1}{9}\left(1+ \frac{G^s_{\rm conn}(x_0)}{G^{\rho\rho}(x_0)}\right) 
\stackrel{x_0\to\infty}{\longrightarrow} -\frac{1}{9},
\end{equation}
the first term on the RHS is positive-definite and can be obtained by selecting isoscalar final states
in $e^+e^-\to{\,\rm hadrons}$.
The second can be obtained on the lattice. To avoid a delicate cancellation between the two terms, 
this way of proceding should be used at distances where the first term is small compared to the second.
We find that $G^{ls}_{\rm disc}(t)$ approaches the asymptotic (\ref{eq:GlsEst}) between 2fm and 4fm.

A direct lattice evaluation of ${\cal R}(x_0)=\frac{1}{9} \frac{G_{\rm
    disc}^{ls}(x_0)}{G^{\rho\rho}(x_0)}$ with up to 1000
configurations shows that this quantity is always zero within
statistical errors, and the error reaches $\frac{1}{9}$ around
$t=1.1\,$fm. Stochastic sources are used and it is essential to use
the same sources for the light and the strange quark in order to
reduce statistical fluctuations.  Assuming that ${\cal R}$ jumps to
the asymptotic $-\frac{1}{9}$ at that point indicates that the
magnitude of the disconnected diagram contribution to $a_\mu^{\rm
  HLO}$ is very unlikely to be larger than $3\%$.

\newpage

\subsection{Large-$N_c$ inspired approach to hadronic light-by-light scattering in the muon $g-2$}
\addtocontents{toc}{\hspace{2cm}{\sl A.~Nyffeler }\par}

\vspace{4mm}

A.~Nyffeler 

\vspace{4mm}

\noindent
Institut f\"ur Kernphysik, Johannes Gutenberg Universit\"at Mainz, Germany\\

\vspace{3mm}
\setcounter{equation}{0}
The large-$N_c$ QCD inspired approach to hadronic light-by-light
scattering (HLbL) in the muon $g-2$ is based on the idea of the
Minimal Hadronic Ansatz (MHA)~\cite{MHA}. The spectrum of physical
states in large-$N_c$ QCD consists of an infinite tower of narrow
resonances in each channel. At leading order in $N_c$, only tree-level
diagrams with the exchanges of resonance states contribute to a given
Green's function which has only poles at the resonance masses, no cuts
from multi-particle intermediate states.  The low-energy and
short-distance behavior of the Green's function is matched with
results rooted in QCD, using Chiral Perturbation Theory and the
Operator Product Expansion (OPE). One then makes an ansatz for the
Green's function as a ratio of two polynomials in several momentum
variables with, in practice, the exchanges of a finite number of
resonances. The MHA assumes that taking the lowest few resonances in
each channel to reproduce low-energy and short-distance constraints
gives already a good description of the Green's function in the real
world. This interpolation works best for Green's functions which are
order parameters of chiral symmetry breaking and for integrals in
Euclidean space. However, one cannot fulfill all short-distance
constraints on Green's functions and form factors with a finite number
of resonances~\cite{Bijnens_et_al_03}.

As example, we consider the Green's function $\langle VVP \rangle$
which is relevant for the evaluation of the pseudoscalar pole and
exchange contribution to HLbL. The MHA ansatz for the form factor with
one on-shell pion and two off-shell photons and with two multiplets of
vector resonances $\rho$ and $\rho^\prime$ (lowest meson dominance
(LMD) + V) reads~\cite{KN_EPJC_01,N_JN_09}
\begin{equation} \label{FF}
{\cal F}_{\pi^0\gamma^*\gamma^*}^{\rm LMD+V}(q_1^2, q_2^2) 
= \frac{F_\pi}{3}\, {q_1^2\,q_2^2\,(q_1^2 + q_2^2) +
   h_1\,(q_1^2+q_2^2)^2 + \bar{h}_2 \,q_1^2\,q_2^2 +
   \bar{h}_5 \,(q_1^2+q_2^2) + \bar{h}_7 
\over (q_1^2-M_{V_1}^2) \, (q_1^2-M_{V_2}^2) \, 
   (q_2^2-M_{V_1}^2) \, (q_2^2-M_{V_2}^2)}  
\end{equation}
where $F_\pi = 92.4~\mbox{MeV}$ is the (charged) pion decay constant
and the poles correspond to the physical resonance masses $M_{V_1} =
M_\rho = 775.49~\mbox{MeV}$ and $M_{V_2} = M_{\rho^\prime} =
1.465~\mbox{GeV}$. The quantities $h_i (\bar{h}_i)$ in the numerator
are the model parameters in the off-shell pion form factor, see
Refs.~\cite{KN_EPJC_01,N_JN_09} for their determination. The ansatz in
Eq.~(\ref{FF}) fulfills all leading and some subleading QCD
short-distance constraints from the OPE. The OPE uniquely fixes the
first term in the numerator, therefore the form factor does not
factorize into functions of $q_1^2$ and $q_2^2$.  The LMD ansatz with
only one multiplet of vector resonances does not simultaneously
fulfill the OPE and the Brodsky-Lepage behavior for the transition
form factor $\lim_{Q^2 \to \infty} {\cal
  F}_{\pi^0\gamma^*\gamma^*}(-Q^2,0) \sim 1/Q^2$. It can be obtained
in Eq.~(\ref{FF}) with $h_1 = 0~\mbox{GeV}^2$.

One can generalize the method by using a resonance Lagrangian which
respects chiral symmetry and fulfills as many short-distance
constraints as possible to fix some of the couplings in the Lagrangian
(Resonance Chiral Theory, R$\chi$T~\cite{Ecker_et_al_PLB_89}). The
Lagrangian approach allows to connect different Green's functions and
one can more easily identify which parameters enter in various
physical processes. In principle, such a Lagrangian framework also
allows to study effects beyond leading order in $N_c$, like the finite
width of resonances and loops of resonances. The R$\chi$T for the odd
intrinsic parity sector has been developed in
Refs.~\cite{Ecker_et_al_PLB_89,RchiT_odd_parity}, sometimes with an
additional multiplet of heavy pseudoscalar mesons $\pi^\prime$ (LMD+P)
or with two multiplets of vector mesons (LMD+V).

Table~\ref{tab:HLbLPS} shows the results for the pseudoscalar pole and
exchange contributions to HLbL within the MHA/R$\chi$T framework. Even
within such a restricted approach, where many low-energy,
short-distance and experimental constraints are built in, there is a
variation of the results, e.g.\ for the pion-pole contribution, of the
order of about 10\%. For the full HLbL contribution, a detailed
analysis of the four-point function $\langle VVVV \rangle$ is
needed~\cite{KN_VVVV}.

\vspace*{-0.3cm}
\begin{table}[h]
\centering
\caption{Pseudoscalar contributions to HLbL in large-$N_c$ QCD
  inspired approaches.}   
\label{tab:HLbLPS}
\renewcommand{\arraystretch}{1.1}
\vspace*{1mm}
\begin{tabular}{|l|c|c|}
\hline
 Model for ${\cal F}_{P^{(*)}\gamma^*\gamma^*}$ [Reference]
 & $a_\mu(\pi^0) \times 10^{11}$ & 
   $a_\mu(\pi^0,\eta,\eta') \times 10^{11}$ \\
\hline
LMD (pole) \cite{KN_02} & 73 & $-$  \\  
LMD+V (pole, $h_2=-10~\mbox{GeV}^2$) \cite{KN_02} & 63(10) & 88(12) \\ 
LMD+V (on-shell FF, constant 2nd FF) \cite{MV_04} & 77(7) & 114(10) \\   
LMD+V (off-shell) \cite{N_JN_09}   & 72(12) & 99(16)   \\   
LMD+P (off-shell) \cite{KaNo_11}    & 65.8(1.2) & $-$  \\ 
LMD+P (pole) \cite{RGL_14}      & 57.5(0.5) &  82.7(2.8) \\ 
LMD+P (off-shell) \cite{RGL_14}     & 66.5(1.9) & 104.3(5.2) \\ 
\hline 
\end{tabular}
\end{table}

\newpage

\subsection{Light-by-light scattering sum rules}
\addtocontents{toc}{\hspace{2cm}{\sl  V.~Pascalutsa}\par}

\vspace{5mm}

V.~Pascalutsa

\vspace{5mm}

\noindent
Institut f\"ur Kernphysik,  Johannes Gutenberg Universit\"at Mainz

\vspace{5mm}
\setcounter{equation}{0}

General considerations based on unitarity and causality (dispersion relations), as well as the Lorentz and electromagnetic gauge symmetries, lead to a number 
of model-independent relations, {\it viz.}~sum rules, for $\gamma\gamma$ system \cite{Pascalutsa:2010sj,Pascalutsa:2012pr}. These sum rules express the low-energy properties of light-light (LbL) interaction, arising due to vacuum fluctuations, with the integrals of classical cross-sections for $\gamma\gamma$ fusion. For example, writing the most general form of the Lagrangian describing the low-energy photon self-interaction:
\begin{equation}
{\mathcal L} = c_1 (F_{\mu\nu} F^{\mu\nu})^2 + c_2 (F_{\mu\nu} \tilde F^{\mu\nu})^2,
\end{equation}
where $c_1$ and $c_2$ are arbitrary constants and 
$F$ ($\tilde F$) is the (dual) electromagnetic field-strength tensor, 
we obtain \cite{Pascalutsa:2010sj}: 
\begin{equation}
c_1\pm c_2  =  \frac{1}{8\pi }\int_{0}^{\infty} d s\,  \frac{ \sigma_{||} (s) \pm \sigma_\perp(s)}{s^2}\,,
\end{equation}
where $\sigma_{||}$ and $\sigma_{\perp}$ are the total cross-sections
for fusion of linearly polarised photons with polarisations oriented 
parallel or  perpendicular to each other; the are functions of the invariant
energy $s$ only. In this way one immediately sees that $c_i$'s are positive-definite, and hence photons always attract at very low energies (the corresponding Hamiltonian is negative-definite). 
Another remarkable relation of this type is the Gerasimov-Drell-Hearn sum rule,
which in the case of the $\gamma\gamma$ system takes the following form~\cite{Gerasimov:1973ja}:
\begin{equation}
\int_{0}^\infty \!\! d s \, \frac{\sigma_0(s)-\sigma_2(s)}{s} = 0\,,
\end{equation}
where $\sigma_0$ and $\sigma_2$ are the total cross-sections
for fusion of photons circularly polarised in the same or opposite
direction, respectively. An extension of these sum rules to virtual
photons was derived in \cite{Pascalutsa:2012pr},
where the constraints put by these sum rules on meson transition 
form factors were discussed too. Some more formal implications of these
sum rules have recently  been addressed in Refs.~\cite{Pauk:2013hxa,Pascalutsa:2014ena}.

The empirical information and specific phenomenological models, used to calculate the hadronic-LbL contribution to $(g-2)_\mu$, can and should be constrained by these sum rules.  An interesting such study is presented by Mike Pennington elsewhere in these mini-proceedings.

\newpage

\subsection{Dispersive approach to the muon's anomalous magnetic moment}
\addtocontents{toc}{\hspace{2cm}{\sl  V.~Pauk}\par}

\vspace{5mm}

V.~Pauk

\vspace{5mm}

\noindent
Johannes Gutenberg University, Mainz \\

\vspace{5mm}
\setcounter{equation}{0}

We present a new dispersive formalism for evaluating the hadronic light-by-light (HLbL) scattering contribution to the anomalous magnetic moment of the muon $a_{\mu}$. It is suggested to represent this contribution as a dispersive integral of the vertex function discontinuity in the virtuality of the external photon. By unitarity this discontinuity is related to the amplitudes of decay and production of hadrons. 

\begin{figure}[h]
\centering
  \includegraphics[width=8.5cm]{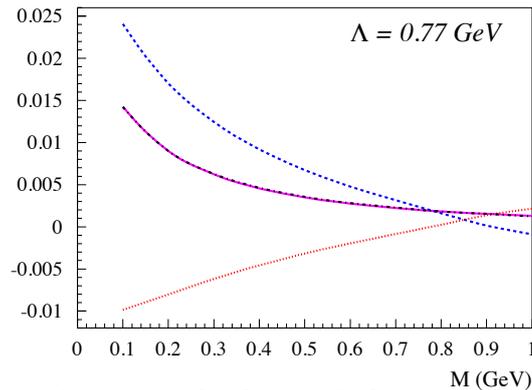}
  \vspace{-3.5cm}
  \caption{The Pauli form factor in the limit of the vanishing external momentum $k\to0$ $F_2(0)$ as function of the pseudoscalar meson mass $M$. The red dotted (blue dashed) line denotes the dispersive evaluation due to the three-particle (two-particle) discontinuity. The black dashed-dotted line denotes the sum of both contributions. The pink solid line is obtained by the direct evaluation of the corresponding two-loop integral.}
  \label{figxxx}
\end{figure}

As a test of the dispersive formalism, we firstly applied it to the case of a scalar two-loop vertex diagram of similar topology as entering the HLbL contribution to $a_{\mu}$ \cite{Pauk:2014jza}. Here, we provide a first realistic application of the proposed formalism to the case of pole exchanges. To define the analytical structure of the light-by-light amplitude in the photon's virtuality we consider the VMD-like approximation with the rho-meson pole exchange. In such an approximation there are two different contributions to the vertex function. For the case of a single meson exchange the discontinuity of the vertex function is defined by two- and three-particle cuts. The contributions of these discontinuities and their sum versus the direct evaluation using the Gegenbauer polynomial technique \cite{Knecht:2001qf222} for the dominant diagram depending on the mass of exchanged meson is shown in Fig. \ref{figxxx}.

A crucial distinctive feature of the dispersion approach is that it allows extension to implement the form factors beyond the simplest monopole or dipole approximations and to include  multi-meson channels. The next important step will be to include the two-pion channel. Moreover, it allows for a more straightforward implementation of the experimental data. The ongoing measurements by the BES-III Collaboration will be a crucial input into the presented dispersive formalism.

\newpage

\subsection{Light-by-light scattering with \lq\lq real'' photons}
\addtocontents{toc}{\hspace{2cm}{\sl M.R.~Pennington}\par}

\vspace{5mm}

 M.R.~Pennington

\vspace{5mm}

\noindent
Theory Center, Thomas Jefferson National Accelerator Facility,\\ Newport News, VA 23606, USA\\

\vspace{5mm}

The development of a dispersive approach to the calculation of hadronic light-by-light scattering~\cite{hoferichter} provides an opportunity for a more realistic assessment of the uncertainties in this contribution to the anomalous magnetic moment of the muon. Electron-positron colliders have over decades published data on the two photon production of hadrons. However, it is only with the high statistics possible with the heavy flavor factories, like Belle and BaBar, that precision data have been taken. The largest cross-sections occur when the electron and positron barely scatter and the photons are consequently very nearly real --- only MeV from massless. Thus we have data on two photon production of $\pi^+\pi^-$, $\pi^0\pi^0$, $K^+K^-$, $K_sK_s$ and $\pi^0\eta$ from Belle~\cite{belle_pic,belle_pin,belle_Kc,belle_Ks,belle_eta}. While these data have no polarization information and limited angular coverage, the close connection, provided by Analyticity, Crossing and Unitarity, with the corresponding meson-meson scattering amplitudes makes an Amplitude Analysis possible, at least for some of these channels in the low energy regime. It is the imaginary parts of these amplitudes in low partial waves at energies below $\sim 1.5$ GeV that are expected to dominate the dispersive calculation of light-by-light scattering for $(g-2)_\mu$.

While the formalism for how to perform such an Amplitude Analysis for two photon production of spinless mesons was set out 25 years ago~\cite{morgan_p,Daphne_p}, and subsequently applied to the data then available~\cite{morgan_p2,boglione,mori}, it is only now with the publication of the Belle results that motivates a new Analysis. Unitarity connects the amplitudes for $\gamma\gamma\to\pi\pi$ and ${\overline K}K$, for instance, in each partial wave with definite spin and isospin with the corresponding partial waves for $\pi\pi\to\pi\pi$ and ${\overline K}K$. With results on low energy $\pi\pi$ scattering extracted from 
precision experimental results from $K\to (\pi\pi)e\nu_e$ decays~\cite{NA48} and the DIRAC experiment~\cite{dirac} at CERN, together with improved extensive dispersive analyses~\cite{KPY,descotes},  meson-meson scattering amplitudes are now under better control than ever before. Armed with this information we fit not only the Belle $\pi\pi$ and ${\overline K}K$ results, but all published data on these channels~\cite{markII}-\cite{
tpc}. This includes both integrated and differential cross-sections. An Amplitude Analysis is then possible up to 1.4 GeV, where the $\pi\pi$ and ${\overline K}K$ channels are deemed to saturate unitarity. A single solution is found with rather restrictive uncertainties in the isospin zero and two channels, and with a larger range in the isovector ${\overline K}K$ channel~\cite{dai_letter,dai_long}. 

For the light-by-light contributions to $(g-2)_\mu$, the contribution of the long-lived pseudoscalar mesons, $\pi^0$, $\eta$ and $\eta'$, can be computed accurately in the narrow resonance approximation~\cite{vanderhaeghen}. However, this approximation is not appropriate for the much shorter-lived $f_0(980)$, $f_2(1270)$, $a_2(1320)$, etc. Indeed, by inputting the results of this Amplitude Analysis there is no need to separate resonances from \lq\lq backgrounds'' from pion and kaon loops. These are all automatically included in our partial wave solutions, at least for real photon scattering.
This encodes our present knowledge of the dominant di-meson production. When combined with robust modeling of the virtuality of the photons, this provides a realistic way of computing the contribution to hadronic light-by-light scattering through $\pi\pi$ and ${\overline K}K$ intermediate states. The Amplitude Analysis presented in Refs.~\cite{dai_long} is a step towards a reliable determination of this key component.

\newpage

\subsection{Fits and related systematics for the hadronic vacuum polarization on the lattice}
\addtocontents{toc}{\hspace{2cm}{\sl  S.~Peris}\par}

\vspace{5mm}

C. Aubin$^{a}$, T. Blum$^{b}$, M. Golterman$^{c}$, K. Maltman$^{d}$ and \underline{S. Peris}$^{e}$

\vspace{5mm}

\noindent
$^{a}$ Dept. of Physics and Engineering Physics, Fordham Univ., Bronx, NY 10458, USA\\
$^{b}$ Physics Dept., Univ. of Connecticut, Storrs, CT 06269, USA\\
$^{c}$ Dept. of Physics and Astronomy, SFSU, San Francisco, CA 94132, USA\\
$^{d}$ Dept. of Mathematics and Statistics, York Univ., Toronto, ON Canada M3J 1P3\\
$^{e}$ Dept. of Physics, UAB, 08193 Bellaterra (Barcelona), Spain\\

\vspace{5mm}

In order to test the systematic error coming from the extrapolation at low $Q^2$ carried out in present lattice determinations of the hadronic vacuum polarization contribution to the muon anomalous magnetic moment, we employ a physically motivated model for the isospin-one non-strange vacuum polarization function $\Pi(Q^2)$ \cite{Golterman:2013vca}. The model is based on the OPAL experimental vector-channel spectral function for energies below the $\tau$ mass and a successful parametrization, including perturbation
theory and a model for quark-hadron duality violations, for higher energies. Using the same covariance
matrix and $Q^2$ values as in a recent lattice simulation, we then generate fake data for $\Pi(Q^2)$. The fake data is then used to extrapolate to low $Q^2$ and evaluate the hadronic vacuum polarization contribution to the muon anomalous magnetic moment, after which the
result is compared to the exact model value. From this comparison we unravel a systematic error much larger than the few-percent total error sometimes claimed for
such extractions in the literature. We find that errors deduced from fits using a
Vector Meson Dominance ansatz are misleading, typically turning out to be much smaller than
the actual discrepancy between the fit and exact model results. The use of a sequence of multipoint Pade
approximants appears to provide a safer fitting strategy \cite{Aubin:2012me}. Alternatively, the use of one-point Pades based on the coefficients of the Taylor expansion of $\Pi(Q^2)$ at $Q^2=0$ could also prove effective, as recently emphasized in Ref.~\cite{Chakraborty:2014mwa}, but only if these coefficients are accurately known, not only for the $s,c$ quarks but also for $u$ and $d$.

\newpage

\subsection{ $\pi^0\rightarrow e^-e^+$ decay implications on $F_{\pi^0\gamma^*\gamma^*}(Q_1^2,Q_2^2)$}
\addtocontents{toc}{\hspace{2cm}{\sl P.~Sanchez-Puertas}\par}

\vspace{5mm}

P.~Masjuan and \underline{P.~Sanchez-Puertas}

\vspace{5mm}

\noindent
PRISMA Cluster of Excellence and Institut f\"ur Kernphysik, Johannes Gutenberg-Universt\"at Mainz, Germany \\

\vspace{5mm}
\setcounter{equation}{0}
The precision to which the $(g-2)_{\mu}$ is measured~\cite{Bennett:2006fi} makes it a very interesting quantity: being sensible to all the SM sectors, it probes our understanding of fundamental physics. 
The precision expected for future experiments demands an accurate theoretical calculation, which is particularly challenging for the $(g-2)_{\mu}^{HLbL}$ contribution. This last contribution 
has been modeled based on large-$N_c$ ideas~\cite{Jegerlehner:2009ryp}, where the $\pi^0$-pole contribution is the leading term. Therefore, the 
$F_{\pi^0\gamma^*\gamma^*}(Q_1^2,Q_2^2)$ description is fundamental. This last  is known at the limits $F_{\pi^0\gamma\gamma}(0,0), F_{\pi^0\gamma^*\gamma}(-\infty,0)$ 
and $F_{\pi^0\gamma^*\gamma^*}(-\infty,-\infty)$ from $\chi PT$~\cite{Adler:1969gk}, pQCD~\cite{Lepage:1980fj} and OPE~\cite{Novikov:1983jt} respectively. However, the intermediate and 
low-energy regimes are not well-understood from first principles and must be modeled~\cite{Jegerlehner:2009ryp}. Here, experimental data on the TFF is crucial. Regretfully, the 
lack of experimental data for the double-virtual case, leaves unconstrained parameters which translate into large-ignored uncertainties for  $(g-2)_{\mu}^{HLbL;\pi^0}$.\\

We propose to supply this lack of information using the $\pi^0\rightarrow e^-e^+$ decay. Proceeding through an 
intermediate $2\gamma$ loop, it probes the $\pi^0\gamma^*\gamma^*$ vertex which involves the double-virtual TFF.
We find that no model (see~\cite{Dorokhov:2009xs} and references therein) is available to reproduce the experimental value~\cite{Abouzaid:2006kkp}. Still, those models with free parameters may be adjusted to yield the closest possible result, 
which, in the better case, lie ultimately $3\sigma$ away from experiment. Nevertheless, these modifications have an impact on the $(g-2)_{\mu}^{HLbL;\pi^0}$ prediction (i.e.: ${\mathcal{O}}(20\%)$ for~\cite{Knecht:2001qf}).\
 
With the aim of improving the TFF description, we extend the systematic approach described in~\cite{Masjuan:2012wy,Escribano:2013kba} for $F_{\pi^0\gamma^*\gamma}(Q^2,0)$ 
based on Pad\'e Approximants (PA)~\cite{Baker}, to the double virtual case. 
Such method is a powerful tool, providing an excellent description for the TFF at the low energies relevant for $(g-2)_{\mu}^{HLbL;\pi^0}$, and allowing for the high-energy 
behavior implementation. We choose two different approaches. The first one (I) is the standard factorization approach~\cite{Knecht:2001qf}, 
$F_{\pi^0\gamma^*\gamma^*}(Q_1^2,Q_2^2) \sim F_{\pi^0\gamma^*\gamma}(Q_1^2,0)\times F_{\pi^0\gamma^*\gamma}(0,Q_2^2)$, given at lowest order by
\begin{equation}
 \label{eq:modelI}
 \textrm{Approach I:} \ \ \ \ \ F_{\pi^0\gamma^*\gamma^*}(Q_1^2,Q_2^2) = \frac{1}{1 + aQ_1^2}\frac{1}{1 + aQ_2^2},
\end{equation}
where $a$ is the TFF slope~\cite{Masjuan:2012wy}. The second one (II) is based on Chisholm Approximants (CA), a natural extension of PA for two variables~\cite{Chisholm}. 
At higher order, is very similar to the construction in~\cite{Babu:1982yz}, to lowest order (OPE constrains $b=0$) we have
\begin{equation}
\label{eq:modelII}
 \textrm{Approach II:} \ \ \ \ \ F_{\pi^0\gamma^*\gamma^*}(Q_1^2,Q_2^2) =  \frac{1}{1+a(Q_1^2 + Q_2^2) + bQ_1^2Q_2^2} \rightarrow \frac{1}{1+a(Q_1^2 + Q_2^2)}.
\end{equation}
Taking the slope from~\cite{Masjuan:2012wy} to 
obtain $a$~\cite{Masjuan:2007ay}, we find  for $\pi^0\rightarrow e^-e^+$ that $BR = 6.36(5)10^{-8}$ ($6.22(7)10^{-8}$) for approach I(II). Therefore, we quote 
\begin{equation}
BR=(6.36(5)\div6.22(7))10^{-8}, 
\end{equation}
where the errors in parenthesis refer 
to the slope statistic and systematic errors, while the band refers to the uncertainty due to the factorization of $F_{\pi^0\gamma^*\gamma^*}(Q_1^2,Q_2^2)$.
This is to date the most precise estimate if either approach I or II is assumed, and incorporates,  for the first time, systematic uncertainties. Compared to the experiment, 
this represents a  $3\sigma$ deviation. A better agreement would be found for a faster decreasing TFF. This contrast with $(g-2)_{\mu}^{HLbL;\pi^0}$, which is enhanced by a slowly decreasing TFF. 
We find $(g-2)_{\mu}^{HLbL;\pi^0} = (5.53(27)\div6.64(28))\times10^{-10}$ for I(II), which reflects the large uncertainty in reconstructing 
the double virtual TFF. This represents an additional non-considered uncertainty, which for~\cite{Jegerlehner:2009ry} reads
\begin{equation}\label{numbererror}
a_{\mu}^{HLbL} = 116(39)\times10^{-11} \rightarrow 116(39)(11)\times10^{-11}.
\end{equation}
To improve in precision, new experimental data on the double virtual TFF is needed. In order to incorpore such information, we intend~\cite{PMPS} to use Chisholm's method. This is 
an approach towards a model-independent reconstruction of the most general TFF. Finally, we remark that the extension to the $\eta$ case is straightforward in this approach and represents another source of ignored systematic error to sum up into Eq.~(\ref{numbererror}).

\newpage

\subsection{Motivation and status of the planned muon $g-2$ experiments}
\addtocontents{toc}{\hspace{2cm}{\sl D.~St\"ockinger}\par}

\vspace{5mm}
D.~St\"ockinger

\vspace{5mm}

\noindent
Institut f\"ur Kern und Teilchenphysik, TU Dresden, Dresden, D-01062, Germany.

\vspace{5mm}
\setcounter{equation}{0}

Magnetic moments in general and the muon anomalous magnetic moment
$a_\mu=(g_\mu-2)/2$ in particular are clean and sensitive probes of
fundamental particles and interactions. After the Brookhaven
measurement, $a_\mu$ is sensitive to all interactions of the Standard
Model of particle physics. The observed deviation from the Standard
Model theory prediction might be due to physics beyond the
Standard Model (BSM), but at the same time it constrains
BSM scenarios. A new generation of $a_\mu$ measurements will further
increase the experimental accuracy and the sensitivity to SM and BSM
physics. The goal of the workshop is to initiate and contribute to
progress on the SM theory prediction of $a_\mu$, and in the following
paragraphs we will give a reminder of the current status and the
motivation for further improvement.

Huge progress has been achieved on the SM theory prediction of $a_\mu$ in the past years. We highlight the 5-loop
QED computation \cite{Kinoshita2012}, the inclusion of high-precision
$e^+e^-$-data into the hadronic vacuum polarization contributions
\cite{Davier,HMNT,Benayoun:2012wc}, the resolution of the
$\tau$-vs.-$e^+e^-$-puzzle \cite{Benayoun:2012wc,JegerlehnerSzafron},
and the exact evaluation of the electroweak contributions after the
Higgs boson mass measurement \cite{Gnendiger:2013pva}.
As a result of this progress, the SM theory prediction has a smaller
uncertainty than the Brookhaven measurement, but the precision of the
hadronic contributions needs to be further improved to match the new
experiments.

One new $a_\mu$ measurement will be carried out at Fermilab \cite{Carey:2009zzb}. It
combines the technique of the Brookhaven experiment with specific
advantages present at Fermilab. Datataking is
expected to start in 2017. A second promising experiment is planned at
J-PARC. It would make use of an entirely complementary strategy and
therefore provide important cross-checks.
Both experiments promise to reduce the uncertainty
by a factor four, down to a level less than half as large as the
current SM theory uncertainties coming from the hadronic vacuum
polarization and hadronic light-by-light contributions. 

Measuring and computing the SM prediction for $a_\mu$ as precisely as
possible is very important also to study hypothetical new physics
scenarios. This statement is independent of whether the current
deviation will increase or decrease.
The importance of $a_\mu$ as a constraint on BSM physics 
is due to two facts. First, different types of BSM physics can
contribute to $a_\mu$ in very different amounts, so $a_\mu$
constitutes a meaningful benchmark  and discriminator between BSM
models. Second, the 
constraints from $a_\mu$ on BSM models are different and
complementary to constraints from other observables from the
low-energy and high-energy frontier.

\begin{figure}
\centerline{\setlength{\unitlength}{1cm}
\begin{picture}(9,5.5)
\epsfysize=6.8cm
\put(0,0.7){\epsfbox{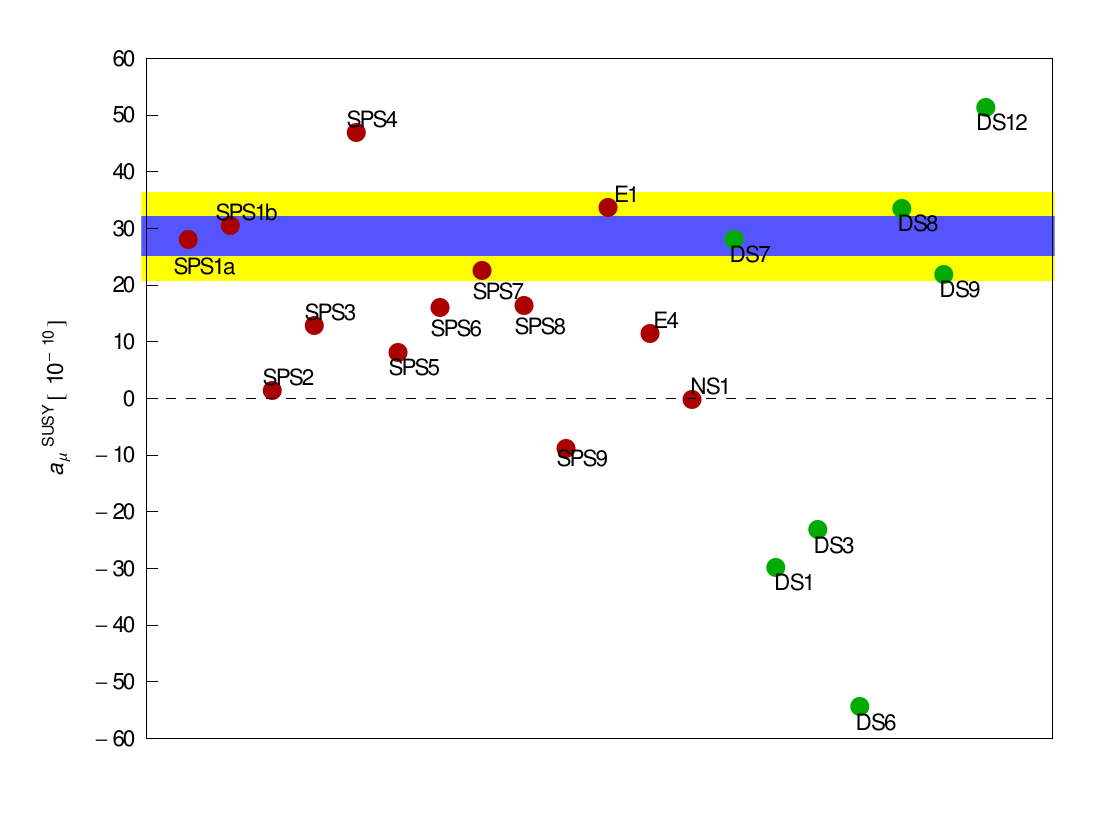}}
\end{picture}}
\caption{\label{fig:susyplots} SUSY contributions to $a_\mu$ for
the SPS and other benchmark points (red), and for the ``degenerate solutions''
from Ref.\ \cite{Adam:2010uz}. The yellow and blue bands are the $\pm 1~\sigma$ errors
from the Brookhaven and the planned Fermilab measurements. } 
\end{figure}

Both aspects can be illustrated within the framework of supersymmetric
models, as shown in Figure \ref{fig:susyplots}. The red points in the
Figure show that the $a_\mu$-predictions of various benchmark
scenarios proposed in the literature scatter widely. Any future
measurement of $a_\mu$ will rule out many of these points,
illustrating the discriminating power of $a_\mu$. The green points in
the Figure illustrate the complementarity of $a_\mu$. In the
hypothetical scenario considered in \cite{Adam:2010uz}, the LHC can
find most supersymmetric particles and measure their masses, and yet
there are several very different choices of supersymmetric parameters
which give an equally good fit to LHC data. The $a_\mu$-predictions of
these ``degenerate solutions'' however, differ, hence allowing to lift
the LHC degeneracies by taking into account $a_\mu$.

\newpage

\section{Summaries of the talks \emph{$g-2$ Quo vadis?} Workshop}

\subsection{The role of radiative corrections in hadronic production}
\addtocontents{toc}{\hspace{2cm}{\sl H.~Czy\.z}\par}

\vspace{5mm}
H.~Czy\.z

\vspace{5mm}

\noindent
Institute of Physics, University of Silesia, Katowice, Poland
\vspace{5mm}
\setcounter{equation}{0}

The constantly improving experimental accuracy in measurements
 of the hadronic cross
section, both with the scan and the radiative return method as well as 
 of the hadron production in two-photon scattering require controlling 
 of the radiative corrections in Monte Carlo generators
 at the unprecedented level (for a review see \cite{Actis:2010gg}).
 The team working on PHOKHARA Monte Carlo
 event generator started the physics program in  2000 \cite{Czyz:2000wh}
 extending the EVA generator \cite{Binner:1999bt} based on structure function
 method to 4$\pi$ final states. It was soon clear that the method is
 not accurate enough especially for the experimental configurations 
 with photon tagging and very sophisticated event selections, thus the group
 decided to switch to fixed order exact matrix elements. 
 The initial state radiative corrections at NLO 
 \cite{Rodrigo:2001jr,Kuhn:2002xg}, universal for all final states,
  were added in
  \cite{Rodrigo:2001kf,Czyz:2002np}. The NLO corrections 
 involving mixed photons
 emission one from initial and one from final states and 
 corresponding virtual corrections were added for $\pi^+\pi^-$ 
 \cite{Czyz:2003ue}, $\mu^+\mu^-$ \cite{Czyz:2004rj} and $K^+K^-$ 
 \cite{Czyz:2010hj} production. Finally the complete NLO radiative corrections
 for $\mu^+\mu^-$ production
 were added in \cite{Campanario:2013uea} and a version of the generator
 suitable for scan measurements was prepared in \cite{Czyz:2013xga}.
 In \cite{Campanario:2013uea} it was confirmed that
  the corrections coming from penta-box diagrams, expected to be small,
 are indeed below 0.1\% for KLOE event selections 
\cite{Babusci:2012rp,Ambrosino:2010bv}
 and below 0.25\%
 for BaBar event selections \cite{Lees:2012cj}.
  It was shown also that they can  potentially reach a level
 of 1-2\% for different event selections. It was an important check
 as the discrepancy between KLOE~\cite{Babusci:2012rp,Ambrosino:2010bv}
 and BaBar \cite{Lees:2012cj} extraction of the pion-pair cross section
  using radiative return method might have been partly caused by using
  a Monte Carlo generator with non-complete radiative corrections.
  It is to be checked in future that the corrections are also small for
  pion-pair production. As some threshold enhancements might occur 
  in penta-box diagrams, in this case the corrections might be slightly
 bigger than the aforementioned corrections to muon-pair production.
 It has to be stressed that at the accuracy better
  than 1\% the size of the radiative corrections can be studied only 
  with realistic event selections as the corrections do strongly depend  
  on them. As a result, using a Monte Carlo generator is indispensable.  
  For the processes $e^+e^-\to e^+e^-+\ {\rm hadrons}$ there exists only
 one Monte Carlo event generator containing radiative
 corrections in integrated form \cite{Druzhinin:2010er}. As this cross
 section is very much sensitive to kinematic variables, it is necessary
  to check to what extent event selections change their size. For this scope
 exclusive radiative corrections are being implemented \cite{Ivashyn:2013uja} 
 in the
 Monte Carlo event generator EKHARA \cite{Czyz:2006dm,Czyz:2010sp}.

\newpage

\subsection{Electromagnetic form factors in Dual-Large $N_c$-QCD}
\addtocontents{toc}{\hspace{2cm}{\sl C.A.~Dominguez}\par}

\vspace{5mm}
C.A.~Dominguez

\vspace{5mm}

\noindent
Centre for Theoretical \& Mathematical Physics, and Department of Physics, University of
Cape Town, Rondebosch 7700, South Africa
\vspace{5mm}
\setcounter{equation}{0}

It is well known that in QCD and for an infinite number of colours, $QCD_\infty$, a typical form factor has the generic form
\begin{equation}
F(s) = \sum_{n=0}^{\infty}
\frac{C_{n}}{(M_{n}^{2} -s)} \;,
\end{equation}
where $s \equiv q^2$ is the momentum transfer squared, and  the masses $M_n$, and the couplings $C_n$ remain unspecified. 
In Dual-$QCD_{\infty}$ they are given by \cite{CAD1}
\begin{equation}
C_{n} = \frac{\Gamma(\beta-1/2)}{\alpha' \sqrt{\pi}} \; \frac{(-1)^n}
{\Gamma(n+1)} \;
\frac{1}{\Gamma(\beta-1-n)} \;, 
\end{equation}
where $\beta$ is a free parameter, and the string tension $\alpha'$ is $\alpha' = 1/2 M_{\rho}{^2}$,  as it enters the rho-meson Regge trajectory
$\alpha_{\rho}(s) = 1 + \alpha ' (s-M_{\rho}^{2})$.

\begin{figure}[ht]
\begin{center}
\includegraphics[width=4.0 in]{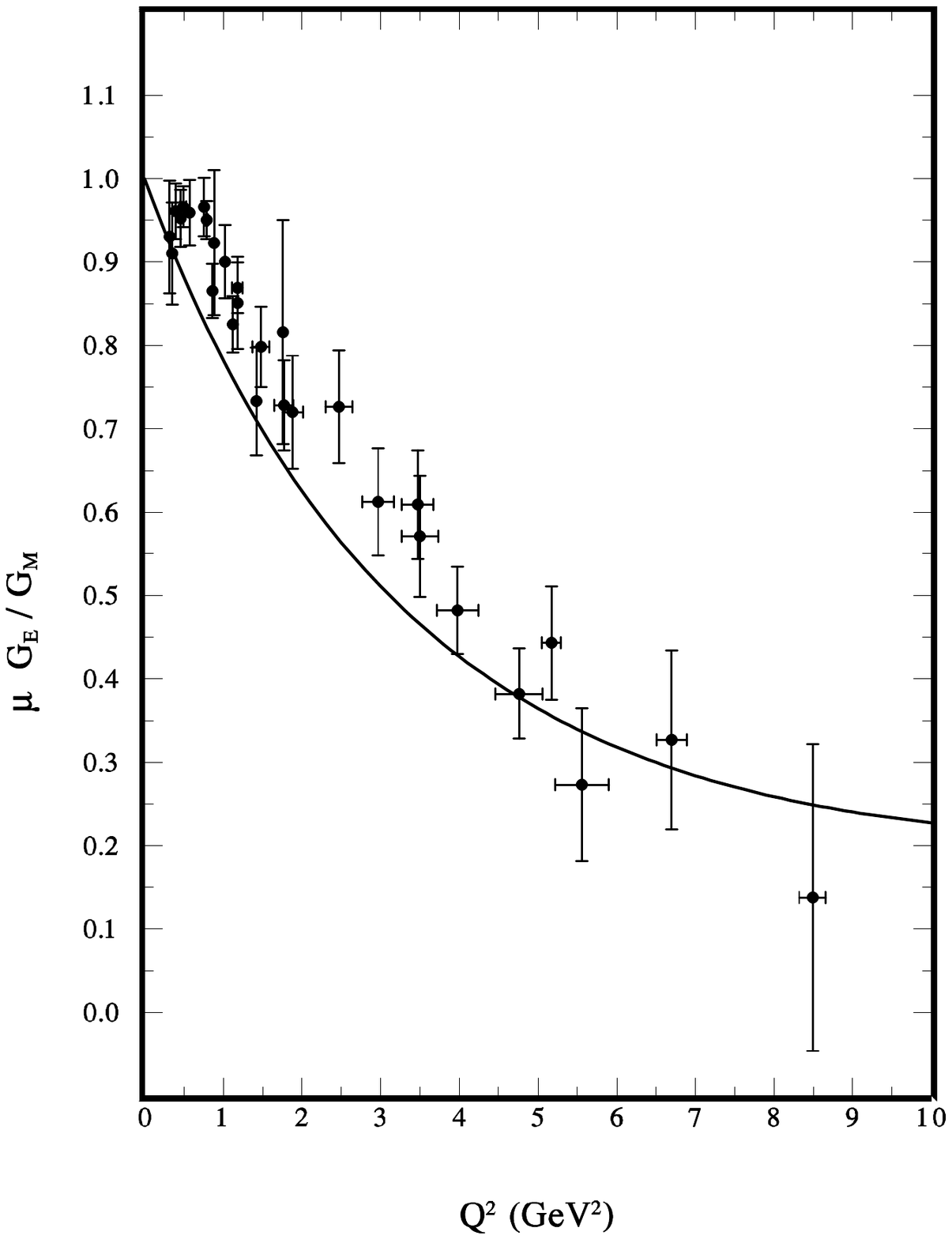}
\caption{Ratio.}
\end{center}
\end{figure}

The mass spectrum is chosen as
$M_{n}^{2} = M_{\rho}^{2} (1 + 2 n)$. This simple formula correctly predicts the first few radial excitations. Other, e.g. non-linear mass formulas could be used \cite{Masjuan}, but this hardly changes the results in the space-like region, and only affects the time-like region behaviour for very large $q^2$. With these choices the form factor becomes an Euler Beta-function, i.e.
\begin{eqnarray}
F(s) &=& \frac{\Gamma(\beta-1/2)}{\sqrt{\pi}} \; \sum_{n=0}^{\infty}\;
\frac{(-1)^{n}}{\Gamma(n+1)} \; \frac{1}{\Gamma(\beta-1-n)} \nonumber\\ 
& \times &\frac{1}
{[n+1-\alpha_\rho(s)]} =\frac{1}{\sqrt{\pi}} \; \frac{\Gamma (\beta-1/2)}{\Gamma(\beta-1)} \nonumber\\ [.2cm]
&\times &
B(\beta - 1,\; 1/2 - \alpha' s)\;,
\end{eqnarray}
where $B(x,y) = \Gamma(x) \Gamma(y)/\Gamma(x+y)$.
 The form factor exhibits asymptotic power behavior in the space-like region, i.e.
\begin{equation}
\lim_{s \rightarrow - \infty} F(s) = (- \alpha' \;s)^{(1-\beta)}\;,
\end{equation}
from which one identifies the free parameter $\beta$ as controlling this asymptotic behaviour. Notice that while each term in Eq.(3) is of the monopole form, the result is not necessarily of this form because it involves a sum over an infinite number of states. The exception occurs for integer values of $\beta$, which leads to a finite sum.
Successful applications are the pion form factor \cite{CAD1}, and the nucleon form factors \cite{CAD2} which after determining the parameters $\beta_1$ and $\beta_2$, corresponding to the form factors $F_1$ and $F_2$ leads to electric and magnetic form factors in excellent agreement with data; in particular the ratio $\mu\, G_E/G_M$ as shown in Fig.2. In addition the form factors of the $\Delta(1236)$ are also well described \cite{CAD3}, as well as radiative decays of mesons \cite{CAD4}. Finally, this model also accounts for fully off-shell three-point functions for arbitrary particles in the vertex. It was shown in \cite{CAD5} that in this case the three-point function factorizes, a result which renders
the model ideal to apply e.g. to form factor evaluations for the light-by-light contribution to the muon $g-2$ anomaly. It should also be mentioned that  Dual-$QCD_\infty$  can be made compatible with the asymptotic logarithmic behaviour expected from perturbative QCD \cite{RB}.
For additional applications of Dual-$QCD_\infty$ see \cite{Ruiz} and many references therein.

\newpage

\subsection{$\gamma\gamma$ physics (experiment)}
\addtocontents{toc}{\hspace{2cm}{\sl S.~Eidelman}\par}

\vspace{5mm}

S.~Eidelman

\vspace{5mm}

\noindent
Budker Institute of Nuclear Physics SB RAS and  \\
Novosibirsk State University, Novosibirsk, Russia 

\vspace{5mm}

The hadronic light-by-light (HLBL) contribution is known to give a large
contribution to the uncertainty of the muon anomalous magnetic 
moment~\cite{lbl1} and its purely theoretical calculations have
strong model dependence. Lately various approaches to the HLBL determination
based on measurements of transition form factors (TFF) were widely discussed,
with  the $P\gamma\gamma$ vertex generally accepted as the most important.
Recently a dispersive formalism for a model-independent evaluation of
the HLBL term was suggested~\cite{lbl2}.  

One is interested in studying the $P\gamma\gamma$
vertex and the related TFF, ${\cal F}_P(q^2_1,q^2_2)$,
at any $q^2_{1(2)}$ and $P=\pi^0,~\eta,~\eta^{\prime}$, where 
the processes studied are $P \to \gamma^{(*)}\gamma^{(*)}$,
$\gamma^{(*)} \to P\gamma^{(*)}$ and $\gamma^{(*)}\gamma^{(*)} \to P$.

In $e^+e^-$ annihilation we study: $e^+e^- \to \gamma^* \to P\gamma$,
~$q^2_1=s>0$ and $q^2_2=0$; 
$e^+e^- \to \gamma^* \to P\gamma^* \to Pl^+l^-$, $l=e,\mu$,
~$q^2_1=s>0$,  $4m^2_l<q^2_2<(\sqrt{s}-m_P)^2$; 
$e^+e^- \to e^+e^-\gamma^{*}\gamma^{*} \to e^+e^-P$
with $q^2_{1(2)}~<~0$.

In VDM (vector dominance model) hadrons are produced via vector mesons,
so any production of vectors $\gamma^* \to V \to P\gamma^{(*)}$ is relevant, 
e.g., $e^+e^- \to V \to P\gamma$ with $q^2_1 \sim m^2_V$ and $q^2_2=0$ 
or $e^+e^- \to V \to Pl^+l^-$ with $q^2_1 \sim m^2_V$ and 
 $4m^2_l~<~q^2_2~<~(m_V-m_P)^2$.

At the $V$ factory radiative decays like $V \to P\gamma$ are 
a copious source  of $P$ decays, e.g., $P \to l^+l^-\gamma$
and $P \to l^+l^-l^+l^-$ can be studied.

Below 1.4 GeV there are detailed studies of the processes
$e^+e^- \to \pi^0\gamma,~\eta\gamma$ by SND~\cite{sndrad1,sndrad2,sndrad3}
and CMD-2~\cite{cmdrad}. Data from the regions beyond the $\rho,~\omega$
and $\phi$ resonances are scarce. There is a single measurement of the 
$\eta$ TFF at 112 GeV$^2$ performed by BaBar~\cite{babarrad}. One expects a 
breakthrough after experiments at VEPP-2000, where recently first results 
above 1.4 GeV were reported by SND~\cite{sndradnew}. For the $\eta^{\prime}$
TFF there are measurements at the peak of the $\phi$ meson (see literature
in Ref.~\cite{KLOEe}) and  at 112 GeV$^2$ from BaBar~\cite{babarrad}.  

Conversion decays to $e^+e^-$ pairs at the peaks of the $\omega~(\rho)$
and $\phi$ mesons were measured by CMD-2~\cite{cmdcon1,cmdcon2,cmdcon3}
and SND~\cite{sndcon1,sndcon2,sndcon3}. Two existing measurements of
conversion decays to muon pairs show bad 
consistence~\cite{dzhelyadin,arnaldi}.  

This work is supported by the Ministry of Education and Science of the
Russian Federation, the RFBR grants 12-02-01032, 13-02-00215   and 
the DFG grant HA 1457/9-1.

\newpage

\subsection{Meson transition form factors at KLOE/KLOE-2}
\addtocontents{toc}{\hspace{2cm}{\sl P.~Gauzzi}\par}

\vspace{5mm}

P.~Gauzzi \\ on behalf of the KLOE-2 Collaboration

\vspace{5mm}

\noindent
Universit\`a di Roma La Sapienza e INFN Sezione di Roma

\vspace{5mm}

At KLOE the Transition Form Factors (TFFs) of the pseudoscalar mesons can
be investigated for time-like $q^2$ by means of the Dalitz decays,
$\phi\to\eta e^+e^-$ and $\phi\to\pi^0 e^+e^-$.
According to Vector Meson Dominance (VMD) the TFFs are parametrized as
$F(q^2)=1/(1-\frac{q^2}{\Lambda^2})$, where $\Lambda$ can be identified
with the mass of the nearest vector meson.   
Different theoretical
models~\cite{Terschlusen:2011pm,Schneider:2012ez,Ivashyn:2011hb}  
predict deviations from VMD for the TFFs of those decays.\\ 
At KLOE we are studying both decays with the 1.7 fb$^{-1}$ of data collected
from 2001 to 2006.
We obtain $Br(\phi\to\eta e^+e^-)=(1.075\pm 0.007\pm
0.038)\times 10^{-4}$, which improves the precision of the measurement of
the CMD-2 and SND experiments.
We extracted the slope from a fit to the $e^+e^-$ invariant mass,
$b_{\eta}=(1.17\pm 0.10^{+0.07}_{-0.11})$ GeV$^{-2}$, in agreement with the VMD
prediction $b_{\eta}\simeq 1$ GeV$^{-2}$.
Our measurement is also consistent with the old SND result
$b_{\eta}=(3.8\pm 1.8)$ GeV$^{-2}$, which has about 50\% uncertainty.
We are also analyzing the $\phi\to\eta\mu^+\mu^-$ and
$\phi\to\eta\pi^+\pi^-$ decays.\\
Concerning $\phi\to\pi^0 e^+e^-$, the branching ratio is known with 25\%
uncertainty from the Novosibirsk measurements, and there are no data
available on the TFF slope.
We selected about 9000 candidate events $\phi\to\pi^0 e^+e^-$: the
measurements of the branching ratio and of $b_{\pi^0}$ are in progress.\\

$\gamma\gamma$ processes ($e^+e^-\to
e^+e^-\gamma^{\star}\gamma^{\star}\to e^+e^-X$) are complementary to
$e^+e^-$ annihilation.
Single and double pseudoscalar production is accessible at the DA$\Phi$NE
energy, $X=\pi^0,\eta$, $X=\pi\pi$. 
The single pseudoscalar production cross-section is related to
$\sigma(\gamma^{\star}\gamma^{\star}\to
P)=\frac{8\pi^2}{m_P}\Gamma(P\to\gamma\gamma)\delta(w^2-m^2_P)\left|F(q^2_1,q^2_2)\right|$,
from which the radiative width $\Gamma(P\to\gamma\gamma)$ and the TFF
$F(q^2_1,q^2_2)$ for space-like $q^2$ can be obtained.
We studied the process $\gamma^{\star}\gamma^{\star}\to\eta$ with the 240
pb$^{-1}$ of data collected off-peak at $\sqrt{s}=1$ GeV, during the
2001-06 data-taking wiyout any specific device to detect the scattered
electrons. 
We exploited both $\eta\to\pi^+\pi^-\pi^0$ and $\eta\to\pi^0\pi^0\pi^0$
decays and we measure $\sigma(e^+e^-\to e^+e^-\eta)=(32.7\pm 1.3\pm 0.7)$ 
pb, from which we extract the radiative width
$\Gamma(\eta\to\gamma\gamma)=(520\pm 20\pm 13)$ eV~\cite{Babusci:2012ik}.\\
With the off-peak sample we are also analyzing the
$\gamma^{\star}\gamma^{\star}\to\pi^0\pi^0$ decay, which is relevant for
the study of the lightest scalar meson $f_0(500)$, and also for the new
dispersive approaches to the calculation of the Light-by-Light scattering
contribution to $(g-2)_{\mu}$~\cite{Colangelo:2014dfa22,Masjuan:2014}.\\
For the KLOE-2 data-taking, among other detectors, two taggers for
$\gamma\gamma$ physics (the Low Energy
Tagger~\cite{Babusci:2009sg}, and the High Energy
Tagger~\cite{Archilli:2010zza}) have been installed.  
With the help of these devices the measurement of the $\pi^0$ radiative
width at about 1\% level, and the measurement of the TFF
$F(q^2,0)$ of the $\pi^0$ with one quasi-real photon and a virtual one, in
the still unexplored range $q^2<0.1$ GeV$^2$, will be
possible~\cite{Babusci:2011bg}.

\newpage

\subsection{$\gamma^{(*)}\gamma^{(*)}\to\pip\pim$ at BESIII}
\addtocontents{toc}{\hspace{2cm}{\sl Y.~Guo}\par}

\vspace{5mm}

Y.~Guo

\vspace{5mm}

\noindent
Institute f\"ur Kernphysik, Johannes Gutenberg-Universit\"at Mainz, Germany\\

\vspace{5mm}

As it is well known, exist $3\sim4$ $\sigma$ deviation for the anomalous magnet
moment of muon ($a_{\mu}$) between experimental and theoretical value. The experimental 
average value based on the measurements from both CERN and E821~\cite{g-2:experimental}, 
which is $a_{\mu}^{\rm exp}=116592089\pm54({}_{\rm stat})\pm 33({}_{\rm sys})\times10^{-11}$. 
The theoretical calculation is composed from QED, weak and hadron contributions. 
The uncertainties from QED and weak contribution is quite small and the main uncertainty 
comes from the hadron contribution, and the calculation of hadronic interaction are 
highly depend on the experimental input~\cite{g-2:review}. The leading hadronic contribution
is the hadronic vacuum polarisation (HVP), and although the hadronic 
light-by-light (HLBL) is contribution is not so large but with large 
uncertainty. Considering there will be an improved measurement of the 
$a_{\mu}$ with the uncertainty 4 times at Fermilab, the 
improvement on the theoretical calculation are highly desired. 

The process $\gamma\gamma\to\pip\pim$ contains the contribution of the charged $\pi$ loop
and also the contribution of the resonances to the hadronic light-by-light part of 
$a_{\mu}$. Recently, a dispersive approach is developed a method which can 
using the experimental information to evaluate the contribution of 
charged $\pi$ process to the $a_{\mu}$~\cite{g-2:LBL:dispersive approach}.
Beside, this process can also be used to study the $\pi^{+}\pi^{-}$ scattering 
effect at the low mass region. Currently, all the measurement about this process are 
performed in the two quasi real photon case, the only measurement at the low
$\pi^{+}\pi^{-}i$ mass region comes from an early measurement at MarkII with large 
uncertainty~\cite{pipi:experiments}. There is no 
information from experimental side about the form factor as a function of the
virtuality of one or two photons. Motivated by this, we launched a study of 
$\gamma^{(*)}\gamma^{(*)}\to\pi^{+}\pi^{-}$ at BESIII, start with one virtual photon 
case.     

In principle, all the data samples collected at BESIII can be used to study this process,
but as the effective cross section for two photon process increase when go to higher 
centre of mass energy, we will use the data samples taken above 4.0 GeV, which corresponding 
to a total luminosity about 3 fb$^{-1}$. 

We studied the possibility of this study using the MC simulation of both 
signal process and background processes. After the selection, we can conclude that the 
background from the vector charmonium or charmonium-like state decay is negligible, 
and the background from continuum process and QED process is also very small. The main 
backgrounds come from the processes $e^{+}e^{-} \to e^{+}e^{-} \mu^{+}\mu^{-}$ and $e^{+}e^{-} \to e^{+}e^{-} \pi^{+}\pi^{-}$ (those not 
come from two-photon process). For the first background channel, this is a QED process,
and there are MC generators such as RADCOR~\cite{RADCOR}, DIAG36~\cite{DIAG36}, 
and so on, but these generators are all developed at high energy region, 
its application in the low energy region still need to be checked. While for the 
second type of background, which has the same final state as our signal channel, 
there is a generator working in progress which can help us
to understand this kind of background~\cite{EKHARA}.

In summary, we studied the possibility of $\gamma^{(*)}\gamma^{(*)}\to \pi^{+}\pi^{-}$ with one
virtual photon case at BESIII. The MC simulation shows we can reach the low $\pi^{+}\pi^{-}$ mass 
spectrum (start from threshold) and also cover a $Q^{2}$ from $0.2$ GeV${}^{2}$ to 
$2.0$ GeV${}^{2}$. In the whole $Q^{2}$ and $\pi^{+}\pi^{-}$ mass region, we may expect about 25000
signal events in data samples above 4.0 GeV at BESIII with a accuracy about 10\% based on 
our understanding of the two main background processes.

\newpage

\subsection{Dispersive approach to hadronic light-by-light scattering: Reconstructing $\gamma^*\gamma^*\to\pi\pi$} 
\addtocontents{toc}{\hspace{2cm}{\sl M.~Hoferichter}\par}

\vspace{5mm}

\underline{M.~Hoferichter}, G.~Colangelo, M.~Procura, and P.~Stoffer

\vspace{5mm}

\noindent
Albert Einstein Center for Fundamental Physics and Institute for Theoretical Physics,
	    Universit\"at Bern, Switzerland\\

\vspace{4mm}

Crucial ingredients for a dispersive analysis of hadronic light-by-light scattering~\cite{Colangelo:2014dfa} are data on $\gamma^*\gamma^*\to \text{hadrons}$, in particular for light pseudoscalars $\pi^0, \eta, \eta'$ and two-meson states $\pi\pi, K\bar K$. The experimental input for the latter is required in terms of partial waves for the helicity amplitudes. While for the on-shell case $\gamma\gamma\to\pi\pi$ data are sufficiently good to allow for a partial-wave analysis~\cite{Dai:2014zta}, this will not be possible for the singly- and doubly-virtual processes in the foreseeable future.   

Therefore, the partial waves for the virtual processes need to be reconstructed again by means of dispersive techniques, see~\cite{GM,Hoferichter:2011wk} for the on-shell case and~\cite{Moussallam13} for the generalization to the $S$-wave of the singly-virtual process. 
In the time-like region of $\gamma^*\gamma^*\to\pi\pi$ the analytic structure of the amplitudes is affected by anomalous thresholds, which can be taken into account according to~\cite{Hoferichter:2013ama}. We point out that the reconstruction of the left-hand cut for $2\pi$ and $3\pi$ intermediate states requires knowledge of processes relevant also for the pion transition form factor~\cite{Kubis}, i.e.\ $\omega,\phi\to\pi^0\gamma^*$ and $\gamma^*\to3\pi$. 

An additional complication concerns the subtraction terms, which become functions of the virtualities $q_i^2$ of the photons. While some of these subtraction functions are genuinely doubly-virtual and can only be extracted from doubly-virtual measurements (or constrained by ChPT at low energies), the full $q_i^2$-dependence of those functions already present in the singly-virtual case can even be reconstructed dispersively. 
Thus, a combination of singly-virtual measurements, ChPT, and possibly (limited) doubly-virtual data should allow for 
a sufficiently accurate determination of the subtraction functions in $\gamma^*\gamma^*\to\pi\pi$.

\newpage
\subsection{R-scan programme at BESIII}
\addtocontents{toc}{\hspace{2cm}{\sl G.~Huang}\par}

\vspace{5mm}

G.~Huang (For the BESIII Collaboration)

\vspace{5mm}

\noindent
University of Science and Technology of China, Hefei 230026, China 

\vspace{5mm}

The BESIII experiment \cite{bes3} has been in operation since 2009.
It locates at the BEPCII $\ee$ collider in Beijing, China,
running in a center-of-mass energy range from 2.0 GeV to 4.6 GeV.
So far the world largest samples of $J/\psi$, $\psi(3686)$, $\psi(3770)$, 
$\psi(4040)$, $\psi(4415)$ have been collected. There are also samples
above open charm threshold for the study of the exotic XYZ states and
$R$ measurement, and samples below 3 GeV for QCD study.

The program of $R$ measurement and QCD study at BESIII has a 3-phase
design. The first phase is a test run, the second the low energy continuum
region, and the third the high mass charmonia region.

The purpose of the test run is mainly for machine study, so that a detail
scan plan can be made based on the performance of BEPCII. With the data 
taken at a few energy points, supposed to cover the whole energy range
as possible, the analysis chain of the $R$ measurement can be established
as well, including parameter tuning of Monte Carlo (MC) generators. In 2012,
data at 4 energies, 2.23, 2.4, 2.8 and 3.4 GeV, were taken, with a total
integrated luminosity 11.5 pb$^{-1}$. Together with high energy data
for XYZ study later on, the first phase of data taking has been essentially
finished.

Since for the BEPCII the priority is to reach its design goal at $\psi(3770)$,
the second phase for $R$ scan was changed to the resonant region, and it
has been completed in 45 days in the beginning of the 2013 - 2014 run. 
Data at 104 points were taken, with the energy range from 3.85 to 4.59 GeV, 
the step size as small as 2 MeV, $\sim$100$k$ observed hadronic events at 
each energy, and total integrated luminosity $\sim$800 pb$^{-1}$.

The next phase to do is the continuum region, and the data taking plan
has been approved by the BESIII Collaboration. Because of lower peak 
luminosity, there will be much fewer points in the low energy range,
roughly, about 20 points, total integrated luminosity around 500 pb$^{-1}$.
The physics topics include but not limited to: $R$ measurement, nucleon
form factor, hyperon form factor, hyperon-pair threshold production,
search for $\ee \to \eta_c$, likely Y(2175), etc. The luminosity at
each point is optimized mainly for proton form factor measurement,
to supersede the BaBar result, and for polarization/phase 
measurement of $\Lambda$, on top of its form factor measurement. 
The low energy data taking would need 1 full run-year of BESIII, 
i.e., 6 months.

With the small amount data from the test run, a number of analyses
has been carried out, including proton form factor measurement
and $\Lambda$ pair production at threshold, but eventually the results will be significantly improved using much larger 
data samples being expected. Efforts are also being paid to extract the 
open charm cross sections, study the high mass charmonium resonances, 
using the high energy scan data.

\newpage
\subsection{Measurement of hadronic cross sections using initial state radiation at BESIII}
\addtocontents{toc}{\hspace{2cm}{\sl B.~Kloss}\par}

\vspace{5mm}

A.~Denig, \underline{B.~Kloss}

\vspace{5mm}

\noindent
Institut f\"ur Kernphysik, Johannes Gutenberg Universt\"at Mainz, Germany

\vspace{5mm}

Cross sections of the form $e^+e^-\rightarrow hadrons$ are an important input for the standard model prediction of the hadronic contribution to the anomalous magnetic moment of the muon $a_\mu$ \cite{Jegerlehner}. The hadronic contribution caused by vacuum polarization can be calculated with a dispersion integral
\begin{equation}
	a_\mu^{hadr} \cong \frac{1}{4\pi^3}\int_{4m_\pi^2}^\infty K(s)\sigma (e^+e^-\rightarrow hadrons)ds
\end{equation}
where $K(s)\propto\frac{1}{s}$ is the so called Kernel Function. The experimental uncertainty in these hadronic cross sections limits the standard model prediction completely. \\
\\
The largest contribution to the absolute value of $a_\mu^{hard}$ comes from cross sections at an energy below 1 GeV, i.e. the $\pi^+\pi^-$ cross section. This one has been measured with high precision at the BaBar, KLOE and CMD2 experiments \cite{BaBar,KLOE,CMD2}. For the error $\Delta a_\mu^{hadr}$ contributions between 1 and 2 GeV get more important, which means the $\pi^+\pi^-\pi^0$, $\pi^+\pi^-\pi^0\pi^0$ and $\pi^+\pi^-\pi^+\pi^-$ final states. Our goal is to measure these cross sections at the BESIII experiment \cite{design and construction of BES-III} with a very high precision.\\
\\
Therefor we want to use the technique of Initial State Radiation \cite{ISR}. If a photon is emitted in the initial state the center of mass energy is lowered by the energy of the emitted photon. So measurements of cross sections at different energies are possible although the collider has a fixed cms energy. By measuring the ISR cross section it is then possible to extract the non-radiative cross section which is the input for the dispersion integral via
\begin{equation}
	\frac{d\sigma_{ISR}(M_{hadrons})}{dM_{hadrons}} = \frac{2M_{hadrons}}{s}\cdot W(s,x,\theta_\gamma) \cdot \sigma(M_{hadrons})
\end{equation}
where $M_{hadrons}$ is the invariant mass of the hadronic system and $W$ the so called Radiator Function which gives the probability that the ISR photon is emitted with a specific energy fraction $x$ and angle $\theta_\gamma$. For the Monte-Carlo prediction we are using the ISR generator PHOKHARA 7.0 \cite{PHOKHARA1,PHOKHARA2,PHOKHARA3}.\\
\\
At a cms energy $s$ = 3.773 GeV a data set of 2916 pb$^{-1}$ \cite{lumi} has been taken at the BESIII experiment where currently the $\pi^+\pi^-$, $\pi^+\pi^-\pi^0$, $\pi^+\pi^-\pi^0\pi^0$ cross sections are under investigation. We hope that we are able to make a contribution to the precise measurement of these hadronic cross sections.

\newpage

\subsection{$\gamma\gamma$ physics at Belle }
\addtocontents{toc}{\hspace{2cm}{\sl Z.Q.~Liu}\par}

\vspace{5mm}

Z.Q.~Liu

\vspace{5mm}

\noindent
Institute of High Energy Physics, Beijing, 100049 \\

\vspace{5mm}

The Belle experiment is an asymmetric $e^+e^-$ experiment, which is designed for $B$ meson CPV
study~\cite{belle1}. Benefit from its high luminosity, $\gamma\gamma$ physics also becomes available, 
which has been investigated both for hadron spectroscopy and meson form factor measurement. With more than 10 years running, Belle has accumulated more than 1000 fb$^{-1}$ data near $\Upsilon(nS)$, 
$n=1,~\cdots,~5$, all of which could be used for $\gamma\gamma$ study.

For the $\gamma\gamma$ physics in a $e^+e^-$ machine, typically there are two categories. One is
for quasi-real photons with small $Q^2$ transfer ($Q^2<<W^2_{\gamma\gamma}$). Usually the
electron (positron) are with small scattering angle, and thus was lost in the beam direction. The
non-tag method (nether electron nor positron was detected) is suitable for quasi-real photons
collision study. The other category is for high virtuality photons with large $Q^2$ transfer 
($Q^2>>W^2_{\gamma\gamma}$). In this case, either one or both electron/positron are with 
large scattering angle, and thus can be captured by the detector. Usually the single-tag (either electron
or positron detected) or double-tag (both electron and positron detected)
method will be employed to identify signal process.

Using quasi-real photons, the $\gamma\gamma\to D\bar{D}$ process has been explored with 395~fb$^{-1}$
data at Belle. We found a resonance (named $Z(3930)$) in the $M(D\bar{D})$ invariant mass distribution, 
and its mass is measured to be $M=3929\pm5\pm2$~MeV/c$^2$, and 
width $\Gamma=29\pm10\pm2$~MeV~\cite{z3930}.
Further study of the angular distribution of final particles suggest a spin assignment $J=2$ for $Z(3930)$,
which is consistent with a excited $P$-wave charmonium state $\chi_{c2}(2P)$. The measured 
two-photon width product branching ratio
$\Gamma_{\gamma\gamma}\cdot\mathcal{B}$$(Z(3930)\to D\bar{D})=0.18\pm0.05\pm0.03$~keV 
also agrees with potential model prediction~\cite{potential}. Another success for hadron spectroscopy search
using quasi-real photons is the observation of a resonance structure $X(3915)$ in 
$\gamma\gamma\to\omega J/\psi$ process, based on a data sample with 694~fb$^{-1}$~\cite{x3915}.
With $\omega\to\pi^+\pi^-\pi^0$ and $J/\psi\to\ell^+\ell^-$, the $X(3915)$ resonance is observed
with $7.7\sigma$ significance. The mass of $X(3915)$ is measured to be 
$M=3915\pm3\pm2$~MeV/c$^2$,  and width $\Gamma=17\pm10\pm3$~MeV. The good news
is that $BABAR$ has confirmed its existence, and further determined the spin-parity of $X(3915)$ 
to be $J^{PC}=0^{++}$~\cite{babar-x3915}. Thus, it suggests that $X(3915)$ might be another 
missing $P$-wave charmonium state $\chi_{c0}(2P)$.

In addition to the conventional hadron spectroscopy, we also try to search for exotic hadrons, such
as four quark states using quasi-real photon at Belle. With a data sample of 870~fb$^{-1}$, the
production cross section of $\gamma\gamma\to \omega\phi,~\phi\phi,~\omega\omega$ was measured
up to 4~GeV~\cite{gg2vv}. Near vector meson pairs production threshold, we observe obvious 
enhancements, which are quite different from existing theoretical calculations, such as tetraquark 
model~\cite{tetra}, $t$-channel factorization model~\cite{t-chan} and one-pion-exchange 
model~\cite{one-pion} and so on. Further spin-parity analysis of these 
enhancements show there are resonant like tensor components ($J=2$), and continuum like scalar 
components ($J=0$). In the higher energy range, we observe $\eta_c/\chi_{c0}/\chi_{c2}\to \phi\phi$,
and the first evidence for charmonium state $\eta_c\to \omega\omega$. The continuum production
cross section of vector meson pairs in higher energy range are also measured, and fitted with a power law 
$1/W^{-n}$ behavior. The fits give $n=7.2\pm0.6,~8.4\pm1.1,~9.1\pm0.6$ for $\omega\phi,~\phi\phi$
and $\omega\omega$, respectively. These measurements agree with pertubative QCD prediction 
($n\sim 8 - 10$) very well~\cite{pQCD}.

The $\pi^0$ time-like transition form factor (TFF) provides an ideal test for QCD asymptotic behavior.
Currently, pertubative QCD predicts $Q^2F(Q^2)=\sqrt{2}f_\pi\simeq0.185$~GeV, when $Q^2\to\infty$.
A recent measurement of $\pi^0$ TFF by $BABAR$ shows deviation in the high $Q^2$ range, 
which is beyond standard QCD prediction and may suggest new physics~\cite{babar-tff}. In this 
situation, Belle perform the same measurement of $\pi^0$ TFF. Using single-tag method, 
Belle extracted $\gamma\gamma^*\to\pi^0$ events in different $Q^2$ range based on a data sample
of 759~fb$^{-1}$~\cite{belle-tff}. Significant $\pi^0$ production was observed both for 
electron-tag and positron-tag
case, and $\gamma\gamma^*\to\pi^0\pi^0$ background was subtracted based on the same data set.
By combining both electron-tag and positron-tag events, Belle finally measured the 
$Q^2$ dependent $\pi^0$ TFF. No rapid growth above $Q^2>9$~GeV$^2$ is observed,
which differs from $BABAR$'s measurement, and agrees with QCD asymptotic prediction~\cite{pi0}.

\newpage

\subsection{Dispersion formalism for $\gamma^{*} \gamma^{*} \rightarrow \pi \pi$ }
\addtocontents{toc}{\hspace{2cm}{\sl P.~Masjuan}\par}

\vspace{5mm}

P.~Masjuan

\vspace{5mm}

\noindent
PRISMA Cluster of Excellence, Institut f\"ur Kernphysik and Helmholtz~Institut~Mainz, Johannes Gutenberg-Universit\"at Mainz,
D-55099 Mainz, Germany \\

\vspace{5mm}

In this talk we present a first step towards a dispersion formalism for the $\gamma^{*} \gamma^{*} \rightarrow \pi \pi$ process. This process is not only interesting by its own, as it will be measured for both one- and two-virtual photons in BESIII and Belle with high precision and encodes at once several interesting aspects (gauge invariance, final-state interactions, form factors), but also for its potential relation to the hadronic light-by-light~\cite{Colangelo:2014dfa2}. 

Our goal is to provide a friendly useful parameterization for such process based on dispersion relations~\cite{Pennington,Drechsel:1999rf,GarciaMartin:2010cw2,Hoferichter:2011wk2,Dai:2014zta2}  while keeping its essential ingredients~\cite{AMV}. This demands identifying the crucial pieces of information that allow for a reliable, albeit not complete, description of the current~\cite{belle_pic2,belle_pin2} and forthcoming data at BESIII~\cite{yguo}. A thoroughly analysis of the most complete works along these lines~\cite{GarciaMartin:2010cw2,Hoferichter:2011wk2}, and taking into account only the $\pi\pi$ channel (neglecting then any inelasticity up to almost $1$ GeV), suggests to neglect the $K\bar{K}$ channel, the left-hand cut contributions beyond the one-pion exchange~\cite{Moussallam:2013una2} and the coupling between the S and D partial waves~\cite{Hoferichter:2011wk2}. 

Keeping that, we construct an unsubtracted dispersion relation as in Ref.~\cite{Drechsel:1999rf} but with both S- and D-waves unitarized using, thanks to the Fermi-Watson theorem, the $\pi\pi$ phase-shift, solutions of which are taken from Ref.~\cite{GarciaMartin:2011cn2}. Since, however, we want to isolate the $\pi\pi$ channel in front of the $K\bar{K}$ one, such face shifts should be modified. For that we use the proposal of Ref.~\cite{Moussallam:2013una2} to define a piecewise function.

Dealing with virtual photons demands to extend the standard formalism for $\gamma \gamma \rightarrow \pi \pi$ which has two independent helicity amplitudes to three (one-virtual photon) and five (two-virtual photons) independent amplitudes. However, not all of the new ones contribute the same way and we identify, among them, the longitudinal-longitudinal components to be the most relevant while neglecting the others. This new amplitude yields an enhancement on the cross section of the same order as the transversal amplitudes. Such enhancement helps to slightly compensate the dramatic decrease of the cross section due to the suppression from kinematics in presence of photon virtualities together with the photon (vector) form factor (and the modification of the Cauchy kernels of the dispersive amplitudes from the soft-photon limits~\cite{Moussallam:2013una2}). We conclude~\cite{AMV} that such cross section can be measured at BESIII.

Other input ingredients are the coupling of the virtual photon with the pion (given by the vector form factor). We use a data driven paramaterization specially suited for low energies from Ref.~\cite{Masjuan:2008fv} as input. And the $f_2(1275)$ tensor resonance, which shows up in the the $\gamma \gamma \rightarrow \pi \pi$ process around $1.2$GeV. Even though we are concerned with the low-energy sector of the process, we also include such effect in our formalism as it was done in Ref.~\cite{Drechsel:1999rf} but including the five helicity components of the amplitude. The problem now is that not all of them are known. A recent sum rule~\cite{Pascalutsa:2010sj2} suggests to consider only the transversal component which, though is not known either, can be parameterized in terms of the $\eta$ and $\eta'$ form factors~\cite{Escribano:2013kba2}.

\newpage
\subsection{Muon $g$-$2$/EDM measurement at J-PARC }
\addtocontents{toc}{\hspace{2cm}{\sl T.~Mibe}\par}

\vspace{4mm}

T.~Mibe for the J-PARC $g$-2/EDM collaboration

\vspace{4mm}

\noindent
Institute of Particle and Nuclear Studies, KEK, Tsukuba, Japan\\

\vspace{2mm}

The J-PARC experiment E34 aims to measure the anomalous magnetic moment ($g$-2) and electric
dipole moment (EDM) of the positive muon with a novel
technique utilizing an ultra-cold muons accelerated to 300 MeV/$c$ and a
66 cm-diameter compact muon storage ring without focusing-electric
field~\cite{E34CDR}. This measurement will be complementary to the previous BNL E821~\cite{Bennett2006}
experiment and upcoming FNAL E989~\cite{FNAL_E989} with the muon beam at the
magic momentum 3.1~GeV/$c$ stored in a 14 m-diameter storage ring. The E34 experiment
aims to achieve the sensitivity down to 0.1 ppm for $g$-2, and $10^{-21}$ $e\cdot$cm for EDM.
The new approach to be used in the E34  removes the necessity of the magic muon momentum,
thanks to the fact that focusing electric field can be turned off.
Such a condition is realized by utilizing a ultra-low emittance beam (ultra-cold muon beam) that is generated by accelerating 
ultra-slow muons from laser-resonant ionization of thermal-velocity muoniums.
The experiment uses a compact muon storage ring with a high precision
magnetic field based on the MRI technology, and a compact detection system with particle tracking capability.

The E34 will launch at high intensity muon beam line (H-line) in the Material and Life science Facility (MLF) 
of J-PARC. 
At H-line, a pulsed positive muon beam with kinetic energy of about 4~MeV is 
stopped in a material and converted to form a muonium ($\mu^+e^-$ bound state).
Recent experiments at TRIUMF~\cite{Bakule2013,S1249-2013} and J-PARC~\cite{J-PARC_Mu_2013B} confirmed that silica aerogel 
serves as an efficient source of thermal-velocity muonium in vacuum.
Muoniums are ionized by intense Deep-UV lasers of wave length 122~nm and 355~nm, generating thermal-velocity muons.
Muons are accelerated by electrostatic field, and then RF accelerators
consisting of RFQ, and three stages of LINAC. The accelerated muon beam is injected 
to a 3~T muon storage magnet. The magnetic field of the storage magnet is carefully designed to guide muon to the 
storage region by making a spiral trajectory. A pulsed magnetic kick is applied to store the muon beam in the storage region
where magnetic field uniformity is carefully controlled and monitored. Positron from
muon decay carries information of the spin direction at decay. A tracking detector consisting of radial vanes of silicon-strip
sensors measures positron track from which spin oscillation due to $g$-2 and EDM is precisely measured.

\newpage
\subsection{Single meson light-by-light contributions to the muon's anomalous magnetic moment }
\addtocontents{toc}{\hspace{2cm}{\sl V.~Pauk}\par}

\vspace{5mm}

V.~Pauk

\vspace{5mm}

\noindent
Johannes Gutenberg University, Mainz \\

\vspace{5mm}

We develop the formalism to provide an improved estimate for the hadronic light-by-light correction 
to the muon's anomalous magnetic moment $a_\mu$, by considering single meson contributions beyond the leading pseudo-scalar mesons. 
We incorporate available experimental input as well as constraints from 
light-by-light scattering sum rules to estimate the effects of axial-vector, scalar, and tensor mesons. The details of these calculations are given in Ref. \cite{Pauk:2014rta}. Here, we give numerical evaluations for the hadronic light-by-light contribution of these states to $a_\mu$. The comparison of our results with the previous estimates is summarized in Table \ref{tab_comp}. 
  
\begin{table}[h]
{\centering \begin{tabular}{|c|c|c|c|c|c|}
\hline
& axial-vectors   &  scalars  & tensors   \\
\hline 
BPP  \cite{BPPxxx}  & \quad $2.5 \pm 1.0$ \quad   & \quad $-7 \pm 2$ \quad & \quad  -  \quad    \\
HKS \cite{HKS,HKxxx}  & \quad $1.7 \pm 1.7$ \quad   & \quad - \quad & \quad  -  \quad    \\
MV \cite{MVxxx} & \quad $22 \pm 5$ \quad   & \quad - \quad & \quad  -  \quad    \\
PdRV \cite{Prades:2009twxxx}  & \quad $15 \pm 10$ \quad   & \quad $-7 \pm 7$ \quad & \quad  -  \quad    \\
N/JN \cite{Jegerlehner:2009ryxxx} & \quad $22 \pm 5$ \quad   & \quad $-7 \pm 2$ \quad & \quad  -  \quad    \\
\hline
this work  & \quad $6.4 \pm 2.0$ \quad   & \quad $-3.1 \pm 0.8$ \quad & \quad  $1.1 \pm 0.1 $  \quad    \\
\hline 
\end{tabular}\par}
\caption{HLbL contribution to $a_\mu$ (in units $10^{-11}$)
due to axial-vector, scalar, and tensor mesons obtained in our work \cite{Pauk:2014rta},
compared with various previous estimates. 
For our scalar meson estimate, we have quoted the value corresponding with $\Lambda_{\mathrm{mon}} = 2$ GeV. }
\label{tab_comp}
\end{table}

The presented formalism allows to further improve on these estimates, once new data for such meson states will become available.

\newpage

\subsection{Meson transition form factors at BESIII  }
\addtocontents{toc}{\hspace{2cm}{\sl C.F.~Redmer}\par}

\vspace{5mm}

C.F.~Redmer for the BESIII Collaboration

\vspace{5mm}

\noindent
Institut f\"{ur} Kernphysik, Johannes Gutenberg-Universit\"{a}t Mainz, Germany

\vspace{5mm}

The BESIII experiment~\cite{BES3H}, operated at the BEPCII $e^+ e^-$ collider in Beijing (China), collects data in a 
center-of-mass energy range from 2.0~GeV to 4.6~GeV. In the past years, the worlds largest samples of $\textrm{J}/\psi, 
\psi^\prime$ and $\psi(3770)$, as well as significant samples at energies above 4~GeV devoted to the study of the exotic 
XYZ states have been acquired.

Based on the data, meson transition form factors (TFF) can be determined in various regions of momentum transfer. 
Time-like TFF can be studied either in the annihilation reaction $e^+ e^- \rightarrow \textrm{P}\gamma$, where the 
momentum transfer is fixed to the center of mass energy of the accelerator, or in Dalitz decays of pseudoscalar and 
vector mesons of the type $P\rightarrow\gamma e^+ e^-$ and $V\rightarrow P e^+ e^-$, respectively. In the case of meson 
decays, the range of momentum transfer is limited by the masses of the involved mesons. Recently, the rare decays of 
$\textrm{J}/\psi\rightarrow P e^+ e^-$, with $P = \pi^0, \eta, \eta^\prime$ have been measured for the first 
time~\cite{jpsidal}. The pseudoscalar mesons have been tagged in their respective decay channels 
$\eta^\prime\rightarrow\pi^+\pi^-\gamma$, $\eta^\prime\rightarrow\pi^+\pi^-\eta$, $\eta\rightarrow\gamma\gamma$, 
$\eta\rightarrow\pi^+\pi^-\pi^0$, and $\pi^0\rightarrow\gamma\gamma$. While background from other $\textrm{J}/\psi$ 
decays has been suppressed by kinematic cuts, a condition on the vertex position of leptonic track pairs has been 
applied to reject background from conversion of photons in the detector material, such as the beam pipe or the inner 
wall of the drift chamber. A good agreement between data and Monte Carlo simulations is found for the selected events. 
The branching ratios determined for $\textrm{J}/\psi\rightarrow \eta^\prime e^+ e^-$ and $\textrm{J}/\psi\rightarrow 
\eta e^+ e^-$ are in agreement with theory predictions~\cite{jpsidalT}. For $\textrm{J}/\psi\rightarrow \pi^0 e^+ e^-$ 
there is some tension between theory and the experimental result. It might be due to the limited statistics or the 
current status of theory calculations. It can be settled after the analysis of the full BESIII data set of approximately 
$1.2\times10^9 \textrm{J}/\psi$ decays.\\

\noindent The $\gamma \gamma$ physics program at the BESIII experiment aims at the measurement of TFF in the 
space-like region. Currently, $2.9\textrm{~fb}^{-1}$ of data taken at the $\psi(3770)$ peak~\cite{beslumi2} are used to 
study of $\pi^0, \eta$ and $\eta^\prime$ mesons. It is planned to extend the analysis to the data sets taken above 4~GeV 
to benefit from higher cross sections and access the to larger ranges of $Q^2$. The intent is to determine meson TFF in 
a range of momentum transfer between 0.3~GeV$^2$ and 10~GeV$^2$, which is not only complementary to the recent results 
from B-factories~\cite{babarbelle}, but also of high relevance for the calculations of hadronic Light-by-Light 
scattering.

The analysis strategy is based on a single-tag technique, where only the produced meson and one of the two 
scattered leptons are reconstructed from detector information. The second lepton is reconstructed from four-momentum 
conservation and required to have a small scattering angle, so that the momentum transfer is small and one of the 
exchanged photons is quasi-real. The ongoing analyses tag the produced pseudoscalar meson in the decay channels 
$\pi^0\rightarrow\gamma\gamma$, $\eta\rightarrow\gamma\gamma$, $\eta\rightarrow\pi^+\pi^-\pi^0$, and 
$\eta^\prime\rightarrow\pi^+\pi^-\eta$. Major sources of background are QED processes such as virtual Compton 
scattering, misidentified hadronic final states, external photon conversion, and on-peak background from two-photon 
processes such as the production of different mesons or and initial state radiation in the signal channel. Conditions 
are being devised to suppress the identified background sources. Current Monte Carlo studies, using the Ekhara event 
generator~\cite{ekhara,2octet}, suggest that the TFF of $\pi^0$ can be extracted with an unprecedented statistical 
accuracy in the range of $0.3 \leq Q^2 [\textrm{GeV}] \leq 1.5$. At larger Q$^2$ the accuracy is compatible with the 
CLEO measurement~\cite{CLEO}. First results are expected soon.

Future prospects of the $\gamma \gamma$ physics program at BESIII comprise the investigation of multi-meson final 
states to study scalar and tensor meson production. As a first step, the investigation of the two-photon production of 
$\pi^+\pi^-$ pairs has been started~\cite{yuping}. Further projects are the measurement of polarization observables, and 
double tagged measurements of $\gamma \gamma$ processes using a dedicated tagging device at smallest scattering angles.

\newpage
\subsection{R measurements at BELLE and perspectives for BELLE II  }
\addtocontents{toc}{\hspace{2cm}{\sl B.A.~Shwartz}\par}

\vspace{5mm}

B.A.~Shwartz,    for Belle collaboration

\vspace{5mm}

\noindent
Institute Budker Institute of Nuclear Physics, Novosibisrsk, Russia\\

\vspace{5mm}

Precise measurements of the total cross sections of the $e^+e^-$
annihilation into hadrons and detail study the final states 
produce important information
about quark interactions, spectroscopy of their bound states
and provide a basis 
for the calculations of the hadronic contributions 
to the fundamental parameters, like muon $(g-2)$ value or
$\alpha(M_Z^2)$.
Huge data samples collected by two B-factories \cite{kekb,pep2} 
opened new reach possibilities to study hadronic cross sections.

Hadronic cross sections are measured by the Belle detector \cite{belle},
operated at KEKB energy asymmetric collider   
using the direct energy scan
in the range from 10.6 to 11.05~GeV as well as by the ISR method.
The main results were obtained for charmed hadrons production:
$ e^+e^- \to D(^*)\overline{D^*}$ \cite{dd};
$e^+e^- \to D^0D^-\pi^+$ \cite{ddpi};
$e^+e^- \to D_s(^*)\overline{D}_s(^*)$ \cite{dsds};
$e^+e^- \to \Lambda_c^+\Lambda_c^-$ \cite{lambda}.
Recently the states Y(4008) and Y(4260) were confirmed in the
$e^+e^- \to \pi^+\pi^-J/\Psi$ the charged charmonium-like state 
$Z(3900)^{\pm}$ was observed in the $\pi J/\Psi$ decay 
\cite{pipijpsi}.
Interesting results were obtained for the processes
$e^+e^- \to \phi \pi^+\pi^-$ and $e^+e^- \to f_0(980)\pi^+\pi-$ 
in the energy range from 1.5 to 3 GeV where parameters of the
$\phi(1680)$ and $Y(2175)$ were measured \cite{phipipi}.
Preliminary results on the $e^+e^- \to \pi^+\pi^-\pi^0$ cross section
in the energy range from 0.7 up to 3.5 GeV when the hard ISR
photon were detected at the large angle were obtained as well.

At present new advanced collider, SuperKEKB, with a luminosity
of $8\times 10^{35}$~cm$^{-2}$s$^{-1}$ is under construction at KEK. 
The Belle~II detector will 
have much better parameters than the Belle. One of the important
task of this upgrade is to provide possibilities for the
precise measurement of the hadronic cross sections.  
At the integrated luminosity of 10~ab$^{-1}$ the equivalent integrated
luminosity obtained with ISR approach will exceeds the
amounts available at VEPP-2000 and BEPC-II colliders in the
energy ranges 1-2~GeV and 2-3~GeV respectively.

\newpage
\subsection{Hadronic cross section measurements in Novosibirsk }
\addtocontents{toc}{\hspace{2cm}{\sl E.~Solodov}\par}

\vspace{5mm}

E.~Solodov

\vspace{5mm}

\noindent
Budker INP, Novosibirsk, Russia

\vspace{5mm}

\begin{figure}[tbh]
\begin{center}
\includegraphics[width=0.7\textwidth]{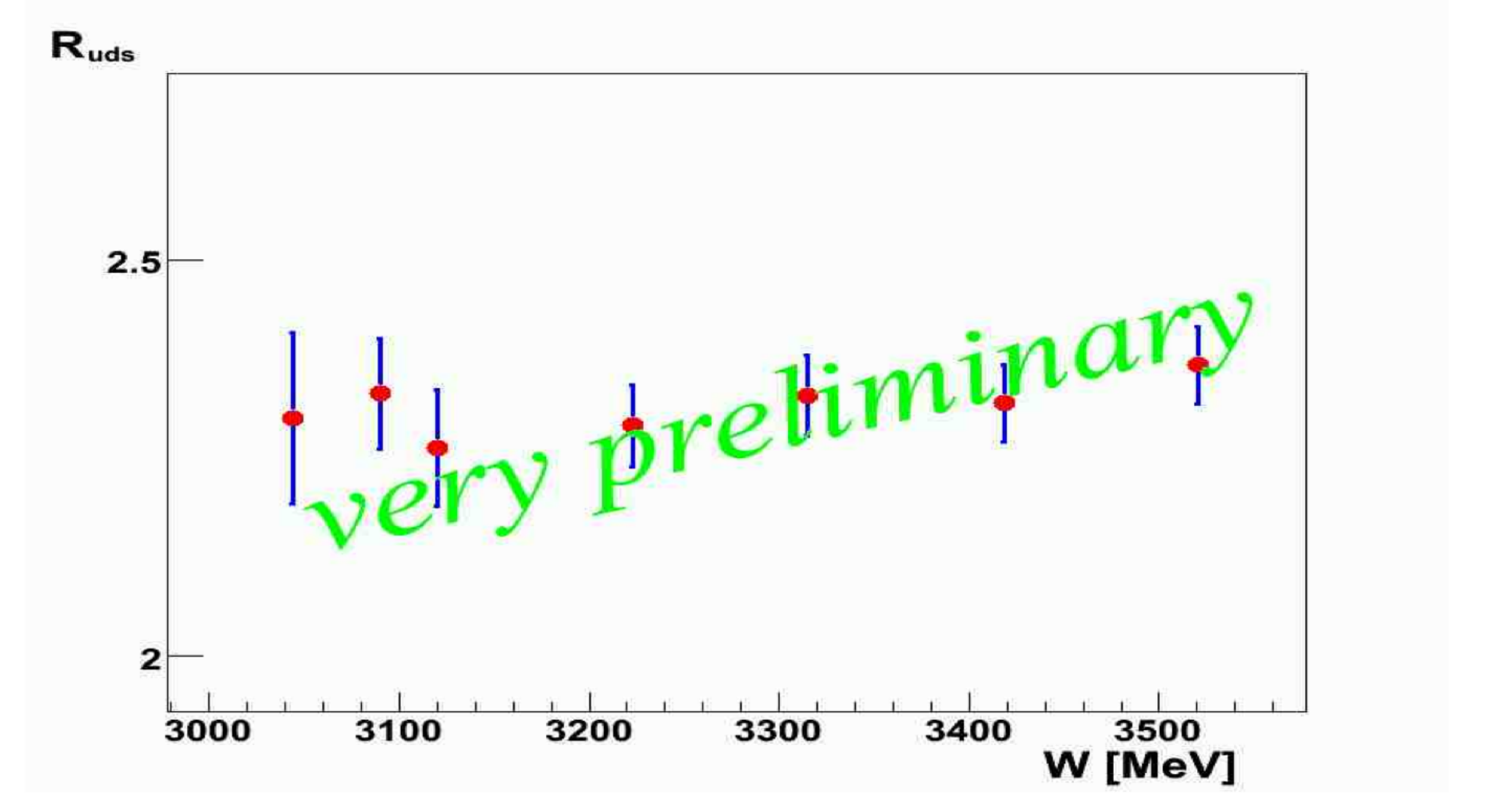}
\vspace{-0.3cm}
\caption
{Preliminary result of energy scan by KEDR
}
\label{vepp4m}
\end{center}
\end{figure}

The $\ee$ VEPP-4M collider inspite of low luminosity has few unuiqe fitures. It has wide energy range from 2 to 10 GeV for the center-of-mass (c.m.) energy, and very precise beam energy measurement and control at the level of 20-30 keV. Energy scan from 1.9 to 3.7 GeV in c.m. has been recently performed. Preliminary data in the 3.1-3.7 GeV are presented in Fig.~\ref{vepp4m}. Nearest plans include energy scan up to 8-10 GeV in c.m.

\begin{figure}[tbh]
\begin{center}
\includegraphics[width=0.49\textwidth]{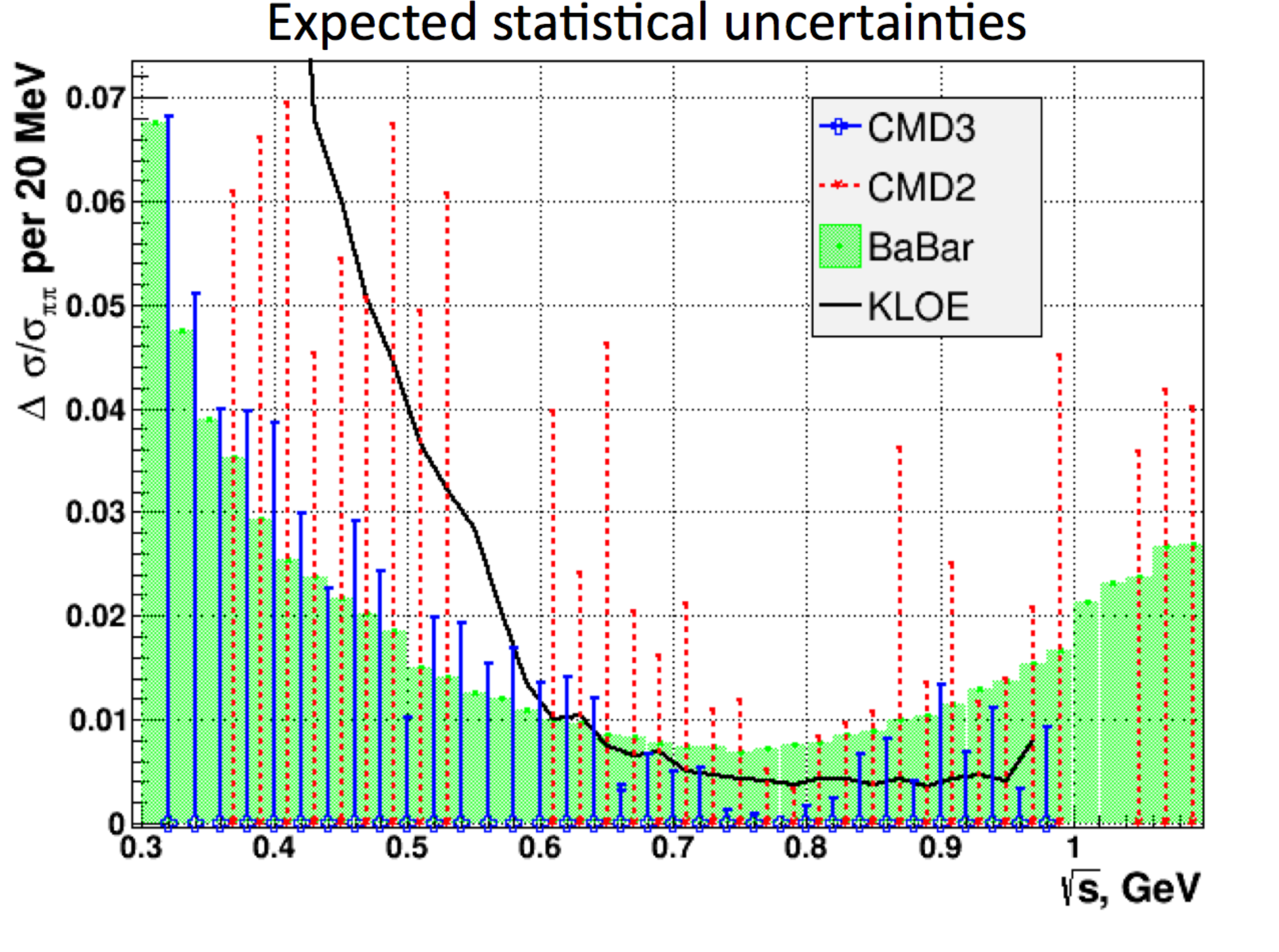}
\includegraphics[width=0.49\textwidth]{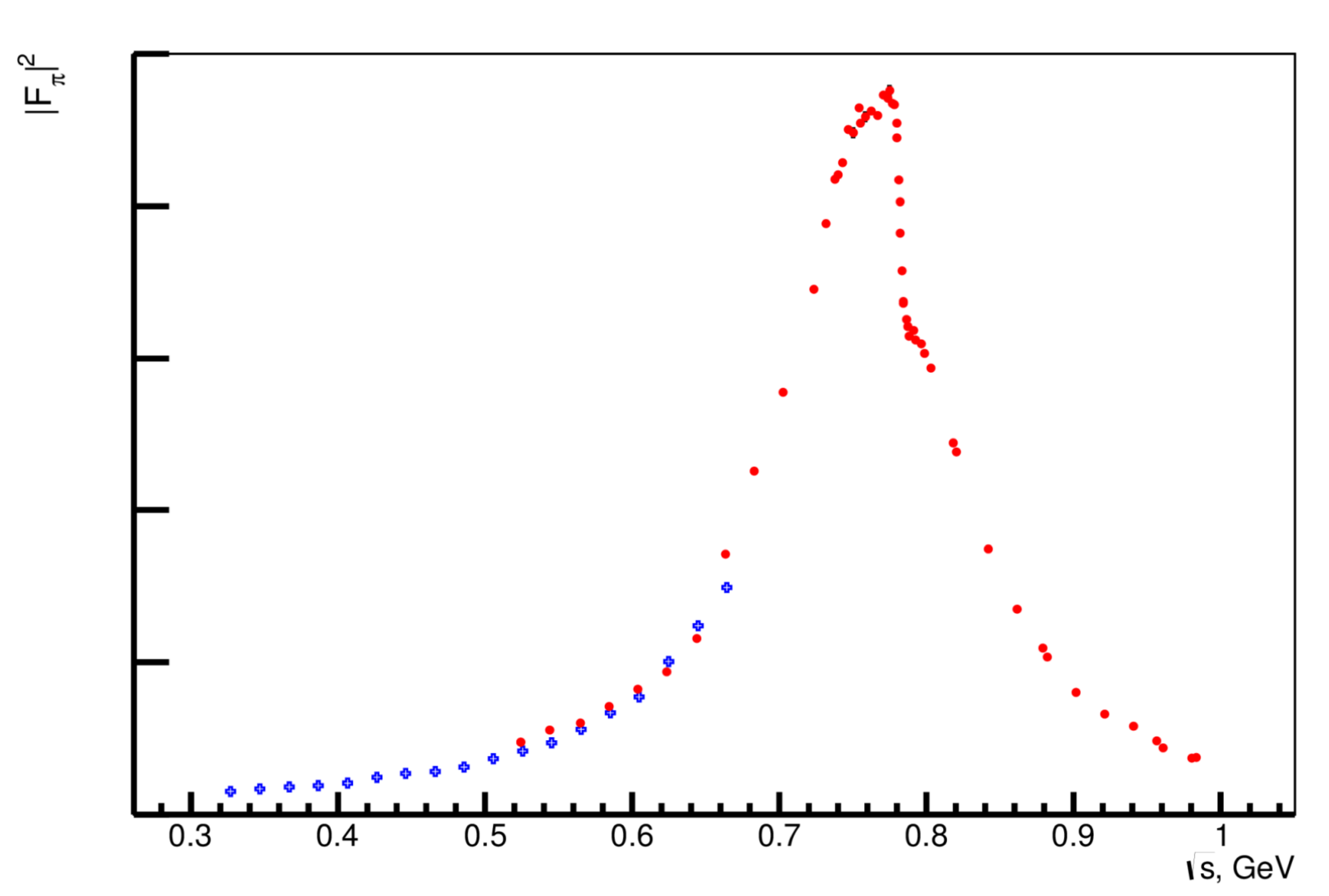}
\caption
{Left: Relative accuracy in $\pi^+\pi^-$ cross section measurement. Right: Preliminary  $\pi^+\pi^-$ form factor measurement by CMD-3. Color shows different separation methods.
}
\label{pipiff}
\end{center}
%
\begin{center}
\vspace{-0.5cm}
\includegraphics[width=0.49\textwidth]{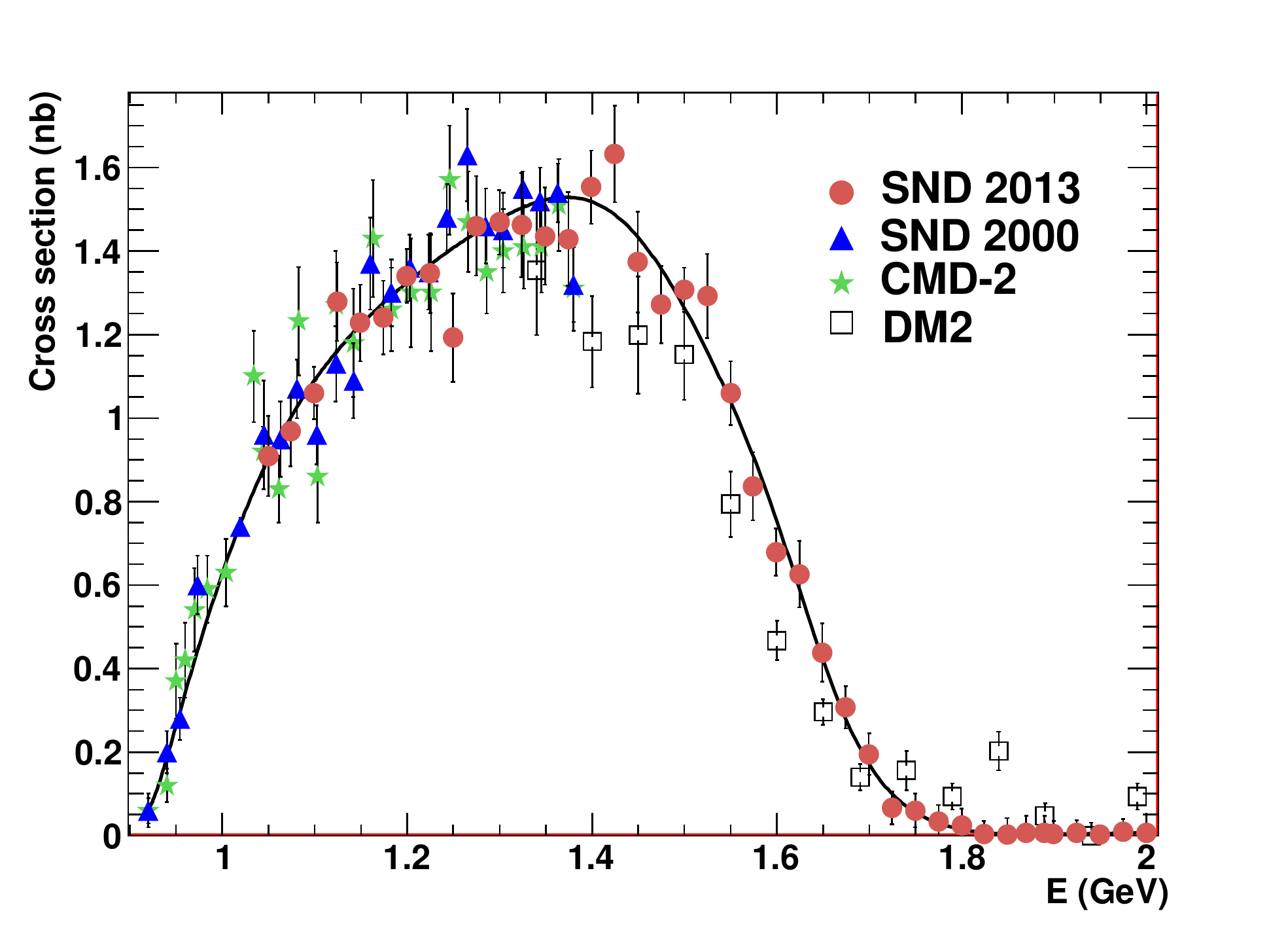}
\includegraphics[width=0.49\textwidth]{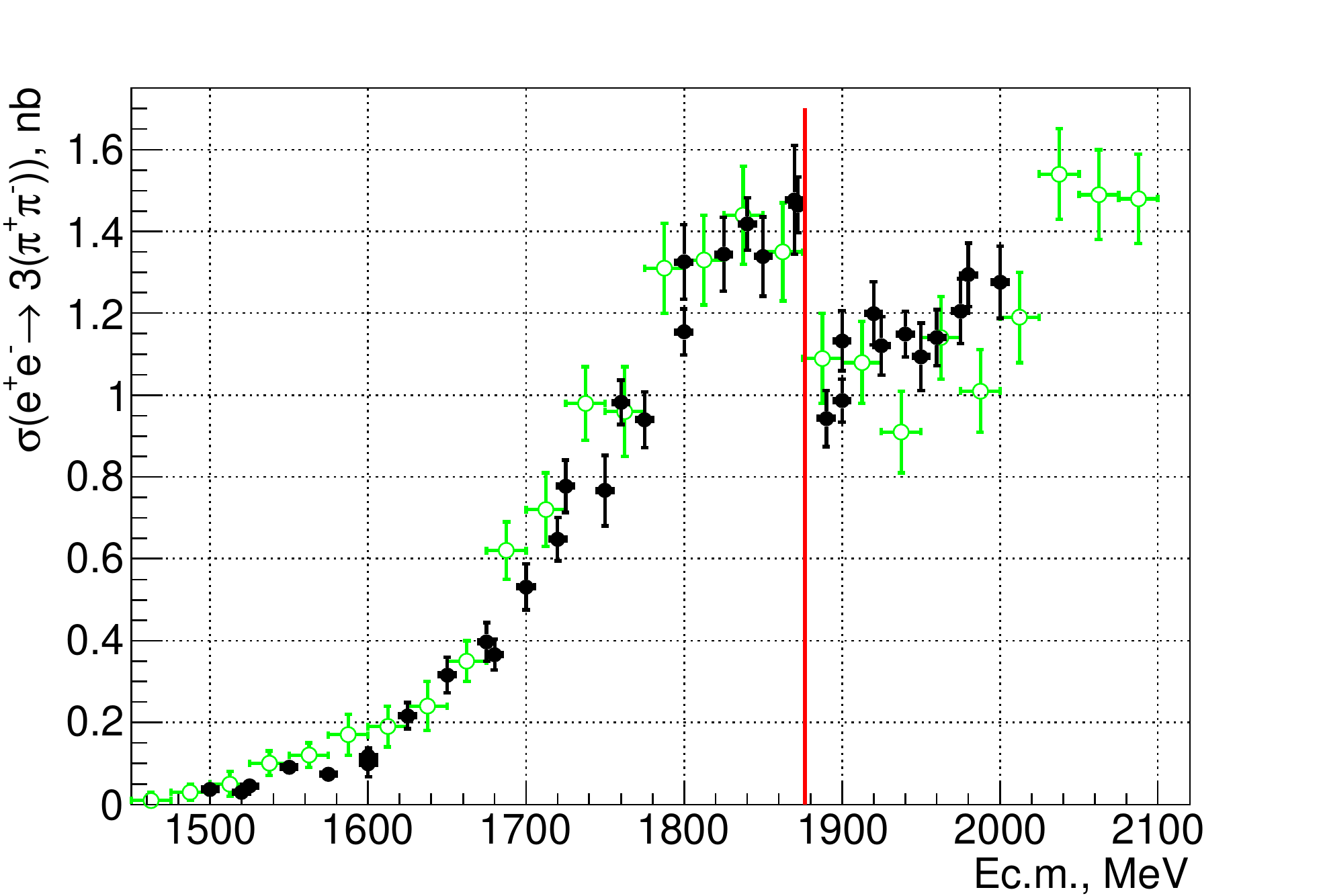}
\caption
{Left: Measurement of the $\ee\to\omega\pi^0\to\pi^0\pi^0\gamma$ process cross section with SND ~\cite{snd}.  Right: Measurement of the 
$\ee\to 3(\pi^+\pi^-)$ process cross section with CMD-3~\cite{cmd} in comparison with BaBar (open points)}
\label{ompi0}
\end{center}
\end{figure}

The $\ee$ VEPP-2000 collider has energy range from 0.32 to 2.0 GeV which was recently scanned with about 70 $pb^{-1}$ integrated luminosity per detector. Data recorded by two detectors, CMD-3 and SND, have statistical power comparable with the world best experiments. One of the most important study is to measure the $\ee\to\pipi$ cross section and extract the pion form factor with better than 1\% systematic uncertainty.
Figure~\ref{pipiff} (Left) shows relative statistical uncertainty for the CMD-3 pion form factor measurements in comparison with other experiments. Figure~\ref{pipiff} (Right) shows overview of the pion form factor for CMD-3 data. Two methods of pions separation, based on momentum measurement in drift chamber or energy deposition in the calorimeter, are shown by color. The overlapped region is used for the systematic uncertainties studies.

The CMD-3 and SND detectors presented preliminary results on the cross section measurements for the $\ee\to\omega\pi^0\to\pi^0\pi^0\gamma$ (published in Ref.~\cite{snd}), $\pi^+\pi^-\eta$, $\pi^+\pi^-\pi^0$, $\pi^+\pi^-\pi^+\pi^-$, $2(\pi^+\pi^-\pi^0)$, $3(\pi^+\pi^-)$ (published in Ref.~\cite{cmd}) and some other processes, which were presented at the talk.
Data analysis is in progress. New data taking with upgraded to about 10 time higher luminosity machine is planned later this year.

\newpage
\subsection{Hadronic vacuum polarisation in $g-2$ and $\alpha_{\rm QED}$}
\addtocontents{toc}{\hspace{2cm}{\sl T.~Teubner}\par}

\vspace{5mm}

T.~Teubner

\vspace{5mm}

\noindent
Department of Mathematical Sciences, University of Liverpool,
Liverpool\ \ L69 3BX, U.K.\\

\vspace{5mm}

The uncertainty of the Standard Model (SM) prediction of the anomalous
magnetic moment of the muon, $a_{\mu}^{\rm SM}$, currently stands at
$\pm 4.9\cdot 10^{-10}$, where the precise number depends on the
details of the compilation. In the combination of all SM contributions
$\pm 4.2\cdot 10^{-10}$ come from
the leading and next-to-leading order hadronic vacuum polarisation (HVP)
contributions~\cite{Hagiwara:2011af}, 
closely followed by the uncertainty in the light-by-light scattering
contributions. As demonstrated in this talk, the prospects to
substantially improve the HVP contributions are good, though this will
require major efforts and should not be taken for granted. 

Recently the calculation of the HVP contributions has been extended to
next-to-next-to leading order, which adds about $1.2\cdot 10^{-10}$ to
the central value of $a_{\mu}^{\rm SM}$ but little to its
uncertainty~\cite{Kurz:2014wya}. 

The HVP contributions are obtained through a dispersion integral with
a well-known kernel function and hadronic cross section data. However,
the understanding, calculation and implementation of radiative
corrections play a paramount role to get the best possible prediction,
and current evaluations contain an uncertainty of about $2\cdot
10^{-10}$ due to uncertainties in the treatment of the radiative
corrections in the hadronic data.

The most pressing issue is to improve the two pion channel. There is a
significant tension between the measurements based on the method of
radiative return from KLOE and from BaBar, see also the related
presentations at this workshop~\cite{SE,AH}. This prohibits a much smaller
uncertainty in the combination with data from the direct scan method, and
there is also the danger of a possible bias depending on the choice of
the data used for $a_{\mu}^{\rm SM}$. A recent analysis from KLOE has
confirmed this picture, see~\cite{SMuller}, but currently the reason for this
discrepancy is unknown. The experiments SND and CMD-3 at Novosibirsk,
BELLE at KEK and BES-III at Beijing have already collected a large
amount of data, and analyses are underway, see the related
presentations at this
workshop~\cite{GS-Novosibirsk,BS-BELLE,BK-ISR-BESIII}. Additional
data sets with high statistics and good systematics will hopefully
supersede or solve the puzzle in the two pion channel, though ideally
one should find out why KLOE and BarBar disagree so markedly. Data
derived from hadronic tau spectral functions have recently found to be
consistent with the $e^+e^- \to hadrons$ data when used in combined
fits based on Hidden Local Symmetry models, and their use slightly
improves the uncertainty of $a_{\mu}^{\rm SM}$, see~\cite{MB-HLS}. 

At higher energies, measurements in many subleading exclusive channels
from BaBar have improved the determination of the HVP contributions in
the region below $2$ GeV, where before one relied on old and
fairly poor inclusive measurements. Numerous further measurements in
subleading channels (multi-pion and channels including $K$s) are
expected from SND, CMD-3, BaBar and also from BELLE and BES-III, see the
presentations~\cite{GS-Novosibirsk,AH,BS-BELLE,BK-ISR-BESIII}. These
will improve the HVP contributions in this region further.
The match with predictions based on perturbative QCD at higher
energies is smooth, but data just above $2$ GeV are still relatively
sparse and/or not very accurate. BES-III will help to constrain this
higher energy region with new data, see~\cite{GH-BESIII}.

Hadronic cross section measurements at higher energies are of particular
importance to precisely determine the running of the QED coupling,
$\alpha_{\rm QED}(q^2)$. Various routines for $\alpha_{\rm QED}(q^2)$
are available and the current precision seems sufficient for the use
in most Monte Carlo codes and to correct the data for the use in
$a_{\mu}^{\rm SM}$ and $\alpha_{\rm QED}(q^2)$ itself. However,
regions of narrow resonances like the $\phi$ need more attention, and
it is important to be aware of the limitations in the use of a running
coupling very close to narrow resonances. The quantity $\alpha_{\rm
  QED}(M_Z^2)$ is important for future precision tests of the Standard
Model. Current best evaluations do not gain as much as previously from
using perturbative QCD instead of data, and the data input will
improve further, especially through BES-III.

With several experiments world-wide, contributing to the measurements of
hadronic cross sections at low energies, the aim to half the error of
$a_{\mu}^{\rm SM}$ in time for the new $g-2$ experiments seems
realistic. Our knowledge of the QED coupling will also further improve
and help to make more stringent tests of the SM possible. To achieve
this, the combined efforts of experimentalists and theorists working
on the related radiative corrections and Monte Carlo programmes is
indispensable.

\newpage

\subsection{$\eta$ and $\eta'$ decays with Crystal Ball at MAMI }
\addtocontents{toc}{\hspace{2cm}{\sl M.~Unverzagt}\par}

\vspace{5mm}

M.~Unverzagt

\vspace{5mm}

\noindent
Institute f\"ur Kernphysik, Johannes Gutenberg-Universit\"at Mainz, Germany\\

\vspace{5mm}

The A2 collaboration at the Institute for Nuclear Physics in Mainz, Germany, performs experiments with Bremsstrahlung photons derived from electrons in the Glasgow-tagging spectrometer~\cite{McGeorge:2007tg}. The Mainz Microtron (MAMI)~\cite{Jankowiak:2006yc,Kaiser:2008zza} accelerator has a maximum electron energy of $E_e = 1604$~MeV. With the Crystal Ball-spectrometer~\cite{Starostin:2001zz} and a forward spectrometer-wall consisting of TAPS-crystals~\cite{Novotny:1991ht} the A2 collaboration studies, e.g., the production of light meson decays and their decays.

One example is the $\eta \to \pi^0 \gamma \gamma$ decay. Its amplitude has first sizable contributions at $O(p^6)$, but the low-energy constants have to be determined from models. Thus, this decay is both a stringent test of $\chi$PT at the order $O(p^6)$ and also of these models. A new analysis of this decay measured with the Crystal Ball at MAMI in 2007 and 2009~\cite{Nefkens:2014zlt} gave $1.2 \cdot 10^3$ $\eta \to \pi^0 \gamma \gamma$ events which is currently the most accurate result in the world. Though it seems that the model by Oset et al.~\cite{Oset:2008hp} is favoured, a conclusive distinction between models can only be made with even higher precision. The preliminary decay width $\Gamma (\eta \to \pi^0 \gamma \gamma) = (0.33 \pm 0.03_{\mathrm{tot}})$~eV determined from the Crystal Ball data~\cite{Nefkens:2014zlt} agrees with all theoretical calculations but disagrees with the competitive preliminary result from the KLOE experiment~\cite{Gauzzi:2012zz} by more than four standard deviations.

The A2 collaboration also contributes to the studies of transition form factors of light mesons. These form factors do not only probe the structure of light mesons but also might give important input to the Standard Model calculations of the light-by-light contribution to the Anomalous Magnetic Moment of the Muon~\cite{Jegerlehner:2009ry}. In 2011, the determination of the $\eta$ transition form factor based on $\sim$1350 $\eta \to e^+ e^- \gamma$ events~\cite{Berghauser:2011zz} was published. An independent analysis of 3 times more data from the Crystal Ball at MAMI experiment gave roughly 20,000 $\eta \to e^+ e^- \gamma$ events~\cite{Aguar-Bartolome:2013vpw1}. The latest result of the A2 collaboration for the $\eta$-transition form factor agrees very well with all earlier measurements. Though the result shows good agreement with theoretical calculations the statistical accuracy does not allow one to rule out any prediction. 

In the next few years the A2 collaboration plans to continue studying decays of light mesons. The statistics on already analysed decays will be improved greatly. The $\eta/\eta' \to 3\pi^0$ and $\eta' \to \eta \pi^0 \pi^0$ decays will be studied as well as pseudoscalar-vector-$\gamma$ transitions like $\eta' \to \omega \gamma$ and $\omega \to \eta \gamma$. Furthermore, it is planned to investigate transition form factors in $\pi^0/\eta/\eta' \to e^+ e^- \gamma$ and $\omega \to \pi^0 e^+ e^-$ decays. $C$- and $CP$-violation will be examined in $\pi^0/\eta \to 3 \gamma$, $\eta \to 2 \pi^0 \gamma$, $\eta \to 3 \pi^0 \gamma$ and $\eta \to 4\pi^0$ decays. The $\pi^0 \to 4 \gamma$ decay is an important background to the $\pi^0 \to 3 \gamma$ decay. It is an allowed but highly suppressed process, and has never been seen yet, but the Standard Model branching ratio of $~10^{-11}$~\cite{Bystritskiy:2006} shows that it might be in reach with the Crystal Ball at MAMI experimental setup.

\newpage
\subsection{Electromagnetic form factors with WASA-at-COSY }
\addtocontents{toc}{\hspace{2cm}{\sl M.~Wolke}\par}

\vspace{5mm}

M.~Wolke \footnote{Supported by the Swedish Research Council (VR).} for the 
 WASA-at-COSY Collaboration

\vspace{5mm}

\noindent
Department of Physics and Astronomy, Uppsala University, Sweden\\

\vspace{5mm}

WASA-at-COSY has accumulated large statistics data sets on fully reconstructed 
$\pi^0$ and $\eta$ decays, which are presently being analyzed.

From measurements in the time--like region experimental uncertainties for the 
$\pi^0$ transition form factor are still rather large, and extrapolations from 
the space--like region at higher energies are model 
dependent \cite{Kampf:2005tz}.
As of now, about $5 \cdot 10^5$ $\pi^0 \rightarrow e^+ e^- \gamma$ events from 
WASA-at-COSY have been analyzed to extract limits on the parameters of a 
hypothetical dark photon \cite{Adlarson:2013eza}.
Already these data have an order of magnitude larger statistics than the 
previous benchmark measurement \cite{MeijerDrees:1992qb}, and we expect 
another order of magnitude in statistics from the analysis of the full WASA 
data set.
A preliminary analysis of the $\pi^0 \rightarrow e^+ e^-$ decay shows that 
this very rare decay can be identified with the WASA detector. 
However, the final statistics is likely to be smaller compared to the KTeV 
measurement \cite{Abouzaid:2006kk}.

$\eta$ decays have been tagged at WASA using both the $p d \rightarrow 
\mbox{}^3\mbox{He}X$ and $pp \rightarrow ppX$ reactions.
Preliminary results have been obtained from the $pd$ data in recent PhD 
theses for $\eta$ decays to $e^+ e^- \gamma$, $e^+ e^- e^+ e^-$, 
$e^+ e^- \pi^+ \pi^-$, and $e^+ e^-$.
In the case of the $\eta$ Dalitz decay we expect the final statistics from the 
$pd$ data to be comparable to the recent CB/TAPS 
result \cite{Aguar-Bartolome:2013vpw}, while the $pp$ data should roughly 
contain an order of magnitude more events.
The latter also holds both for the $\eta$ double Dalitz and the $\eta 
\rightarrow \pi^+ \pi^- e^+ e^-$ decays, in the case of which preliminary 
branching ratios have been extracted from the $pd$ data. 
Within the limited statistical accuracy of the data analyzed so far the values 
are in good agreement with the KLOE results \cite{KLOE2:2011aa}.
Preliminarily, for the very rare $\eta \rightarrow e^+ e^-$ decay we get from 
the analysis of slightly more than $10\,\%$ of the data taken in 
proton--proton collisions an upper limit ($90\,\%$ c.l.) of 
$< 4.6 \cdot 10^{-6}$, below the experimental limit published 
in \cite{HADES:2011ab}.

\newpage

\section{List of participants}

\subsection{Hadronic contributions to the muon anomalous magnetic moment Workshop}

\begin{flushleft}
\begin{itemize}
\item Maurice Benayoun, LPNHE des Universites Paris 6 et 7, {\tt benayoun@in2p3.fr}
\item Johan Bijnens, Lund University, {\tt  bijnens@thep.lu.se}
\item Tom Blum, University of Connecticut, {\tt tblum@phys.uconn.edu}
\item Irinel Caprini, Institue of Physics and Nuclear Engineering, Bucharest, {\tt caprini@theory.nipne.ro}
\item Gilberto Colangelo,  University of Bern, {\tt gilberto@itp.unibe.ch}
\item Henryk Czy\.z, University of Silesia, {\tt henryk.czyz@us.edu.pl}
\item Achim Denig, Universit\"at Mainz, {\tt denig@kph.uni-mainz.de}
\item Cesareo A. Dominguez, University of Cape Town, {\tt cesareo.dominguez@uct.ac.za}
\item Simon Eidelman, Novosibirsk State University, {\tt eidelman@mail.cern.ch }
\item Christian S. Fischer, JLU Giessen, {\tt Christian.Fischer@theo.physik.uni-giessen.de}
\item Antony Francis, Universit\"at Mainz, {\tt francis@kph.uni-mainz.de}
\item Mikhail Gorshteyn, Universit\"at Mainz, {\tt gorshtey@kph.uni-mainz.de}
\item Maarten Golterman, San Francisco State University, {\tt maarten@stars.sfsu.edu}
\item Vera Guelpers, Universit\"at Mainz, {\tt guelpers@kph.uni-mainz.de}
\item Andreas Hafner, Universit\"at Mainz, {\tt hafner@kph.uni-mainz.de }
\item Christoph Hanhart, Forschungszentrum J\"ulich, {\tt c.hanhart@fz-juelich.de}
\item Masashi Hayakawa, University of Nagoya, {\tt hayakawa@eken.phys.nagoya-u.ac.jp}
\item Gregorio Herdoiza, Universidad Autonoma de Madrid, {\tt gregorio.herdoiza@uam.es}
\item Martin Hoferichter, University of Bern, {\tt hoferichter@itp.unibe.ch}
\item Hanno Horch, Universit\"at Mainz, {\tt horch@kph.uni-mainz.de}
\item Taku Izubuchi, Brookhaven National Laboratory, {\tt izubuchi@quark.phy.bnl.gov}
\item Karl Jansen, DESY Zeuthen, {\tt karl.jansen@desy.de}
\item Fred Jegerlehner, Humbold-Universit\"at Berlin, {\tt fjegerlehner@gmail.com}
\item Andreas J\"uttner, University of Southampton, {\tt juttner@soton.ac.uk}
\item Benedikt Kloss, Universit\"at Mainz, {\tt kloss@uni-mainz.de}
\item Marc Knecht, CPT Marseille, {\tt Marc.Knecht@cpt.univ-mrs.fr}
\item Bastian Kubis, University of Bonn, {\tt kubis@hiskp.uni-bonn.de}
\item Andrzej Kupsc, Uppsala University, {\tt Andrzej.Kupsc@physics.uu.se}
\item Lauren Lellouch, CPT Marseille, {\tt laurent.lellouch@cern.ch}
\item Kim Maltman, York University, {\tt kmaltman@yorku.ca}
\item Bill Marciano, Brookhaven National Laboratory, {\tt marciano@bnl.gov}
\item Marina Marinkovic, University of Southampton, {\tt mmarina@physik.hu-berlin.de}
\item Pere Masjuan, Universit\"at Mainz, {\tt masjuan@kph.uni-mainz.de}
\item Harvey B. Meyer, Universit\"at Mainz, {\tt meyer@kph.uni-mainz.de}
\item Andreas Nyffeler, Universit\"at Mainz, {\tt nyffeler@kph.uni-mainz.de}
\item Vladimir Pascalutsa, Universit\"at Mainz, {\tt vladipas@googlemail.com} 
\item Massimo Passera, INFN Padova, {\tt massimo.passera@pd.infn.it}
\item Vladyslav Pauk, Universit\"at Mainz, {\tt pauk@kph.uni-mainz.de}
\item Michael R. Pennington, Jefferslon Laboratory, {\tt michaelp@jlab.org}
\item Santiago Peris, Universitat Autonoma de Barcelona, {\tt peris@ifae.es}
\item Antonin Portelli, University of Southampton, {\tt a.portelli@soton.ac.uk}
\item Christoph F. Redmer, Universit\"at Mainz, {\tt redmer@uni-mainz.de}
\item Pablo Sanchez-Puertas, Universit\"at Mainz, {\tt sanchezp@uni-mainz.de}
\item Eigo Shintani, Universit\"at Mainz, {\tt shintani@kph.uni-mainz.de}
\item Dominik St\"ockinger, TU Dresden, {\tt Dominik.Stoeckinger@tu-dresden.de}
\item Thomas Teubner, University of Liverpool, {\tt thomas.teubner@liverpool.ac.uk}
\item Marc Unverzagt, Universit\"at Mainz, {\tt unvemarc@kph.uni-mainz.de}
\item Marc Vanderhaeghen, Universit\"at Mainz, {\tt marcvdh@kph.uni-mainz.de}
\item Georg von Hippel, Universit\"at Mainz, {\tt hippel@kph.uni-mainz.de}
\item Richard Williams, University of Giessen, {\tt richard.williams@uni-graz.at}
\end{itemize}
\end{flushleft}

\subsection{$g-2$: Quo Vadis Workshop}

\begin{flushleft}
\begin{itemize}
\item Nils Asmussen, Universit\"at Mainz, {\tt nils.asmussen@students.uni-mainz.de }
\item Sabato Stefano Caiazza, Universit\"at Mainz, {\tt caiazza@kph.uni-mainz.de}
\item Henryk Czy\.z, University of Silesia, {\tt henryk.czyz@us.edu.pl}
\item Achim Denig, Universit\"at Mainz, {\tt denig@kph.uni-mainz.de}
\item Cesareo A. Dominguez, University of Cape Town, {\tt cesareo.dominguez@uct.ac.za}
\item Simon Eidelman, Novosibirsk State University, {\tt eidelman@mail.cern.ch }
\item Paolo Gauzzi, Sapienza Universit\'a di Roma e INFN, {\tt paolo.gauzzi@roma1.infn.it}
\item Mikhail Gorshteyn, Universit\"at Mainz, {\tt gorshtey@kph.uni-mainz.de}
\item Wolfgang Gradl, Universit\"at Mainz, {\tt gradl@kph.uni-mainz.de }
\item Yuping Guo, Universit\"at Mainz, {\tt guo@kph.uni-mainz.de }
\item Andreas Hafner, Universit\"at Mainz, {\tt hafner@kph.uni-mainz.de }
\item Gregorio Herdoiza, Universidad Autonoma de Madrid, {\tt gregorio.herdoiza@uam.es}
\item Martin Hoferichter, University of Bern, {\tt hoferichter@itp.unibe.ch}
\item Guangshun Huang, University of Science and Technology of China, {\tt hgs@ustc.edu.cn}
\item Haiming Hu, IHEP, {\tt huhm@ihep.ac.cn}
\item Fred Jegerlehner, Humbold-Universit\"at Berlin, {\tt fjegerlehner@gmail.com}
\item Benedikt Kloss, Universit\"at Mainz, {\tt kloss@uni-mainz.de }
\item Andrzej Kupsc, Uppsala University, {\tt Andrzej.Kupsc@physics.uu.se }
\item Zhiqing Liu,  Universit\"at Mainz, {\tt liuz@uni-mainz.de}
\item Peter Lukin, Budker Institute of Nuclear Physics, {\tt P.A.Lukin@inp.nsk.su} 
\item Bill Marciano, Brookhaven National Laboratory, {\tt marciano@bnl.gov}
\item Pere Masjuan, Universit\"at Mainz, {\tt masjuan@kph.uni-mainz.de}
\item Harvey B. Meyer, Universit\"at Mainz, {\tt meyer@kph.uni-mainz.de}
\item Tsutomu Mibe, IPNS, KEK, {\tt mibe@post.kek.jp}
\item Andreas Nyffeler, Universit\"at Mainz, {\tt nyffeler@kph.uni-mainz.de}
\item Vladimir Pascalutsa, Universit\"at Mainz, {\tt vladipas@googlemail.com} 
\item Vladyslav Pauk, Universit\"at Mainz, {\tt pauk@kph.uni-mainz.de}
\item Santiago Peris, Universitat Autonoma de Barcelona, {\tt peris@ifae.es}
\item Christoph F. Redmer, Universit\"at Mainz, {\tt redmer@kph.uni-mainz.de }
\item Martin Ripka, Universit\"at Mainz, {\tt ripka@uni-mainz.de }
\item Naohito Saito, KEK-IPNS/J-PARC, {\tt Naohito.Saito@kek.jp}
\item Pablo Sanchez-Puertas, Universit\"at Mainz, {\tt sanchezp@kph.uni-mainz.de }
\item Sven Schumann, Universit\"at Mainz, {\tt schumanns@kph.uni-mainz.de }
\item Boris Shwartz, Budker Institute of Nuclear Physics, {\tt B.A.Shwartz@inp.nsk.su} 
\item Evgeny Solodov, Budker Institute of Nuclear Physics, {\tt E.P.Solodov@inp.nsk.su} 
\item Thomas Teubner, University of Liverpool, {\tt thomas.teubner@liverpool.ac.uk}
\item Marc Unverzagt, Universit\"at Mainz, {\tt unvemarc@kph.uni-mainz.de }
\item Marc Vanderhaeghen, Universit\"at Mainz, {\tt marcvdh@kph.uni-mainz.de }
\item Yaqian Wang, Universit\"at Mainz, {\tt wangy@kph.uni-mainz.de }
\item Zhiyong Wang, IHEP, CAS {\tt wangzy@ihep.ac.cn}
\item Hartmut Wittig, Universit\"at Mainz, {\tt wittig@kph.uni-mainz.de } 
\item Magnus Wolke, Uppsala University, {\tt magnus.wolke@fysast.uu.se}
\item Jakub Zaremba, Institute of Nuclear Physics, Krakow, {\tt jakub.zaremba@ifj.edu.pl}
\end{itemize}
\end{flushleft}


\begin{thebibliography}{99}
\bibitem{vanderBij:2014mxa}
  J.~J.~van der Bij, H.~Czy?, S.~Eidelman, G.~Fedotovich, T.~Ferber, V.~Ivanov, A.~Korobov and Z.~Liu {\it et al.},
  arXiv:1406.4639 [hep-ph].
\bibitem{Czyz:2013zga}
  P.~Masjuan (ed.), G.~Venanzoni (ed.), H.~Czy\.z, A.~Denig, M.~Vanderhaeghen, G.~Venanzoni, A.~Denig and S.~Eidelman {\it et al.},
  arXiv:1306.2045 [hep-ph].
\bibitem{Bijnens:2014fya}
  J.~Bijnens, R.~Escribano, S.~Fang, S.~Giovannella, W.~Gradl, C.~Hanhart, B.~Kubis, S.~Leupold and M.~F.~M.~Lutz {\it et al.},
  arXiv:1403.6380 [hep-ph].
\end{thebibliography}

\begin{thebibliography}{99}

 \bibitem{Bennett:2006fi} 
  G.~W.~Bennett {\it et al.}  [Muon G-2 Collaboration],
  Phys.\ Rev.\ D {\bf 73}, 072003 (2006)
  \bibitem{Jegerlehner:2009rya} 
  F.~Jegerlehner and A.~Nyffeler,
  Phys.\ Rept.\  {\bf 477}, 1 (2009)
  [arXiv:0902.3360].
 \bibitem{Hagiwara:2011af1} 
  K.~Hagiwara, R.~Liao, A.~D.~Martin, D.~Nomura and T.~Teubner,
  J.\ Phys.\ G {\bf 38}, 085003 (2011)
  [arXiv:1105.3149].
  \bibitem{BDDJ}
  M.~Benayoun, P.~David, L.~DelBuono, F.~Jegerlehner,
  Eur.\ Phys.\ J.\ C {\bf 72}, 1848 (2012) [arXiv:1106.1315],
  Eur.\ Phys.\ J.\ C {\bf 73}, 2453 (2013) [arXiv:1210.7184].
  \bibitem{HLS}
   M.~Harada and Y.~Yamawaki, Phys.\ Rept.\  {\bf 381}, 1 (2003)
 [hep-ph/0302103].
\end{thebibliography}

\begin{thebibliography}{99}

\bibitem{deRafael}
  E.~de Rafael,
  Phys.\ Lett.\ B {\bf 322} (1994) 239
  [hep-ph/9311316].

\bibitem{HKS}
  M.~Hayakawa, T.~Kinoshita and A.~Sanda,
  Phys.\ Rev.\ Lett.\  {\bf 75} (1995) 790
  [hep-ph/9503463];
%
  Phys.\ Rev.\ D {\bf 54} (1996) 3137
  [hep-ph/9601310].

  M.~Hayakawa and T.~Kinoshita,
  Phys.\ Rev.\ D {\bf 57} (1998) 465 [hep-ph/9708227]
   [Erratum-ibid.\ D {\bf 66} (2002) 019902, [hep-ph/0112102]].

\bibitem{BPP}
  J.~Bijnens, E.~Pallante and J.~Prades,
  Phys.\ Rev.\ Lett.\  {\bf 75} (1995) 1447
   [Erratum-ibid.\  {\bf 75} (1995) 3781]
  [hep-ph/9505251];
%
  Nucl.\ Phys.\ B {\bf 474} (1996) 379
  [hep-ph/9511388];
%
  Nucl.\ Phys.\ B {\bf 626} (2002) 410
  [hep-ph/0112255].

\bibitem{KN}
  M.~Knecht and A.~Nyffeler,
  Phys.\ Rev.\ D {\bf 65} (2002) 073034
  [hep-ph/0111058].

\bibitem{PdRV}
  J.~Prades, E.~de Rafael and A.~Vainshtein,
  (Advanced series on directions in high energy physics. 20)
  [arXiv:0901.0306 [hep-ph]].

\bibitem{BP}
  J.~Bijnens and J.~Prades,
  Mod.\ Phys.\ Lett.\ A {\bf 22} (2007) 767
  [hep-ph/0702170].

\bibitem{JNreview}
  F.~Jegerlehner and A.~Nyffeler,
  Phys.\ Rept.\  {\bf 477} (2009) 1
  [arXiv:0902.3360 [hep-ph]].

\bibitem{MV}
  K.~Melnikov and A.~Vainshtein,
  Phys.\ Rev.\ D {\bf 70} (2004) 113006
  [hep-ph/0312226].

\bibitem{KNprague}
  K.~Kampf and J.~Novotny,
  Phys.\ Rev.\ D {\bf 84} (2011) 014036
  [arXiv:1104.3137 [hep-ph]].

\bibitem{RGL}
  P.~Roig {\it et al.}, 
  Phys.\ Rev.\ D {\bf 89} (2014) 073016
  [arXiv:1401.4099 [hep-ph]].

\bibitem{BGLP}
  J.~Bijnens {\it et al.}, 
  JHEP {\bf 0304} (2003) 055
  [hep-ph/0304222].

\bibitem{talk}
  J.~Bijnens and M.~Zahiri-Abyaneh,
  EPJ Web Conf.\  {\bf 37} (2012) 01007
  [arXiv:1208.3548].

\bibitem{Mehranthesis}
  M.~Zahiri~Abyaneh,
  arXiv:1208.2554 [hep-ph],
master thesis.

\bibitem{MesonNet13}
J.~Bijnens in
  M.~J.~Amaryan, {\it et al.},
  ``MesonNet 2013 International Workshop. Mini-proceedings,''
  arXiv:1308.2575 [hep-ph].

\bibitem{BR}
J.~Bijnens and J.~Relefors, to be published.

\bibitem{Ramsey-Musolf1}
  K.~T.~Engel, H.~H.~Patel and M.~J.~Ramsey-Musolf,
  Phys.\ Rev.\ D {\bf 86} (2012) 037502
  [arXiv:1201.0809 [hep-ph]].

\bibitem{Ramsey-Musolf2}
  K.~T.~Engel and M.~J.~Ramsey-Musolf,
  arXiv:1309.2225 [hep-ph].

\end{thebibliography}

\vspace{-2mm}

\begin{thebibliography}{99}
\vspace{-1mm}
\bibitem{Davier:2009} M. Davier, A. Hoecker, B. Malaescu, C.Z. Yuan and  Z. Zhang, Eur. Phys. J. C {\bf 66} (2010) 1, arXiv:0908.4300.

\bibitem{Caprini:2014} 
B. Ananthanarayan, I. Caprini, D. Das and I.S. Imsong,  Phys. Rev. D {\bf 89} (2014) 036007, arXiv:1312.5849.
  
  \bibitem{BABAR}
  B. Aubert {\it et al.}  [BABAR Collaboration],
  Phys.\ Rev.\ Lett.\  {\bf 103} (2009) 231801, arXiv:0908.3589.

\bibitem{SND} M.N. Achasov  {\it et al.} [SND Collaboration],
   J.Exp.Theor.Phys. 103 (2006) 380-384, Zh. Eksp. Teor. Fiz. 130 (2006) 437-441, hep-ex/0605013.
 
\bibitem{CMD22}
  R.R. Akhmetshin {\it et al.}  [CMD-2 Collaboration], JETP Lett. {\bf 84}, 413 (2006);
  Phys.\ Lett.\ B {\bf 648} (2007) 28, hep-ex/0610021.


\bibitem{KLOE3} D. Babusci  {\it et al.}  [KLOE and KLOE-2 Collaborations],
	 Phys. Lett. B {\bf 720} (2013) 336, arXiv:1212.4524.

\bibitem{Abbas:2010EPJA}
  G. Abbas, B. Ananthanarayan, I. Caprini, I.S.~Imsong and S.~Ramanan,
  Eur.\ Phys.\ J.\  A {\bf 45}, 389 (2010), arXiv:1004.4257.
\end{thebibliography}

\begin{thebibliography}{99}
\bibitem{Colangelo:2014dfa}
  G.~Colangelo, M.~Hoferichter, M.~Procura and P.~Stoffer,
  arXiv:1402.7081 [hep-ph].

\bibitem{BPP95} 
  J.~Bijnens, E.~Pallante and J.~Prades,
  Nucl.\ Phys.\ B {\bf 474} (1996) 379
  [hep-ph/9511388].

\bibitem{Mandelstam} 
  S.~Mandelstam,
  { Phys.\ Rev.\ Lett.}\ {\bf 4} (1960) 84.

\bibitem{Stern:1993rg}
  J.~Stern, H.~Sazdjian and N.~H.~Fuchs,
  Phys.\ Rev.\ D {\bf 47} (1993) 3814
  [hep-ph/9301244].

\end{thebibliography}

\begin{thebibliography}{99}
\bibitem{1} S. Bodenstein, C. A. Dominguez and K. Schilcher, Phys. Rev. D {\bf 85}, 014029 (2012).
\bibitem{2}S. Bodenstein, C. A. Dominguez, K. Schilcher and H. Spiesberger, Phys. Rev. D {\bf 88}, 014005 (2013). 
\bibitem{3} ETM Collaboration, F.Burger et al. JHEP 1402 (2014) 099, and K. Jansen, private communication.
 \bibitem{4} HPQCD Collaboration,  B. Chakraborty et al. arXiv: 1403.177.
 
\end{thebibliography}

\begin{thebibliography}{99}
\bibitem{babar}
J.P. Lees et al.,  Phys. Rev. D89 (2014) 092002.
\bibitem{bellea}
J. Crnkovic, Talk at the XII Workshop on Tau Lepton Physics,
Nagoya, Japan, September 2012.
\bibitem{kloe}
D. Babusci et al.,  Phys. Lett. B720 (2013) 336.
\bibitem{ISRe}
V.P. Druzhinin et al., Rev. Mod. Phys. 83 (2011) 1545.
\bibitem{fedor}
F.~Ignatov, PoS EPS-HEP2013, 350 (2014).
\bibitem{snde}
M.N. Achasov et al., Phys. Rev. D 88 (2013) 054013.
\bibitem{cmd3}
R.R. Akhmetshin et al., Phys. Lett. B 723 (2013) 82.
\bibitem{ebes1}
J.Z. Bai et al., Phys. Rev. Lett. 84 (2000) 594.
\bibitem{ebes2}
J.Z. Bai et al., Phys. Rev. Lett. 88 (2002) 101802.
\bibitem{ebes3}
M. Ablikim et al., Phys. Rev. Lett. 97 (2006) 262001.
\bibitem{ebes4}
M. Ablikim et al., Phys. Lett. B677 (2009) 239.
\end{thebibliography}

\begin{thebibliography}{99}

\bibitem{Goecke:2011pe}
  T.~Goecke, C.~S.~Fischer and R.~Williams,
  Phys.\ Lett.\ B {\bf 704} (2011) 211
  [arXiv:1107.2588 [hep-ph]].

\bibitem{Goecke:2013fpa}
  T.~Goecke, C.~S.~Fischer and R.~Williams,
  PoS ConfinementX {\bf } (2012) 231
  [arXiv:1302.5252 [hep-ph]].


\bibitem{Fischer:2010iz} 
  C.~S.~Fischer, T.~Goecke and R.~Williams,
  Eur.\ Phys.\ J.\ A {\bf 47}, 28 (2011)
  [arXiv:1009.5297 [hep-ph]].

\bibitem{Goecke:2010if} 
  T.~Goecke, C.~S.~Fischer and R.~Williams,
  Phys.\ Rev.\ D {\bf 83}, 094006 (2011)
  [Erratum-ibid.\ D {\bf 86}, 099901 (2012)]
  [arXiv:1012.3886 [hep-ph]].

\bibitem{Goecke:2012qm} 
  T.~Goecke, C.~S.~Fischer and R.~Williams,
  Phys.\ Rev.\ D {\bf 87}, no. 3, 034013 (2013)
  [arXiv:1210.1759 [hep-ph]].


\end{thebibliography}

\begin{thebibliography}{99}
\setlength{\itemsep}{0cm}

\bibitem{davier}
M.~Davier {\em et al.}, Europ. Phys. J. C {\bf 71}, 1515 (2011).

\bibitem{prades}
J. Prades et al., arXiv:0901.0306, (2009).

\bibitem{brodsky}
S.J. Brodsky and E. de Rafael, Phys. Rev. {\bf 168}, 1620 (1968).

\bibitem{kloe:all}
F.~Ambrosino {\em et al.} (KLOE~Collaboration), Phys. Lett. B 670, 285 (2009),  
Phys.Lett. B700,  102-110 (2011), 
Phys.Lett. B720, 336-343 (2013). 

\bibitem{babar:data}
B.~Aubert {\em et al.} (\babar~Collaboration), PRD {\bf 70}, 072004 (2004), PRD {\bf 71}, 052001 (2005), 
PRD {\bf 73}, 052003 (2006), 
PRD {\bf 73}, 012005 (2006), 
PRD {\bf 76}, 092006 (2007), 
PRD {\bf 76}, 092005 (2007), 
PRD {\bf 76}, 012008 (2007), 
PRD {\bf 77}, 092002 (2008), 
PRL {\bf 103}, 231801 (2009),
PRD {\bf 85}, 112009 (2012),
PRD {\bf 86}, 032013 (2012),
PRD {\bf 88}, 032013 (2013).

\bibitem{babar:kskl}
J.P.Lees {\em et al.} (\babar~Collaboration), PRD {\bf 89}, 092002 (2014).

\bibitem{cmd2:kskl}
R.R. Akhmetshin {\em et al.} (CMD-2~Collaboration) Phys.Lett. B466, 385 (1999), Erratum-ibid. B508, 217-218 (2001).

 \end{thebibliography}

\begin{thebibliography}{99}
\setlength{\itemsep}{0cm}
\bibitem{gilberto}
  G.~Colangelo, M.~Hoferichter, M.~Procura and P.~Stoffer,
  arXiv:1402.7081 [hep-ph].
\bibitem{eta2gammagamma}
C.~Hanhart, A.~Kupsc, U.-G.~Mei\ss ner, F.~Stollenwerk and A.~Wirzba,
  Eur.\ Phys.\ J.\ C {\bf 73} (2013) 2668
  [arXiv:1307.5654 [hep-ph]].
\bibitem{stollenwerk}
 F.~Stollenwerk, C.~Hanhart, A.~Kupsc, U.-G.~Mei\ss ner and A.~Wirzba,
  Phys.\ Lett.\ B {\bf 707} (2012) 184
  [arXiv:1108.2419 [nucl-th]].
  \bibitem{kloe}
   D.~Babusci {\it et al.}  [KLOE Collaboration],
  Phys.\ Lett.\ B {\bf 718} (2013) 910
  [arXiv:1209.4611 [hep-ex]].
  \bibitem{mainz}
   P.~Aguar-Bartolome {\it et al.}  [A2 Collaboration],
  Phys.\ Rev.\ C {\bf 89} (2014) 044608
  [arXiv:1309.5648 [hep-ex]].
\bibitem{babar}
  B. Aubert et al., Phys. Rev. D {\bf 76} (2007) 092005.
  \bibitem{simon_private}
  S. Eidelman, talk presented at this conference and private communication.
     
 \end{thebibliography}

\vspace{-3mm}
\begin{thebibliography}{99}
\bibitem{Mishima:2013ama} 
  G.~Mishima,
  arXiv:1311.7109 [hep-ph].
\bibitem{Aoyama:2012wj} 
  T.~Aoyama, M.~Hayakawa, T.~Kinoshita and M.~Nio,
  Phys.\ Rev.\ Lett.\  {\bf 109}, 111807 (2012)
  [arXiv:1205.5368 [hep-ph]].
\bibitem{Fael:2014nha} 
  M.~Fael and M.~Passera,
  arXiv:1402.1575 [hep-ph].
\bibitem{Melnikov:2014lwa} 
  K.~Melnikov, A.~Vainshtein and M.~Voloshin,
  arXiv:1402.5690 [hep-ph].
\bibitem{Eides:2014swa} 
  M.~I.~Eides,
  arXiv:1402.5860 [hep-ph].
\bibitem{Hayakawa:2014tla} 
  M.~Hayakawa,
  arXiv:1403.0416 [hep-ph].
\end{thebibliography}

\begin{thebibliography}{99}
\setlength{\itemsep}{0cm}

\bibitem{Shintani:2010ph}
  E.~Shintani, S.~Aoki, H.~Fukaya, S.~Hashimoto, T.~Kaneko, T.~Onogi and N.~Yamada,
  Phys.\ Rev.\ D {\bf 82} (2010) 074505
  [arXiv:1002.0371 [hep-lat]].

\bibitem{Herdoiza:2014jta}
  G.~Herdo\'iza, H.~Horch, B.~J\"ager and H.~Wittig,
  PoS LATTICE {\bf 2013} (2014) 444.

\bibitem{Horch:2013lla}
  H.~Horch, G.~Herdo\'iza, B.~J\"ager, H.~Wittig, M.~Della Morte and A.~J\"uttner,
  PoS LATTICE {\bf 2013} (2013) 304
  [arXiv:1311.6975 [hep-lat]].

\bibitem{Francis:2013fzp}
  A.~Francis, B.~J\"ager, H.~B.~Meyer and H.~Wittig,
  Phys.\ Rev.\ D {\bf 88} (2013) 054502
  [arXiv:1306.2532 [hep-lat]].

\bibitem{Jegerlehner:2011mw}
  F.~Jegerlehner,
  Nuovo Cim.\ C {\bf 034S1} (2011) 31
  [arXiv:1107.4683 [hep-ph]].

\bibitem{Davier:2010nc}
  M.~Davier, A.~Hoecker, B.~Malaescu and Z.~Zhang,
  Eur.\ Phys.\ J.\ C {\bf 71} (2011) 1515
   [Erratum-ibid.\ C {\bf 72} (2012) 1874]
  [arXiv:1010.4180 [hep-ph]].

\bibitem{Hagiwara:2011afg}
  K.~Hagiwara, R.~Liao, A.~D.~Martin, D.~Nomura and T.~Teubner,
  J.\ Phys.\ G {\bf 38} (2011) 085003
  [arXiv:1105.3149 [hep-ph]].

\bibitem{DellaMorte:2011aa1}
  M.~Della Morte, B.~J\"ager, A.~J\"uttner and H.~Wittig,
  JHEP {\bf 1203} (2012) 055
  [arXiv:1112.2894 [hep-lat]].

 \end{thebibliography}

\begin{thebibliography}{99}

\bibitem{Blum:2002ii}
T.~Blum.
\newblock {\em Phys.Rev.Lett.}, 91:052001, 2003.

\bibitem{Gockeler:2003cw}
M.~Gockeler et~al.
\newblock {\em Nucl. Phys.}, B688:135--164, 2004.

\bibitem{Aubin:2006xv}
C.~Aubin and T.~Blum.
\newblock {\em Phys. Rev.}, D75:114502, 2007.

\bibitem{Feng:2011zk}
Xu~Feng, Karl Jansen, Marcus Petschlies, and Dru~B. Renner.
\newblock {\em Phys.Rev.Lett.}, 107:081802, 2011.

\bibitem{Boyle:2011hu}
Peter Boyle, Luigi~Del Debbio, Eoin Kerrane, and James Zanotti.
\newblock {\em Phys.Rev.}, D85:074504, 2012.

\bibitem{DellaMorte:2011aa2}
Michele Della~Morte, Benjamin Jager, Andreas Juttner, and Hartmut Wittig.
\newblock {\em JHEP}, 1203:055, 2012.

\bibitem{Aubin:2012me1}
Christopher Aubin, Thomas Blum, Maarten Golterman, and Santiago Peris.
\newblock {\em Phys.Rev.}, D86:054509, 2012.

\bibitem{deDivitiis:2012vs}
G.M. de~Divitiis, R.~Petronzio, and N.~Tantalo.
\newblock {\em Phys.Lett.}, B718:589--596, 2012.

\bibitem{Feng:2013xsa}
Xu~Feng, Shoji Hashimoto, Grit Hotzel, Karl Jansen, Marcus Petschlies, et~al.
\newblock {\em Phys.Rev.}, D88:034505, 2013.

\bibitem{Ji:2001wha}
X.~D. Ji and C.~W. Jung.
\newblock {\em Phys. Rev. Lett.}, 86:208, 2001.

\bibitem{Meyer:2011um}
Harvey~B. Meyer.
\newblock {\em Phys.Rev.Lett.}, 107:072002, 2011.

\bibitem{Feng:2013xqa}
  X.~Feng, S.~Hashimoto, G.~Hotzel, K.~Jansen, M.~Petschlies and D.~B.~Renner,
\newblock  arXiv:1311.0652.





\end{thebibliography}

\vspace*{-2mm}

\begin{thebibliography}{9}

\bibitem{g3pi}
  M.~Hoferichter, B.~Kubis and D.~Sakkas,
  Phys.\ Rev.\ D {\bf 86} (2012) 116009  [arXiv:1210.6793 [hep-ph]].

\vspace*{-0.5mm}

\bibitem{V3pi}
  F.~Niecknig, B.~Kubis and S.~P.~Schneider,
  Eur.\ Phys.\ J.\ C {\bf 72} (2012) 2014
  [arXiv:1203.2501 [hep-ph]].

\vspace*{-0.5mm}

\bibitem{KLOE:phi}
  A.~Aloisio {\it et al.}  [KLOE Collaboration],
  Phys.\ Lett.\ B {\bf 561} (2003) 55
   [Erratum-ibid.\ B {\bf 609} (2005) 449]
  [hep-ex/0303016].

\vspace*{-0.5mm}

\bibitem{omegaTFF}
  S.~P.~Schneider, B.~Kubis and F.~Niecknig,
  Phys.\ Rev.\ D {\bf 86} (2012) 054013
  [arXiv:1206.3098 [hep-ph]].

\vspace*{-0.5mm}

\bibitem{NA60}
  R.~Arnaldi {\it et al.}  [NA60 Collaboration],
  Phys.\ Lett.\ B {\bf 677} (2009) 260
  [arXiv:0902.2547 [hep-ph]];
%
  G.~Usai  [NA60 Collaboration],
  Nucl.\ Phys.\  A {\bf 855} (2011) 189.

\vspace*{-0.5mm}

\bibitem{pi0TFF}
  M.~Hoferichter, B.~Kubis, S.~Leupold, F.~Niecknig and S.~P.~Schneider, 
  {\it in preparation}.

\end{thebibliography}

\begin{thebibliography}{99}
\bibitem{r1}
A. Kurz, T. Liu, P. Marquard and M. Steinhauser, arXiv:1407.0264.
\bibitem{r2}
J. Beringer et al. (Particle Data Group), Phys. Rev. D86, 010001 (2012).
\bibitem{r3}
A. Czarnecki, W. Marciano and A. Vainshtein, Phys. Rev. D67, 073006 (2003).
\bibitem{r4}
 A. Czarnecki and W. Marciano, Phys. Rev. D64, 013014 (2001).
\bibitem{r5}
M. Pospelov, Phys. Rev. D80, 095002 (2009).
\end{thebibliography}

\vspace{-5mm}
\begin{thebibliography}{99}
\vspace{-3mm}

  \bibitem{Jegerlehner:2009ryc} 
  F.~Jegerlehner and A.~Nyffeler,
  Phys.\ Rept.\  {\bf 477}, 1 (2009)
  [arXiv:0902.3360 [hep-ph]];
  J.~Prades, E.~de Rafael and A.~Vainshtein,
  (Advanced series on directions in high energy physics. 20)
  [arXiv:0901.0306 [hep-ph]].

  \bibitem{tHooft:1973jzc} 
  G.~'t Hooft,
  Nucl.\ Phys.\ B {\bf 72}, 461 (1974).

\bibitem{nyffelerc}
Talk given by A.~Nyffeler, in these proceedings.

\bibitem{ourpaper}
 P.~Masjuan and P.~Sanchez-Puertas,   {\it{In preparation}}.      

\bibitem{colangelo}
Talk given by G.~Colangelo, in these proceedings.

  \bibitem{Masjuan:2007ayc} 
  P.~Masjuan and S.~Peris,
  JHEP {\bf 0705}, 040 (2007)
  [arXiv:0704.1247 [hep-ph]];
  P.~Masjuan and S.~Peris,
  Phys.\ Lett.\ B {\bf 663}, 61 (2008)
  [arXiv:0801.3558 [hep-ph]].
 
  \bibitem{Bakerc}
  G.~A.~Baker and P.~Graves-Morris,   Encyclopedia of Mathematics and its Applications,   Cambridge Univ. Press, 1996;   P.~Masjuan Queralt,
  arXiv:1005.5683 [hep-ph].
 
 \bibitem{Masjuan:2008cpc}
  P.~Masjuan, J.~J.~Sanz-Cillero and J.~Virto,
  Phys.\ Lett.\ B {\bf 668} (2008) 14
  [arXiv:0805.3291 [hep-ph]];
  P.~Masjuan and S.~Peris,
  Phys.\ Lett.\ B {\bf 686}, 307 (2010)
  [arXiv:0903.0294 [hep-ph]];
  P.~Masjuan and J.~J.~Sanz-Cillero,
  Eur.\ Phys.\ J.\ C {\bf 73} (2013) 2594
  [arXiv:1306.6308 [hep-ph]].
 

  \bibitem{Knecht:2001qfc} 
  M.~Knecht and A.~Nyffeler,
  Phys.\ Rev.\ D {\bf 65}, 073034 (2002)
  [hep-ph/0111058].

  \bibitem{Masjuan:2012gcc} 
  P.~Masjuan, E.~Ruiz Arriola and W.~Broniowski,
  Phys.\ Rev.\ D {\bf 85}, 094006 (2012)
  [arXiv:1203.4782 [hep-ph]];
  P.~Masjuan, E.~Ruiz Arriola and W.~Broniowski,
  Phys.\ Rev.\ D {\bf 87}, 014005 (2013)
  [arXiv:1210.0760 [hep-ph]].

\bibitem{Eckerc}
  G.~Ecker, P.~Masjuan and H.~Neufeld,
  Phys.\ Lett.\ B {\bf 692} (2010) 184
  [arXiv:1004.3422 [hep-ph]];
  G.~Ecker, P.~Masjuan and H.~Neufeld,
  Eur.\ Phys.\ J.\ C {\bf 74} (2014) 2748
  [arXiv:1310.8452 [hep-ph]].


   \bibitem{Masjuan:2012wyc} 
  P.~Masjuan,
  Phys.\ Rev.\ D {\bf 86}, 094021 (2012)
  [arXiv:1206.2549 [hep-ph]].

\bibitem{Masjuan:2012qnc} 
  P.~Masjuan and M.~Vanderhaeghen,
  arXiv:1212.0357 [hep-ph].

\bibitem{Pabloc}
Talk given by P.~Sanchez-Puertas, in these proceedings.

 \bibitem{Melnikov:2003xdc} 
  K.~Melnikov and A.~Vainshtein,
  Phys.\ Rev.\ D {\bf 70}, 113006 (2004)
  [hep-ph/0312226].
 



\end{thebibliography}

\begin{thebibliography}{99}
\setlength{\itemsep}{0cm}

\bibitem{Bernecker:2011gh}
  D.~Bernecker and H.~B.~Meyer,
  Eur.\ Phys.\ J.\ A {\bf 47} (2011) 148.
    
 \end{thebibliography}

\vspace*{-1cm}
\begin{thebibliography}{99}
\setlength{\itemsep}{0cm}


\bibitem{MHA}
  B.~Moussallam and J.~Stern,
  hep-ph/9404353; 
%
  B.~Moussallam,
  Phys.\ Rev.\ D {\bf 51} (1995) 4939; 
%
  Nucl.\ Phys.\ B {\bf 504} (1997) 381; 
%
  S.~Peris, M.~Perrottet and E.~de Rafael,
  JHEP {\bf 9805} (1998) 011; 
%
  M.~Knecht {\it et al.}, 
  Phys.\ Rev.\ Lett.\  {\bf 83} (1999) 5230. 
  

\bibitem{Bijnens_et_al_03}
  J.~Bijnens {\it et al.},  
  JHEP {\bf 0304} (2003) 055. 


\bibitem{KN_EPJC_01}
  M.~Knecht and A.~Nyffeler,
  Eur.\ Phys.\ J.\ C {\bf 21} (2001) 659. 


\bibitem{N_JN_09}
  A.~Nyffeler,
  Phys.\ Rev.\ D {\bf 79} (2009) 073012; 
%
  F.~Jegerlehner and A.~Nyffeler,
  Phys.\ Rept.\  {\bf 477} (2009) 1. 


\bibitem{Ecker_et_al_PLB_89}
  G.~Ecker {\it et al.}, 
  Phys.\ Lett.\ B {\bf 223} (1989) 425.


\bibitem{RchiT_odd_parity}
  E.~Pallante and R.~Petronzio,  
  Nucl.\ Phys.\ B {\bf 396} (1993) 205; 
%
  J.~Prades,
  Z.\ Phys.\ C {\bf 63} (1994) 491 
   [Erratum-ibid.\ C {\bf 11} (1999) 571]; 
%
  B.~Ananthanarayan and B.~Moussallam,
  JHEP {\bf 0205} (2002) 052; 
%
  P.~D.~Ruiz-Femenia, A.~Pich and J.~Portoles,
  JHEP {\bf 0307} (2003) 003; 
%
  K.~Kampf and B.~Moussallam,
  Eur.\ Phys.\ J.\ C {\bf 47} (2006) 723; 
%
  V.~Mateu and J.~Portoles,
  Eur.\ Phys.\ J.\ C {\bf 52} (2007) 325; 
%
  K.~Kampf and J.~Novotny,
  Phys.\ Rev.\ D {\bf 84} (2011) 014036; 
%
  H.~Czy\.z {\it et al.}, 
  Phys.\ Rev.\ D {\bf 85} (2012) 094010; 
%
  P.~Roig and J.~J.~Sanz-Cillero,
  Phys.\ Lett.\ B {\bf 733} (2014) 158. 


\bibitem{KN_02}
  M.~Knecht and A.~Nyffeler,
  Phys.\ Rev.\ D {\bf 65} (2002) 073034. 
    

\bibitem{MV_04}
  K.~Melnikov and A.~Vainshtein,
  Phys.\ Rev.\ D {\bf 70} (2004) 113006. 


\bibitem{KaNo_11}
  K.~Kampf and J.~Novotny, quoted in Ref.~\cite{RchiT_odd_parity}. 


\bibitem{RGL_14}
  P.~Roig, A.~Guevara and G.~L.~Castro,
  Phys.\ Rev.\ D {\bf 89} (2014) 073016. 


\bibitem{KN_VVVV}
  M.~Knecht and A.~Nyffeler, work in progress. 


 \end{thebibliography}

\begin{thebibliography}{99}
\setlength{\itemsep}{0cm}


\bibitem{Pascalutsa:2010sj}
  V.~Pascalutsa and M.~Vanderhaeghen,
  Phys.\ Rev.\ Lett.\  {\bf 105} (2010) 201603.
  
\bibitem{Pascalutsa:2012pr}
  V.~Pascalutsa, V.~Pauk and M.~Vanderhaeghen,
  Phys.\ Rev.\ D {\bf 85} (2012) 116001.
  
  \bibitem{Gerasimov:1973ja}
  S.~B.~Gerasimov and J.~Moulin,
  Yad.\ Fiz.\  {\bf 23} (1976) 142
   [Nucl.\ Phys.\ B {\bf 98} (1975) 349].
  
 
   

\bibitem{Pauk:2013hxa}
  V.~Pauk, V.~Pascalutsa and M.~Vanderhaeghen,
  Phys.\ Lett.\ B {\bf 725} (2013) 504.
    
\bibitem{Pascalutsa:2014ena}
  V.~Pascalutsa,
  arXiv:1402.4973 [nucl-th].

 \end{thebibliography}

\begin{thebibliography}{99}
\setlength{\itemsep}{0cm}

\bibitem{Pauk:2014jza} 
  V.~Pauk and M.~Vanderhaeghen,
  arXiv:1403.7503 [hep-ph].
  
\bibitem{Knecht:2001qf222} 
  M.~Knecht and A.~Nyffeler,
  Phys.\ Rev.\ D {\bf 65}, 073034 (2002)
  [hep-ph/0111058].


\end{thebibliography}

\begin{thebibliography}{99}
\bibitem{hoferichter}G. Colangelo, M. Hoferichter, M. Procura, and P. Stoffer, {arXiv:1402.7081 [hep-ph]}.
\bibitem{belle_pic}T. Mori {\it et al.} [Belle], Phys. Rev. {\bf D75} (2007) 051101, {arXiv:0610038 [hep-ex]},
J. Phys. Soc. Jap. {\bf 76} (2007) 074102, {arXiv:0704.3538 [hep-ph]}.
\bibitem{belle_pin}K. Abe {\it et al.} [Belle],  {arXiv:0711.1926 [hep-ex]};

S. Uehara {\it et al.} [Belle], Phys. Rev. {\bf D78} (2008) 052004, {arXiv:0805.3387 [hep-ex]};

S. Uehara {\it et al.} [Belle], Phys. Rev. {\bf D79} (2009) 052009, {arXiv:0903.3697 [hep-ex]}.
\bibitem{belle_Kc}K. Abe {\it et al.} [Belle], Eur. Phys. J. {\bf C32} (2004) 323, {arXiv:0309077 [hep-ex]}.

\bibitem{belle_Ks}S. Uehara {\it et al.} [Belle], PTEP (2013) 123C01,  {arXiv:1307.7457 [hep-ex]}.
\bibitem{belle_eta}S. Uehara {\it et al.} [Belle], Phys.\ Rev.\ {\bf D80} (2009) 032001.

\bibitem{morgan_p}D. Morgan and M.R. Pennington, Z. Phys. {\bf C37} (1988) 431.
\bibitem{Daphne_p}M.R. Pennington, {\it DA$\Phi$NE Physics Handbook}, ed. L. Maiani, G. Pancheri and N. Paver (INFN, Frascati, 1992) pp. 379-418;
{\it Second DA$\Phi$NE Physics Handbook}, ed. L. Maiani {\it et al.} (INFN, Frascati, 1995) pp. 169-190.
\bibitem{morgan_p2}D. Morgan and M.R. Pennington, Z. Phys. {\bf C48} (1990) 623.

\bibitem{boglione}M. Boglione and M.R. Pennington, Eur. Phys. J. {\bf C9} (1999) 11, {arXiv:9812258 [hep-ph]}.
\bibitem{mori}M.R. Pennington, T. Mori, S. Uehara, and Y. Watanabe, Eur. Phys. J. {\bf C56} (2008) 1, {arXiv:0803.3389 [hep-ph]}.


\bibitem{NA48}J. R. Batley {\it et al.} [NA48/2], {Eur. Phys. J. {\bf C70} (2010) 635}.
\bibitem{dirac}L.~Afanasev [DIRAC], AIP Conf.Proc. 814(2006) 220.
     

\bibitem{KPY}
R. Garc\'ia-Mart\'in, R. Kami\'nski, J. R. Pel\'aez, J. Ruiz de Elvira, and F. J. Yndur\'ain, Phys. Rev. {\bf D83} (2011) 074004, {arXiv:1102.2183 [hep-ph]};
Phys. Rev. {\bf D77} (2008) 054015, {arXiv:0710.1150 [hep-ph]};

R. Kami\'nski, J. R. Pel\'aez, and F. J. Yndur\'ain, Phys. Rev. {\bf D74} (2006) 014001, Erratum-ibid. {\bf D74} (2006) 079903, {arXiv:0603170 [hep-ph]};

J.R. Pel\'aez and F. J. Yndur\'ain, Phys. Rev. {\bf D71} (2005) 074016, {arXiv:0411334 [hep-ph]}.

\bibitem{descotes}P. Buttiker, S. Descotes-Genon and B. Moussallam, Eur. Phys. J. {\bf C33} (2004) 409, {arXiv:0310283 [hep-ph]}.
\bibitem{markII}J. Boyer {\it et al.} [Mark II], {Phys. Rev. {\bf D42} (1990) 1350}.
\bibitem{cello}J. Harjes, {Ph.D. thesis}, submitted to the University of Hamburg;
H.J. Behrend {\it et al.} [CELLO], {Z. Phys. {\bf C56} (1992) 381}.
\bibitem{cb}H. Marsiske {\it et al.} [Crystal Ball], {Phys. Rev. {\bf D41} (1990) 3324};
J.K. Bienlein {\it et al.} [Crystal Ball],
{Proc. {\it IX Int. Workshop on Photon-Photon Collisions}},
San Diego 1992, ed. D. Caldwell and H.P. Paar (World Scientific, 1992), pp. 241.
\bibitem{argus}H. Albrecht {\it et al.} [ARGUS], {Z. Phys. {\bf C48} (1990) 183}.
\bibitem{cello_K}H.J. Behrend {\it et al.} [CELLO], {Z. Phys. {\bf C31} (1989) 91}.
\bibitem{tasso}M. Althoff {\it et al.} [TASSO], {Phys. Lett. {\bf B121} (1983) 216}, {Z. Phys. {\bf C29} (1986) 189}.
\bibitem{tpc}H. Aihara {\it et al.} [TPC], {Phys. Rev. Lett. {\bf 57} (1986) 404}.
\bibitem{dai_letter} Ling-Yun Dai and M.~R.~Pennington, arXiv:1403.7414 [hep-ph].
\bibitem{dai_long}Ling-Yun Dai and M.~R.~Pennington, arXiv:1404.7424 [hep-ph].
\bibitem{vanderhaeghen}V. Pascalutsa, V. Pauk and M. Vanderhaeghen, Phys. Rev. {\bf D85} (2012) 116001, {arXiv:1204.0740 [hep-ph]};

A. Nyffeler {arXiv:1312.4804 [hep-ph]}.


\end{thebibliography}

\begin{thebibliography}{99}

\bibitem{Golterman:2013vca}
  M.~Golterman, K.~Maltman and S.~Peris,
  Phys.\ Rev.\ D {\bf 88}, 114508 (2013)
  [arXiv:1309.2153 [hep-lat]].

\bibitem{Aubin:2012me}
  C.~Aubin, T.~Blum, M.~Golterman and S.~Peris,
  Phys.\ Rev.\ D {\bf 86}, 054509 (2012)
  [arXiv:1205.3695 [hep-lat]].



\bibitem{Chakraborty:2014mwa}
  B.~Chakraborty, C.~T.~H.~Davies, G.~C.~Donald, R.~J.~Dowdall, J.~Koponen, G.~P.~Lepage and T.~Teubner,
  arXiv:1403.1778 [hep-lat].




\end{thebibliography}

\begin{thebibliography}{99}
\setlength{\itemsep}{0cm}

\bibitem{Bennett:2006fi}
  G.~W.~Bennett {\it et al.}  [Muon G-2 Collaboration],
  Phys.\ Rev.\ D {\bf 73} (2006) 072003
  [hep-ex/0602035].
  
\bibitem{Jegerlehner:2009ryp}
  F.~Jegerlehner and A.~Nyffeler,
  Phys.\ Rept.\  {\bf 477} (2009) 1
  [arXiv:0902.3360 [hep-ph]].

\bibitem{Adler:1969gk}
  S.~L.~Adler,
  Phys.\ Rev.\  {\bf 177} (1969) 2426;
  J.~S.~Bell and R.~Jackiw,
  Nuovo Cim.\ A {\bf 60} (1969) 47.

\bibitem{Lepage:1980fj}
  G.~P.~Lepage and S.~J.~Brodsky,
  Phys.\ Rev.\ D {\bf 22} (1980) 2157.
  
\bibitem{Novikov:1983jt}
  V.~A.~Novikov, M.~A.~Shifman, A.~I.~Vainshtein, M.~B.~Voloshin and V.~I.~Zakharov,
  Nucl.\ Phys.\ B {\bf 237} (1984) 525.

 \bibitem{Dorokhov:2009xs}
  A.~E.~Dorokhov, M.~A.~Ivanov and S.~G.~Kovalenko,
  Phys.\ Lett.\ B {\bf 677} (2009) 145
  [arXiv:0903.4249 [hep-ph]].
  
  \bibitem{Abouzaid:2006kkp}
  E.~Abouzaid {\it et al.}  [KTeV Collaboration],
  Phys.\ Rev.\ D {\bf 75} (2007) 012004
  [hep-ex/0610072].
  
  \bibitem{Knecht:2001qf}
  M.~Knecht and A.~Nyffeler,
  Phys.\ Rev.\ D {\bf 65} (2002) 073034
  [hep-ph/0111058].
  
\bibitem{Masjuan:2012wy}
  P.~Masjuan,
  Phys.\ Rev.\ D {\bf 86} (2012) 094021
  [arXiv:1206.2549 [hep-ph]].

\bibitem{Escribano:2013kba}
  R.~Escribano, P.~Masjuan and P.~Sanchez-Puertas,
  Phys.\ Rev.\ D {\bf 89} (2014) 034014
  [arXiv:1307.2061 [hep-ph]];
  EPJ Web Conf.\  {\bf 73} (2014) 03004
  [arXiv:1402.4951 [hep-ph]].

  
 \bibitem{Baker}
  G.~A.~Baker and P.~Graves-Morris, 
  Encyclopedia of Mathematics and its Applications, 
  Cambridge Univ. Press, 1996
  
 \bibitem{Chisholm}
  J.~S.~R.~Chisholm,
  Mathematics of Computation {\bf 27} (1973) 841-848.

 \bibitem{Babu:1982yz}
  K.~S.~Babu and E.~Ma,
  Phys.\ Lett.\ B {\bf 119} (1982) 449.

\bibitem{Masjuan:2007ay}
  P.~Masjuan and S.~Peris,
  JHEP {\bf 0705} (2007) 040
  [arXiv:0704.1247 [hep-ph]].

\bibitem{PMPS}
  P.~Masjuan and P.~Sanchez-Puertas,
  {\it{In preparation}}.      

 \end{thebibliography}

\begin{thebibliography}{99}

\bibitem{Kinoshita2012}
  T.~Aoyama, M.~Hayakawa, T.~Kinoshita and M.~Nio,
  Phys.\ Rev.\ Lett.\  {\bf 109} (2012) 111808
  [arXiv:1205.5370 [hep-ph]].
  
\bibitem{Davier}
  M.~Davier, A.~Hoecker, B.~Malaescu and Z.~Zhang,
  Eur.\ Phys.\ J.\ C {\bf 71} (2011) 1515
  [Erratum-ibid.\ C {\bf 72} (2012) 1874]
  [arXiv:1010.4180 [hep-ph]].
  
\bibitem{HMNT}
  K.~Hagiwara, R.~Liao, A.~D.~Martin, D.~Nomura and T.~Teubner,
  J.\ Phys.\ G G {\bf 38} (2011) 085003
  [arXiv:1105.3149 [hep-ph]].

\bibitem{Benayoun:2012wc}
  M.~Benayoun, P.~David, L.~DelBuono and F.~Jegerlehner,
  Eur.\ Phys.\ J.\ C {\bf 73} (2013) 2453
  [arXiv:1210.7184 [hep-ph]].
\bibitem{JegerlehnerSzafron}
  F.~Jegerlehner and R.~Szafron,
  Eur.\ Phys.\ J.\ C {\bf 71} (2011) 1632
  [arXiv:1101.2872 [hep-ph]].
\bibitem{Gnendiger:2013pva}
  C.~Gnendiger, D.~St{\"o}ckinger and H.~St{\"o}ckinger-Kim,
  Phys.\ Rev.\ D {\bf 88} (2013) 053005
  [arXiv:1306.5546 [hep-ph]].

\bibitem{Carey:2009zzb}
  R.~M.~Carey, K.~R.~Lynch, J.~P.~Miller, B.~L.~Roberts, W.~M.~Morse, Y.~K.~Semertzides, V.~P.~Druzhinin and B.~I.~Khazin {\it et al.},
  FERMILAB-PROPOSAL-0989.

\bibitem{Adam:2010uz}
  C.~Adam, J.~-L.~Kneur, R.~Lafaye, T.~Plehn, M.~Rauch and D.~Zerwas,
  Eur.\ Phys.\ J.\ C {\bf 71} (2011) 1520
  [arXiv:1007.2190 [hep-ph]].
\end{thebibliography}

\begin{thebibliography}{99}

\bibitem{Actis:2010gg}
  S.~Actis {\it et al.}  [Working Group on Radiative Corrections and Monte Carlo Generators for Low Energies Collaboration],
  Eur.\ Phys.\ J.\ C {\bf 66} (2010) 585
  [arXiv:0912.0749 [hep-ph]].

\bibitem{Czyz:2000wh}
  H.~Czyz and J.~H.~Kuhn,
  Eur.\ Phys.\ J.\ C {\bf 18} (2001) 497
  [hep-ph/0008262].

\bibitem{Binner:1999bt}
  S.~Binner, J.~H.~Kuhn and K.~Melnikov,
  Phys.\ Lett.\ B {\bf 459} (1999) 279
  [hep-ph/9902399].
\bibitem{Rodrigo:2001jr}
  G.~Rodrigo, A.~Gehrmann-De Ridder, M.~Guilleaume and J.~H.~Kuhn,
  Eur.\ Phys.\ J.\ C {\bf 22} (2001) 81
  [hep-ph/0106132].
\bibitem{Kuhn:2002xg}
  J.~H.~Kuhn and G.~Rodrigo,
  Eur.\ Phys.\ J.\ C {\bf 25} (2002) 215
  [hep-ph/0204283].

\bibitem{Rodrigo:2001kf}
  G.~Rodrigo, H.~Czyz, J.~H.~Kuhn and M.~Szopa,
  Eur.\ Phys.\ J.\ C {\bf 24} (2002) 71
  [hep-ph/0112184].

\bibitem{Czyz:2002np}
  H.~Czyz, A.~Grzelinska, J.~H.~Kuhn and G.~Rodrigo,
  Eur.\ Phys.\ J.\ C {\bf 27} (2003) 563
  [hep-ph/0212225].

\bibitem{Czyz:2003ue}
  H.~Czyz, A.~Grzelinska, J.~H.~Kuhn and G.~Rodrigo,
  Eur.\ Phys.\ J.\ C {\bf 33} (2004) 333
  [hep-ph/0308312].

\bibitem{Czyz:2004rj}
  H.~Czyz, A.~Grzelinska, J.~H.~Kuhn and G.~Rodrigo,
  Eur.\ Phys.\ J.\ C {\bf 39} (2005) 411
  [hep-ph/0404078].

\bibitem{Czyz:2010hj}
  H.~Czyz, A.~Grzelinska and J.~H.~Kuhn,
  Phys.\ Rev.\ D {\bf 81} (2010) 094014
  [arXiv:1002.0279 [hep-ph]].

\bibitem{Campanario:2013uea}
  F.~Campanario, H.~Czy\.z, J.~Gluza, M.~Gunia, T.~Riemann, G.~Rodrigo and V.~Yundin,
  JHEP {\bf 1402} (2014) 114
  [arXiv:1312.3610 [hep-ph]].

\bibitem{Czyz:2013xga}
  H.~Czy\.z, M.~Gunia and J.~H.~K\"uhn,
  JHEP {\bf 1308} (2013) 110
  [arXiv:1306.1985, arXiv:1306.1985 [hep-ph]].

\bibitem{Babusci:2012rp}
  D.~Babusci {\it et al.}  [KLOE Collaboration],
  Phys.\ Lett.\ B {\bf 720} (2013) 336
  [arXiv:1212.4524 [hep-ex]].

\bibitem{Ambrosino:2010bv}
  F.~Ambrosino {\it et al.}  [KLOE Collaboration],
  Phys.\ Lett.\ B {\bf 700} (2011) 102
  [arXiv:1006.5313 [hep-ex]].
\bibitem{Lees:2012cj}
  J.~P.~Lees {\it et al.}  [BaBar Collaboration],
  Phys.\ Rev.\ D {\bf 86} (2012) 032013
  [arXiv:1205.2228 [hep-ex]].
\bibitem{Druzhinin:2010er}
  V.~P.~Druzhinin, L.~A.~Kardapoltsev and V.~A.~Tayursky,
  arXiv:1010.5969 [hep-ph].

\bibitem{Ivashyn:2013uja}
  S.~Ivashyn and H.~Czy\.z,
  Acta Phys.\ Polon.\ B {\bf 44} (2013) 11,  2275.

\bibitem{Czyz:2006dm}
  H.~Czyz and E.~Nowak-Kubat,
  Phys.\ Lett.\ B {\bf 634} (2006) 493
  [hep-ph/0601169].

\bibitem{Czyz:2010sp}
  H.~Czyz and S.~Ivashyn,
  Comput.\ Phys.\ Commun.\  {\bf 182} (2011) 1338
  [arXiv:1009.1881 [hep-ph]].

\end{thebibliography}

\begin{thebibliography}{99}
\bibitem{CAD1}C.~A. Dominguez, Phys. Lett. B \textbf {512}, 331 (2001).
\bibitem{Masjuan} P. Masjuan, E. Ruiz-Arriola and W. Broniowski, EPJ Web Conf. \textbf{73} (2014) 04021.
\bibitem{CAD2}C. A. Dominguez, and T. Thapedi, JHEP \textbf {0410}, 003 (2004).
\bibitem{CAD3} C. A. Dominguez,and R. R\"{o}ntsch,JHEP \textbf {0710}, 085 (2007).
\bibitem{CAD4}C. A. Dominguez, Phys. Rev. D \textbf{28}, 2314 (1983); Mod. Phys. Lett. A \textbf{2}, 983 (1987).
\bibitem{CAD5} R. A. Bryan, C. A. Dominguez and B. J. VerWest, Phys. Rev. C \textbf{22}, 160 (1983).
\bibitem{RB} E. Ruiz-Arriola and W. Broniowski, Phys. Rev. D \textbf{78}, 034031 (2008).
\bibitem{Ruiz} P. Masjuan, E. Ruiz-Arriola and W. Broniowski, Phys. Rev. D \textbf{87}, 014005 (2013).
\end{thebibliography}

\begin{thebibliography}{99}
\bibitem{lbl1}
J. Prades, E. de Rafael, A. Vainshtein, arXiv: 0901.0306.
\bibitem{lbl2}
G. Colangelo et al., arXiv: 1402.7081.
\bibitem{sndrad1}
M.N. Achasov et al., Eur. Phys. J. C12 (2000) 25. 
\bibitem{sndrad2}
M.N. Achasov et al., Phys. Lett. B559 (2003) 171. 
\bibitem{sndrad3}
M.N. Achasov et al., Phys. Rev. D76 (2007) 077101. 
\bibitem{cmdrad}
R.R. Akhmetshin et al., Phys. Lett. B605 (2005) 26.
\bibitem{babarrad}
B. Aubert et al., Phys. Rev. D74 (2006) 012002.
\bibitem{sndradnew}
M.N. Achasov et al., arXiv:1312.7078.
\bibitem{KLOEe}
F. Ambrosino et al., Phys. Lett. B648 (2007) 267.
\bibitem{cmdcon1}
R.R.~Akhmetshin et al., Phys. Lett. B501 (2001) 191.
\bibitem{cmdcon2}
R.R.~Akhmetshin et al., Phys. Lett. B503 (2001) 237.
\bibitem{cmdcon3}
R.R.~Akhmetshin et al., Phys. Lett. B613 (2005) 29.
\bibitem{sndcon1}
M.N. Achasov et al., Phys. Lett. B504 (2001) 275.
\bibitem{sndcon2}
M.N. Achasov et al.,JETP Lett. 75 (2002) 449.
\bibitem{sndcon3}
M.N. Achasov et al., JETP 107 (2008) 61.
\bibitem{dzhelyadin}
R.I. Dzhelyadin et al., Phys. Lett. B102 (1981) 296.
\bibitem{arnaldi}
R. Arnaldi et al., Phys. Lett. B677 (2009) 260.
\bibitem{babarpi}
B. Aubert et al., Phys. Rev. D80 (2009) 052002.
\bibitem{bellepi}
S. Uehara et al., Phys. Rev. D86 (2012) 092007.
\bibitem{babaretac}
J.P. Lees et al., Phys. Rev. D81 (2010) 052010. 
\bibitem{babareta}
P. del Amo Sanchez et al., Phys. Rev. D84 (2011) 052001.
 
\end{thebibliography}

\begin{thebibliography}{1}
\bibitem{Terschlusen:2011pm}
Terschlusen C and Leupold S 2012 {\em Prog.Part.Nucl.Phys.\/} {\bf 67} 401--405

\bibitem{Schneider:2012ez}
Schneider S~P, Kubis B and Niecknig F 2012 {\em Phys.Rev.\/} {\bf D86} 054013

\bibitem{Ivashyn:2011hb}
Ivashyn S 2012 {\em Prob.Atomic Sci.Technol.\/} {\bf 2012N1} 179--182

\bibitem{Babusci:2012ik}
Babusci D {\em et~al.\/} (KLOE-2 Collaboration) 2013 {\em JHEP\/} {\bf 1301} 119

\bibitem{Colangelo:2014dfa22}
Colangelo G {\em et~al.\/} 2014  (\textit{Preprint} 1402.7081)

\bibitem{Masjuan:2014}
Masjuan P 2014 {\em these proceedings\/}

\bibitem{Babusci:2009sg}
Babusci D {\em et~al.\/} 2010 {\em Nucl.Instrum.Meth.\/} {\bf A617} 81--84

\bibitem{Archilli:2010zza}
Archilli F {\em et~al.\/} 2010 {\em Nucl.Instrum.Meth.\/} {\bf A617} 266--268

\bibitem{Babusci:2011bg}
Babusci D {\em et~al.\/} 2012 {\em Eur.Phys.J.\/} {\bf C72} 1917

\end{thebibliography}

\begin{thebibliography}{99}
\setlength{\itemsep}{0cm}
\bibitem{g-2:experimental}
G.~W.~Bennett, et al., [Muon $g-2$ Collaboration],
Phys.\ Rev.\ D {\bf 73} (2006) 072003.
\bibitem{g-2:review}
F.~Jegerlehner and A.~Nyffeler,
Phys.\ Rep.\ {\bf 477} (2009) 1-110.
\bibitem{g-2:LBL:dispersive approach}
G.~Colangelo, M.~Hoferichter, M.~Procura, and P.~Stoffer,
arxiv:1402.7081.
\bibitem{pipi:experiments}
J.~Boyer, et al., [Mark II Collaboration],
Phys.\ Rev.\ D {\bf 42} (1990) 1350;
H.~J.~Behrend, et al., [CELLO Collaboration],
Z.\ Phys.\ C {\bf 56} (1992) 381;
T.~Mori, et al., [Belle Collaboration],
J.\ Phys.\ Soc.\ Jpn. {\bf 76} (2007) 074102;
\bibitem{RADCOR}
F.~A.~Berends, P.~H.~Daverveldt, and R.~Kleiss,
Comput.\ Phys.\ Commun. {\bf 40} (1986) 271-284.
\bibitem{DIAG36}
F.~A.~Berends, P.~H.~Daverveldt, and R.~Kleiss,
Comput.\ Phys.\ Commun. {\bf 40} (1986) 309-326.
\bibitem{EKHARA}  
H.~Czyz and E.~Nowak-Kubat, 
Acta\ Phys.\ Polon.\ {\bf B36}, (2005) 2425;
H.~Czyz and E.~Nowak-Kubat, 
Phys.\ Lett.\ {\bf B634}, (2006) 493.    
 \end{thebibliography}

\vspace{-3mm}
\begin{thebibliography}{9}
\vspace{-1mm}
\bibitem{Colangelo:2014dfa}
  G.~Colangelo, M.~Hoferichter, M.~Procura and P.~Stoffer,
  arXiv:1402.7081 [hep-ph].

\bibitem{Dai:2014zta}
  L.-Y.~Dai and M.~R.~Pennington,
  arXiv:1404.7524 [hep-ph].
  
\bibitem{GM}
  R.~Garc\'ia-Mart\'in and B.~Moussallam,
  Eur.\ Phys.\ J.\ C {\bf 70} (2010) 155
  [arXiv:1006.5373 [hep-ph]].

\bibitem{Hoferichter:2011wk}
  M.~Hoferichter, D.~R.~Phillips and C.~Schat,
  Eur.\ Phys.\ J.\ C {\bf 71} (2011) 1743
  [arXiv:1106.4147 [hep-ph]].
  
\bibitem{Moussallam13} 
  B.~Moussallam,
   Eur.\ Phys.\ J.\ C {\bf 73} (2013)  2539
  [arXiv:1305.3143 [hep-ph]].
  
\bibitem{Hoferichter:2013ama}
  M.~Hoferichter, G.~Colangelo, M.~Procura and P.~Stoffer,
  arXiv:1309.6877 [hep-ph].
  
\bibitem{Kubis}
  B.~Kubis, these proceedings.


\end{thebibliography}

\begin{thebibliography}{99}
\setlength{\itemsep}{0cm}

\bibitem{bes3} M. Ablikim et al. (BESIII Collaboration), 
Nucl. Instr. Meth. A 614, 345 (2010).

\end{thebibliography}

\begin{thebibliography}{99}
\setlength{\itemsep}{0cm}
	\bibitem{Jegerlehner} F. Jegerlehner, Acta Phys.Polon.$\bf{B38}$(2007)3021
	\bibitem{BaBar} BABAR Collaboration, Phys.Rev.Lett.$\bf{103}$:231801 (2009)
	\bibitem{KLOE} KLOE Collaboration, Phys.Lett.$\bf{B700}$:102-110 (2011)
	\bibitem{CMD2} CMD2 Collaboration, Phys.Lett.$\bf{B648}$:28 (2007)	
	\bibitem{design and construction of BES-III} M. Ablikim et al., Nuclear Instruments and Methods in Physics Research A $\bf{614}$:345-399 (2010)
	\bibitem{ISR} V. Druzhinin, S. Eidelman, S. Serednyakov and E. Solodov, Rev. Mod. Phys. $\bf{83}$, 1545 (2011)
	\bibitem{PHOKHARA1} H. Czyz, J.H. KŸhn, A. Wapienik, Phys.Rev.$\bf{D77}$:114005 (2008)
	\bibitem{PHOKHARA2} H. Czyz, A. Grzelinska, J.H. KŸhn, Phys.Rev.$\bf{D81}$:094014 (2010)
	\bibitem{PHOKHARA3} H. Czyz, J.H. KŸhn,, Phys.Rev.$\bf{D80}$:034035 (2009)	
	\bibitem{lumi} BESIII Collaboration, \begin{it}Measurement of the integrated luminosities of the data taken by BESIII at $\sqrt{s}$= 3.650 and 3.773 GeV\end{it}, Chinese Physics C$\bf{37}$, 123001 (2013)
    
 \end{thebibliography}

\begin{thebibliography}{99}
\setlength{\itemsep}{0cm}
\bibitem{belle1} A. Abashian {\em et al.} (Belle Collaboration), Nucl. Instrum.
Methods Phys. Res., Sect. A {\bf 479}, 117 (2002); S. Kurokawa and E. Kikutani, Nucl. Instrum. Methods
Phys. Res., Sect. A {\bf 499}, 1 (2003), and other papers included in this volume.

\bibitem{z3930} S. Uehara {\em et al.} (Belle Collaboration), Phys. Rev. Lett. {\bf 96}, 082003 (2006).

\bibitem{potential} E. Eichten, K. Gottfried, T. Kinoshita, K. D. Lane, and T. M. Yan, 
Phys. Rev. D {\bf 17}, 3090 (1978); Phys. Rev. D {\bf 21}, 203 (1980).

\bibitem{x3915} S. Uehara {\em et al.} (Belle Collaboration), Phys. Rev. Lett. {\bf 104}, 092001 (2010).

\bibitem{babar-x3915} J. P. Lees {\em et al.} ($BABAR$ Collaboration), Phys. Rev. D {\bf 86}, 072002 (2012).

\bibitem{gg2vv} Z. Q. Liu {\em et al.} (Belle Collaboration), Phys. Rev. Lett. {\bf 108}, 232001 (2012).

\bibitem{tetra} N. N. Achasov and G. N. Shestakov, Usp. Fiz. Nauk {\bf 161},
53 (1991) [Sov. Phys. Usp. {\bf 34}, 471 (1991)].
    
\bibitem{t-chan} G. Alexander, A. Levy, and U. Maor, Z. Phys. C {\bf 30}, 65
(1986).

\bibitem{one-pion} N. N. Achasov, V. A. Karnakov, and G. N. Shestakov, Z.
Phys. C {\bf 36}, 661 (1987).  

\bibitem{pQCD} V. L. Chernyak, arXiv:0912.0623.

\bibitem{babar-tff} B. Aubert {\em et al.} ($BABAR$ Collaboration), Phys. Rev. D {\bf 80}, 052002 (2009).

\bibitem{belle-tff} S. Uehara {\em et al.} (Belle Collaboration), Phys. Rev. D {\bf 86}, 092007 (2012).

\bibitem{pi0} G. P. Lepage and S. J. Brodsky, Phys. Rev. D {\bf 22}, 2157 (1980).
\end{thebibliography}

\begin{thebibliography}{99}

\bibitem{Colangelo:2014dfa2}
  G.~Colangelo, M.~Hoferichter, M.~Procura and P.~Stoffer,
  arXiv:1402.7081 [hep-ph].

\bibitem{Pennington}D. Morgan and M.R. Pennington, Z. Phys. {\bf C37} (1988) 431; D. Morgan and M.R. Pennington, Z. Phys. {\bf C48} (1990) 623.

\bibitem{Drechsel:1999rf}
  D.~Drechsel, M.~Gorchtein, B.~Pasquini and M.~Vanderhaeghen,
  Phys.\ Rev.\ C {\bf 61} (1999) 015204
  [hep-ph/9904290].

\bibitem{GarciaMartin:2010cw2}
  R.~Garcia-Martin and B.~Moussallam,
  Eur.\ Phys.\ J.\ C {\bf 70} (2010) 155
  [arXiv:1006.5373 [hep-ph]].

\bibitem{Hoferichter:2011wk2}
  M.~Hoferichter, D.~R.~Phillips and C.~Schat,
  Eur.\ Phys.\ J.\ C {\bf 71} (2011) 1743
  [arXiv:1106.4147 [hep-ph]].
    
\bibitem{Dai:2014zta2}
  L.~-Y.~Dai and M.~R.~Pennington,
  arXiv:1404.7524 [hep-ph].

\bibitem{AMV}
N.~Asmusen, P.~Masjuan, and M.~Vanderhaeghen, \emph{ in Preparation}.

  
\bibitem{belle_pic2}T. Mori {\it et al.} [Belle], Phys. Rev. {\bf D75} (2007) 051101, {arXiv:0610038 [hep-ex]},
J. Phys. Soc. Jap. {\bf 76} (2007) 074102, {arXiv:0704.3538 [hep-ph]}.

\bibitem{belle_pin2}K. Abe {\it et al.} [Belle],  {arXiv:0711.1926 [hep-ex]}; S. Uehara {\it et al.} [Belle], Phys. Rev. {\bf D78} (2008) 052004, {arXiv:0805.3387 [hep-ex]}; S. Uehara {\it et al.} [Belle], Phys. Rev. {\bf D79} (2009) 052009, {arXiv:0903.3697 [hep-ex]}.

\bibitem{yguo}
Talk given by Y.~Guo, in these proceedings.



\bibitem{Moussallam:2013una2}
  B.~Moussallam,
  Eur.\ Phys.\ J.\ C {\bf 73} (2013) 2539
  [arXiv:1305.3143 [hep-ph]].

\bibitem{GarciaMartin:2011cn2}
  R.~Garcia-Martin, R.~Kaminski, J.~R.~Pelaez, J.~Ruiz de Elvira and F.~J.~Yndurain,
  Phys.\ Rev.\ D {\bf 83} (2011) 074004
  [arXiv:1102.2183 [hep-ph]].

\bibitem{Masjuan:2008fv} 
  P.~Masjuan, S.~Peris and J.~J.~Sanz-Cillero,
  Phys.\ Rev.\ D {\bf 78}, 074028 (2008)
  [arXiv:0807.4893 [hep-ph]].

\bibitem{Pascalutsa:2010sj2}
  V.~Pascalutsa and M.~Vanderhaeghen,
  Phys.\ Rev.\ Lett.\  {\bf 105} (2010) 201603.

\bibitem{Escribano:2013kba2}
  R.~Escribano, P.~Masjuan and P.~Sanchez-Puertas,
  Phys.\ Rev.\ D {\bf 89} (2014) 034014
  [arXiv:1307.2061 [hep-ph]];
 


\end{thebibliography}

\vspace{-3mm}
\begin{thebibliography}{99}
\setlength{\itemsep}{0cm}
\vspace{-1mm}
\bibitem{E34CDR}
J-PARC E34 conceptual design report, \textbf{2011}.

\bibitem{Bennett2006}                                                                                           
G.W.~Bennett et al., Phys. Rev. D {\bf 73}, 072003 (2006).                                                      

\bibitem{FNAL_E989}                                                                                           
in these proceedings.

\bibitem{Bakule2013}
P. Bakule et al., Prog. Theor. Exp. Phys., 103C01 (2013).

\bibitem{S1249-2013}
Submitted to Prog. Theor. Exp. Phys. (2014)

\bibitem{J-PARC_Mu_2013B}
To be appear in KEK-MSL report (2013)
 
 \end{thebibliography}

\begin{thebibliography}{99}
\setlength{\itemsep}{0cm}

\bibitem{Pauk:2014rta} 
  V.~Pauk and M.~Vanderhaeghen,
  arXiv:1401.0832 [hep-ph].

\bibitem{BPPxxx}
J.~Bijnens, E.~Pallante, J.~Prades,
Phys.\ Rev.\ Lett.\  {\bf {75}}, 1447  (1995) 
[Erratum-ibid.\  {\bf {75}}, 3781 (1995)];
Nucl.\ Phys.\ B {\bf {474}}, 379 (1996);
[Erratum-ibid.\ {\bf {626}}, 410 (2002)]

\bibitem{HKSxxx}
M.~Hayakawa, T.~Kinoshita, A.~I.~Sanda,
Phys.\ Rev.\ Lett.\  {\bf 75} (1995) 790;
Phys.\ Rev.\ D {\bf {54}}, 3137  (1996). 

\bibitem{HKxxx}
M.~Hayakawa, T.~Kinoshita,
Phys.\ Rev.\ D {\bf 57}, 465  (1998) 
[Erratum-ibid.\ D {\bf {66}}, 019902 (2002)]

\bibitem{MVxxx}
K.~Melnikov, A.~Vainshtein,
Phys.\ Rev.\ D {\bf {70}}, 113006 (2004).  


\bibitem{Prades:2009twxxx}
  J.~Prades, E.~de Rafael, A.~Vainshtein,
  [arXiv:0901.0306 [hep-ph]].
  
   \bibitem{Jegerlehner:2009ryxxx} 
  F.~Jegerlehner and A.~Nyffeler,
  Phys.\ Rept.\  {\bf {477}}, 1 (2009); 
  F.~Jegerlehner,
  Springer Tracts Mod.\ Phys.\  {\bf {226}} (2008) 1.
    
 \end{thebibliography}

\begin{thebibliography}{99}
\setlength{\itemsep}{0cm}

\bibitem{BES3H} M.~Ablikim et al., (BESIII Collaboration), Nucl. Instr. Meth. \textbf{A 614}, 345 (2010).
\bibitem{jpsidal} M.~Ablikim et al., (BESIII Collaboration), Phys. Rev. \textbf{D 89}, 092008 (2014).
\bibitem{jpsidalT} J.~Fu, H.~B.~Li, X.~Qin, and M.~Z.~Yang, Mod. Phys. Lett. \textbf{A 27}, 1250223 (2012).
\bibitem{beslumi2} M.~Ablikim et al., (BESIII Collaboration), Chin. Phys. \textbf{C 37}, 123001 (2013).
\bibitem{babarbelle} B.~Aubert et al. [BABAR~Collab.], Phys. Rev. \textbf{D80}, 052002 (2009).\\
                     S.~Uehara et al. [Belle Collaboration], Phys. Rev. \textbf{D86}, 092007 (2012).
\bibitem{ekhara} H.~Czy\.{z}, S.~Ivashyn, Nucl. Phys. B (Proc. Suppl.) 225-227, 131-134 (2012).
\bibitem{2octet} H.~Czy\.{z}, S.~Ivashyn, A.~Korchin, O.~Shekhovtsova, Phys. Rev. \textbf{D85}, 094010 (2012).
\bibitem{CLEO} J.~Gronberg et al. [CLEO Collaboration], Phys. Rev. \textbf{D 57}, 33 (1998).
\bibitem{yuping} Y.~Guo, \emph{$\gamma\gamma\rightarrow\pi\pi$ at BESIII}, in these proceedings.

    
 \end{thebibliography}

\begin{thebibliography}{99}
\setlength{\itemsep}{0cm}


\bibitem{kekb} S.~Kurokawa {\it et al.},
 Nucl.\ Instr.\ and Meth.\ A {\bf 499} (2003) 1. 
%
\bibitem{pep2} PEP-II Conceptual Design Report, SLAC-372, LBL-PUB-5303,
CALT-68-1715, UCRL-ID-106426, UC-IIRPA-91-01 (1991).
%
\bibitem{belle} A.~Abashian {\it et al.}, 
Nucl.\ Instr.\ and Meth.\ A {\bf 479} (2002) 117.
%
\bibitem{dd} G.~Pakhlova {\it et al.} (Belle Collaboration), 
Phys.\ Rev.\ Lett.\ {\bf 98} (2007) 092001.
%
\bibitem{ddpi}  G.~Pakhlova {\it et al.} (Belle Collaboration), 
Phys.\ Rev.\ Lett.\ {\bf 100} (2008) 062001.
%
\bibitem{dsds} G.~Pakhlova {\it et al.} (Belle Collaboration),  
Phys.\ Rev.\ D {\bf 83} (2011) 011101.
%
\bibitem{lambda} G.~Pakhlova {\it et al.} (Belle Collaboration),
Phys.\ Rev.\ Lett.\ {\bf 101} (2008) 172001.
%
\bibitem{pipijpsi} Z.~Q.~Liu {\it et al.} (Belle Collaboration), 
Phys.\ Rev.\ Lett.\ {\bf 110} (2013) 252002.
%
\bibitem{phipipi} C.~P.~Shen  {\it et al.} (Belle Collaboration),  
Phys.\ Rev.\ D {\bf 80} (2009) 031101.
%
\bibitem{belle2-tdr} T.~Abe {\it et al.}, 
Belle II technical design report, KEK Report 2010-1 (2010).

 \end{thebibliography}

\begin{thebibliography}{99}
\setlength{\itemsep}{0cm}

\bibitem{snd}
  M.N.Achasov {\it et.al.} (SND Collaboration)
  Phys.\ Rev.\  D {\bf 88} (2013) 054013;
   
\bibitem{cmd}
R.R.Akhmetshin {\it et.al.} (CMD-3 Collaboration)
Phys.\ Lett.\ B {\bf 723} (2013) 82.

\end{thebibliography}

\begin{thebibliography}{99}
\setlength{\itemsep}{0cm}

\bibitem{Hagiwara:2011af}
  K.~Hagiwara, R.~Liao, A.~D.~Martin, D.~Nomura and T.~Teubner,
  J.\ Phys.\ G {\bf 38} (2011) 085003
  {\tt [arXiv:1105.3149 [hep-ph]]}.

\bibitem{Kurz:2014wya}
  A.~Kurz, T.~Liu, P.~Marquard and M.~Steinhauser,
  {\tt arXiv:1403.6400 [hep-ph]}.

\bibitem{SE}
Talks given by S.~Eidelman, in these proceedings.

\bibitem{AH}
Talks given by A.~Hafner, in these proceedings.

\bibitem{SMuller}
S.E.~M\"uller, in {\tt arXiv:1406.4639 [hep-ph]}.

\bibitem{GS-Novosibirsk}
Talk given by G.~Solodov, in these proceedings.

\bibitem{BS-BELLE}
Talk given by B.~Shwartz, in these proceedings.

\bibitem{BK-ISR-BESIII}
Talk given by B.~Kloss, in these proceedings.

\bibitem{MB-HLS}
Talk given by M.~Benayoun, in these proceedings.

\bibitem{GH-BESIII}
Talk given by G.~Huang, in these proceedings.

\end{thebibliography}

\begin{thebibliography}{99}
\setlength{\itemsep}{0cm}

\bibitem{McGeorge:2007tg}
  J.~C.~McGeorge, J.~D.~Kellie, J.~R.~M.~Annand, J.~Ahrens, I.~Anthony, A.~Clarkson, E.~F.~McNicoll and P.~S.~Lumsden {\it et al.},
  Eur.\ Phys.\ J.\ A {\bf 37} (2008) 129
  [arXiv:0711.3443 [nucl-ex]].

\bibitem{Jankowiak:2006yc}
  A.~Jankowiak,
  Eur.\ Phys.\ J.\ A {\bf 28S1} (2006) 149.

\bibitem{Kaiser:2008zza}
  K.~H.~Kaiser, K.~Aulenbacher, O.~Chubarov, M.~Dehn, H.~Euteneuer, F.~Hagenbuck, R.~Herr and A.~Jankowiak {\it et al.},
  Nucl.\ Instrum.\ Meth.\ A {\bf 593} (2008) 159.

\bibitem{Starostin:2001zz}
  A.~Starostin {\it et al.}  [Crystal Ball Collaboration],
  Phys.\ Rev.\ C {\bf 64} (2001) 055205.

\bibitem{Novotny:1991ht}
  R.~Novotny [TAPS Collaboration],
  IEEE Trans.\ Nucl.\ Sci.\  {\bf 38} (1991) 379.

\bibitem{Nefkens:2014zlt}
  B.~M.~K.~Nefkens {\it et al.}  [A2 at MAMI Collaboration],
  arXiv:1405.4904 [hep-ex], submitted to Phys. Rev. C.
    
\bibitem{Oset:2008hp}
  E.~Oset, J.~R.~Pelaez and L.~Roca,
  Phys.\ Rev.\ D {\bf 77} (2008) 073001
  [arXiv:0801.2633 [hep-ph]].
  
\bibitem{Gauzzi:2012zz}
  P.~Gauzzi [KLOE-2 Collaboration],
  J.\ Phys.\ Conf.\ Ser.\  {\bf 349} (2012) 012002.
 
\bibitem{Jegerlehner:2009ry}
  F.~Jegerlehner and A.~Nyffeler,
  Phys.\ Rept.\  {\bf 477} (2009) 1
  [arXiv:0902.3360 [hep-ph]].
 
\bibitem{Berghauser:2011zz}
  H.~Berghauser, V.~Metag, A.~Starostin, P.~Aguar-Bartolome, L.~K.~Akasoy, J.~R.~M.~Annand, H.~J.~Arends and K.~Bantawa {\it et al.},
  Phys.\ Lett.\ B {\bf 701} (2011) 562.
    
\bibitem{Aguar-Bartolome:2013vpw1}
  P.~Aguar-Bartolome {\it et al.}  [A2 Collaboration],
  Phys.\ Rev.\ C {\bf 89} (2014) 044608
  [arXiv:1309.5648 [hep-ex]].
  
\bibitem{Bystritskiy:2006}
 Yu.~M.~Bystritskiy, V.~V.~Bytev and E.~A.~Kuraev, Phys.\ Rev.\ D {\bf 73} (2006) 054021.
    
 \end{thebibliography}

\begin{thebibliography}{99}
\setlength{\itemsep}{0cm}

\bibitem{Kampf:2005tz}
 K.~Kampf, M.~Knecht, J.~Novotn\'y, 
 Eur.\ Phys.\ J.\ C {\bf 46} (2006) 191.

\bibitem{Adlarson:2013eza}
 P.~Adlarson {\it et al.}  [WASA-at-COSY Collaboration],
 Phys.\ Lett.\ B {\bf 726} (2013) 187.

\bibitem{MeijerDrees:1992qb}
 R.~Meijer Drees {\it et al.} [SINDRUM--I Collaboration],
 Phys.\ Rev.\ D {\bf 45} (1992) 1439.

\bibitem{Abouzaid:2006kk}
E.~Abouzaid {\it et al.} [KTeV Collaboration], 
 Phys.\ Rev.\ D {\bf 75} (2007) 012004.

\bibitem{Aguar-Bartolome:2013vpw}
P.~Aguar--Bartolom\'e {\it et al.} [A2 Collaboration],
 Phys.\ Rev.\ C {\bf 89} (2014) 044608.

\bibitem{KLOE2:2011aa}
F.~Ambrosino {\it et al.} [KLOE-2 Collaboration],
 Phys.\ Lett.\ B {\bf 702} (2011) 324; Phys.\ Lett.\ B {\bf 675} (2009) 283.

\bibitem{HADES:2011ab}
G.~Agakishiev {\it et al.} [HADES Collaboration],
 Eur.\ Phys.\ J.\ A {\bf 48} (2012) 64.

\end{thebibliography}
\end{document}